\documentclass[a4paper,fleqn]{cas-sc}
\usepackage[T1]{fontenc}

\usepackage{amsmath}
\usepackage{amssymb}
\usepackage{amsthm}
\usepackage{mathtools}
\usepackage{bm}

\usepackage{graphicx}
\usepackage{array}
\usepackage{booktabs}
\usepackage{longtable}
\usepackage{microtype}
\usepackage{placeins}

\usepackage{xcolor}
\newif\ifrevision
\revisionfalse
\protected\long\def\rev#1{{\ifrevision\color{red}\fi #1}}

\usepackage[authoryear,round]{natbib}
\usepackage{hyperref}
\hypersetup{
	colorlinks=true,
	linkcolor=blue,
	citecolor=blue,
	urlcolor=blue
}

\numberwithin{equation}{section}

\newcommand{\RR}{\mathbb{R}}

\newcommand{\II}{\mathbf I}
\newcommand{\QQ}{\mathbf Q}
\newcommand{\EE}{\mathbf E}
\newcommand{\FF}{\mathbf F}
\newcommand{\CCt}{\mathbb C}
\newcommand{\AAA}{\mathsf A}
\newcommand{\PP}{\mathsf P}

\newcommand{\Tr}{\operatorname{Tr}}
\newcommand{\tr}{\operatorname{tr}}
\newcommand{\dev}{\operatorname{dev}}
\newcommand{\Stab}{\operatorname{Stab}}
\newcommand{\sym}{\operatorname{sym}}

\newcommand{\dd}{\mathrm d}
\newcommand{\kb}{k_{\mathrm B}}

\newcommand{\cO}{\mathcal O}
\newcommand{\cF}{\mathcal F}
\newcommand{\cV}{\mathcal V}
\newcommand{\cM}{\mathcal M}
\newcommand{\cI}{\mathcal I}
\newcommand{\cH}{\mathcal H}

\newcommand{\cW}{\mathcal W}
\newcommand{\cD}{\mathcal D}
\newcommand{\cT}{\mathcal T}
\newcommand{\cS}{\mathcal S}

\newcommand{\norm}[1]{\left\lVert #1\right\rVert}

\begin{document}

\shorttitle{Acoustic signatures of structural phase transitions}    

\shortauthors{P. Zhu et al.}  

\title [mode = title]{A Symmetry-Stratified Framework for Phase Transitions:
Thermodynamic States, Critical Dynamics, and Spectral Mechanical Signatures}  



%

\author[1]{Pengfei Zhu}[orcid=0000-0003-0431-1538] 
\cormark[1] 
\ead{pengfei.zhu@bam.de} 
\credit{Methodology, Investigation, Visualization, Writing -- original draft, Writing -- review \& editing}

\author[2,3]{Leiming Du} 
\cormark[1] 
\credit{Investigation, Writing -- review \& editing}

\cortext[1]{Corresponding authors}

\affiliation[1]{ organization={Bundesanstalt für Materialforschung und -prüfung (BAM)}, addressline={Unter den Eichen 87}, city={Berlin}, postcode={12205}, country={Germany} }

\affiliation[2]{ organization={Max Planck Institute for Sustainable Materials}, addressline={Max-Planck-Straße 1}, city={Düsseldorf}, postcode={40237}, country={Germany} }

\affiliation[3]{ organization={\rev{Department of Microelectronics, Delft University of Technology}}, addressline={Mekelweg 4}, city={Delft}, postcode={2628 CD}, country={Netherlands}}


\begin{abstract}[]
\rev{Structural symmetry, thermodynamic phase stability, and elastic-wave response provide related but distinct descriptions of a transforming solid. We formulate a symmetry-stratified framework that connects these descriptions through a finite-dimensional internal state and a parent-invariant free energy. Stabilizers classify structural states, whereas free-energy comparisons and stability conditions determine equilibrium branches and dissipative evolution. Equivariance ensures that the frozen elastic tangent inherits the structural stabilizer. Equality requires additional identifiability conditions. Little-group representations and degenerate perturbation theory then determine whether an acoustic doublet splits and the first nonzero splitting order $p$. On a selected continuous branch with order-parameter exponent $\beta$, a regular, observable acoustic gap scales as $|t|^{p\beta}$, provided critical self-energies and structural relaxation do not dominate the measurement. Explicit tensorial and cubic-to-tetragonal realizations yield linear and quadratic splitting channels, together with directionally protected degeneracies. Separate numerical benchmarks examine free-energy reconstruction, coexistence, elastic compatibility, finite-temperature spectra, and controlled critical readouts. Reanalysis of NiTi measurements illustrates effective elastic degeneracy, thermal-history dependence, and the predictive value of independently measured phase fractions. Most importantly, the framework identifies the constitutive and measurement assumptions required to infer structural changes from mechanical spectra.}
\end{abstract}



\maketitle


\section{Introduction}
\label{sec:scope}

Phase transitions are among the most fundamental mechanisms by which
materials acquire new structures, properties, and functional responses~\citep{Levitas1997}.
They arise in a broad range of physical systems, including magnetic
ordering, ferroelectricity, crystallization,
martensitic transformations, and structural instabilities in solids~\citep{Shi2013,AndersonBlount1965,Bhattacharya2004}.
These phenomena are characterized by changes in an underlying order parameter,
the emergence or disappearance of symmetry, and the reorganization
of the free-energy landscape~\citep{KirkpatrickThirumalai2015}. In solid materials, such changes alter the
constitutive symmetry, generate transformation strains, modify
elastic anisotropy, and change the propagation and polarization of
elastic waves~\citep{Nakanishi1980}. A fundamental objective is
therefore to establish how structural transformations emerge from
thermodynamic principles and how their signatures can be identified
through measurable mechanical responses.

The classical thermodynamic theory of phase transitions provides
the starting point for this problem. The Gibbs and Ehrenfest
descriptions characterize equilibrium transformations through
changes in thermodynamic potentials and their derivatives~\citep{Gibbs1878}. Landau
theory subsequently introduced a systematic description of
symmetry-breaking transitions in terms of order parameters and
symmetry-admissible free-energy expansions
\citep{ChaikinLubensky1995}. Nevertheless, conventional Landau models are commonly
formulated for a selected order parameter and a prescribed symmetry
class. Different phases or material systems may require different
representations, invariant expansions, and constitutive assumptions.
Moreover, a change in the symmetry of an order parameter does not
necessarily constitute a thermodynamic phase transition.
Statistical mechanics extends this picture by connecting phase
transitions to collective fluctuations and the thermodynamic limit.
The theory of critical phenomena and the renormalization group
explain why microscopically different systems may exhibit identical
critical exponents, and why mean-field predictions can fail near
continuous transitions~\citep{Kadanoff2009}. Dynamic critical phenomena further classify
relaxation and transport according to the conservation laws and
reversible or dissipative couplings of the relevant slow variables
\citep{HohenbergHalperin1977}.
However, a universality class does not by itself specify the
instantaneous material point group, the complete elastic tangent,
or the directional spectral response of a mechanically probed solid.
In particular, knowledge of an order-parameter exponent does not
automatically determine the exponent measured through an acoustic
mode splitting.

\begin{figure}[pos=!htbp]
	\centering
	\includegraphics[width=\linewidth]{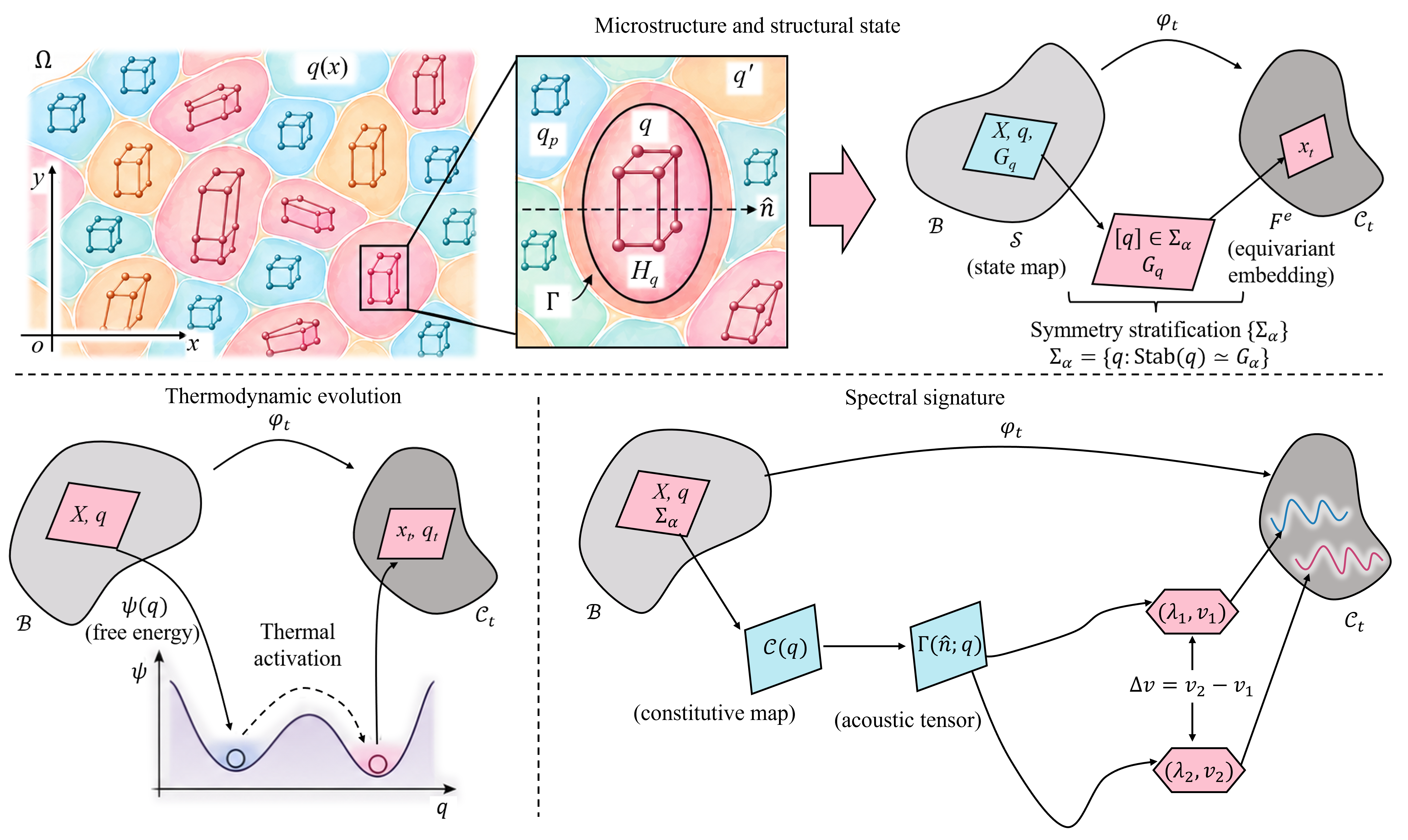}
	\caption{\textbf{Schematic illustration of the symmetry-stratified framework for structural phase transitions and their mechanical signatures} The upper panels illustrate the heterogeneous microstructure of a transforming solid, in which local structural states \(q(\mathbf X,t)\) belong to distinct symmetry strata characterized by their stabilizer subgroups \(G_q\). The magnified region highlights a representative structural variant, while the accompanying configuration-space diagram depicts the mapping from the reference body \(\mathcal B\) to the current configuration \(\mathcal C_t\) through a symmetry-stratified state representation. The lower-left panel illustrates the thermodynamic description, in which a parent-invariant free-energy landscape governs the stability, coexistence, and evolution of structural states under thermal activation or external driving. The lower-right panel establishes the mechanical connection between structural symmetry and observable acoustic spectra: the evolving internal state determines the tangent elastic modulus \(\mathbb C(q)\), from which the direction-dependent acoustic tensor \(\boldsymbol{\Gamma}(\hat{\mathbf n};q)\) and its polarization eigenpairs are obtained. Symmetry constraints and constitutive coupling determine whether initially degenerate acoustic branches remain protected or undergo splitting, providing a spectral signature of the underlying structural transformation. The schematic distinguishes structural classification, thermodynamic phase selection, and mechanical observability within a unified constitutive framework.}
	\label{fig10}
\end{figure}

A complementary line of development originates from continuum
thermodynamics and phase-field modeling. Ginzburg--Landau
functionals introduce gradient contributions that penalize spatial
variations of the order parameter, while Allen--Cahn and
Cahn--Hilliard-type evolution equations describe, respectively,
representative nonconserved and conserved dissipative dynamics
\citep{AllenCahn1979,CahnHilliard1958}.
More detailed approaches, including
phase-field-crystal models, retain aspects of crystalline periodicity
and couple structural ordering to elastic and defect-related
phenomena \citep{Elder2010,Elder2007,Emmerich2012}.
Their limitations are model dependent, and difficult to construct a
general correspondence between the local symmetry of an evolving
state, the symmetry of its elastic tangent, and experimentally
observable acoustic degeneracies.
In solid mechanics, structural phase transformations have also
been investigated through nonlinear elasticity, crystallographic
compatibility, energy minimization, and the theory of martensitic
microstructures. Variational models characterize transformations
through multiple energy wells and deformation gradients related
by crystallographic symmetry~\citep{VattreDenoual2019}. They have provided fundamental
insight into twinning, coherent interfaces, transformation
compatibility, and the reversibility of shape-memory materials
\citep{JamesHane2000,Ball2004,Bhattacharya2004}.
Group-theoretic classifications of ferroelastic transitions
similarly identify admissible symmetry-breaking pathways and
transformation variants~\citep{Troster2014}. However, crystallographic
classification, energy-well geometry, and nonlinear compatibility
do not independently establish the critical fluctuation dynamics
or the spectral observability of a given structural transition.
Furthermore, constitutive descriptions employing separate
phase-specific elastic laws may reproduce different phases without
explicitly deriving their changing material symmetries from a
common evolving internal state.
These developments suggest that several logically distinct
questions are often addressed separately.
Although substantial connections between these approaches already
exist, a remaining challenge is to formulate their common
symmetry structure without conflating their respective physical
criteria.

The present work addresses these issues through a
symmetry-stratified constitutive framework for phase transitions
with local, symmetry-bearing internal variables (see Figure~\ref{fig10}). Building on
Landau-type thermodynamics, the theory of critical phenomena,
and the internal-tensor description of evolving material symmetry
developed in Ref.~\citep{SourceManuscript}, we introduce a
finite-dimensional representation of the parent symmetry group
in which different structural states belong to distinct
group orbits and stabilizer strata. Two mechanical realizations are established, including an isotropic parent with a
symmetric traceless structural tensor and a cubic-to-tetragonal
ferroelastic transformation. The numerical simulation and an experimental NiTi measurements were used to validate the prediction capability of the proposed models.

The central mechanical contribution is to connect this
state-space description to the elastic and acoustic operators.
For a parent-equivariant constitutive law, the structural state
constrains the symmetry of the tangent stiffness, while the
Christoffel operator translates these restrictions into
direction-dependent wave speeds and polarizations. Representation
theory identifies symmetry-protected degeneracies, and
degenerate perturbation theory determines the first nonvanishing
order at which a structural perturbation becomes spectrally
observable.

\section{State variables, thermodynamics, and evolution}
\label{sec:framework}
The formulation uses two levels of description. The state representation
determines the symmetry of a given configuration. The free energy and
evolution law determine which configurations can occur under specified
mechanical and thermal controls.

\subsection{Parent symmetry and structural state}
\label{sec:geometry}

Let $\Omega\subset\RR^d$ be a reference body, and let $G_0$ be a compact
parent point group or compact internal symmetry group. For broken spatial
translations, spatially modulated phases, and noncompact groups, additional
structure must be supplied. The following construction concerns a compact
group acting on a finite-dimensional local state~\citep{SartoriValente2003}.

Choose a finite-dimensional real vector space with orthogonal representation
\begin{equation}
	\cV=\bigoplus_{a=1}^m V_a,\qquad
	\rho:G_0\longrightarrow O(\cV),\qquad
	\bm\eta(\mathbf X,t)=(\bm\eta_1,\ldots,\bm\eta_m).
	\label{eq:representation}
\end{equation}
For complex fields, take their realification, or equivalently allow a unitary
representation and complex-valued invariants. Under a material
transformation $g$, the state changes as
\begin{equation}
	\bm\eta\mapsto\rho(g)\bm\eta.
\end{equation}
\rev{A structural tensor, magnetic vector, complex condensate, or coupled
multiplet is a particular representation. In the mechanical formulations
below, $G_0$ is a material point group represented by orthogonal spatial
matrices, and $g$ denotes that spatial action in expressions such as
$g\EE g^T$ and $g\hat{\mathbf n}$. An abstract internal symmetry group
requires a separately specified action on the mechanical variables;
its representation on $\cV$ alone does not define these transformations.} Spatial transformation rules for
$\nabla\bm\eta$ must rotate both the order-parameter indices and the spatial
derivative index. In a material description, a superposed spatial observer
rotation leaves referential structural coordinates unchanged; frame
indifference and parent-group covariance are distinct requirements.

Throughout, $\rho(g_1g_2)=\rho(g_1)\rho(g_2)$ is a left action.
For a second-order tensor we accordingly use the active material action
$\QQ\mapsto g\QQ g^T$ and $\EE\mapsto g\EE g^T$.
This convention makes successive rotations consistent with the
conjugation rule for stabilizers used below.

Define the stabilizer and group orbit by
\begin{align}
	H_{\bm\eta}&=\Stab_{G_0}(\bm\eta)
	=\{g\in G_0:\rho(g)\bm\eta=\bm\eta\},\label{eq:stab}\\
	\cO_{\bm\eta}&=\{\rho(g)\bm\eta:g\in G_0\}
	\simeq G_0/H_{\bm\eta}.\label{eq:orbit}
\end{align}
Stabilizers conjugated by the parent group correspond to the same
intrinsic symmetry type. The space decomposes into orbit types,
\begin{equation}
	\cV=\bigcup_{[H]}\cV_{(H)},\qquad
	\cV_{(H)}=\{\bm\eta:H_{\bm\eta}\text{ is conjugate to }H\}.
	\label{eq:orbits}
\end{equation}
The orbit space $\cV/G_0$ encodes intrinsic amplitudes and invariant
shape variables but suppresses the material orientation of an orbit.
The symmetry class is unchanged under a parent-group reorientation: if $\bm\eta_2=\rho(g)\bm\eta_1$, then
	$H_{\bm\eta_2}=gH_{\bm\eta_1}g^{-1}$. Thus a state path contained in one
	orbit cannot change intrinsic symmetry type.
For any $h\in H_{\bm\eta_1}$,
	$\rho(ghg^{-1})\bm\eta_2
	=\rho(g)\rho(h)\bm\eta_1=\bm\eta_2$.
	Conversely, conjugation by $g^{-1}$ maps any stabilizer of
	$\bm\eta_2$ into the stabilizer of $\bm\eta_1$.
Consequently three paths must be distinguished:
(1) orbit motion: $\bm\eta(t)=\rho(g(t))\bm\eta_*$,
(2)	within-stratum deformation: $[H_{\bm\eta(t)}]=[H], \bm\eta(t)\notin\cO_{\bm\eta_*}$,
(3) symmetry-class change: $[H_{\bm\eta(t_-)}]\ne[H_{\bm\eta(t_+)}]$.
The group label can change abruptly when eigenvalues meet even if
$\bm\eta(t)$ and all smooth constitutive functions remain continuous.
It follows that an orbit-type crossing is not, by itself, a thermodynamic singularity.

If separate variables $\bm\eta_a$ carry independent structural information,
the stabilizer of their direct sum is
\begin{equation}
	H_{\bm\eta}=\bigcap_a H_{\bm\eta_a}.
	\label{eq:intersection}
\end{equation}
The instantaneous symmetry of a mechanically loaded state can be even
smaller, because the strain, prestress, gradients, and boundary conditions
must also be preserved. For a bulk state with strain $\EE$, for example,
\begin{equation}
	H_{\mathrm{loaded}}
	=H_{\bm\eta}\cap H_{\EE}\cap H_{\bm\xi},
	\label{eq:loaded}
\end{equation}
when $\bm\xi$ denotes all other symmetry-bearing internal variables.
Equations \eqref{eq:stab}--\eqref{eq:loaded} should not be confused with
an assertion that a specimen has the same global symmetry as one of its
local material points.

For the isotropic parent $G_0=SO(3)$, a transparent specialization is
\begin{equation}
	\QQ\in\mathrm{Sym}_0(3)
	=\{\mathbf A\in\RR^{3\times3}:\mathbf A^T=\mathbf A,
	\ \tr\mathbf A=0\},\qquad
	\QQ\mapsto g\QQ g^T.
	\label{eq:Qspace}
\end{equation}
With
\begin{equation}
	J_2=\tr\QQ^2,\quad J_3=\tr\QQ^3,
	\quad 6J_3^2\leq J_2^3,
	\label{eq:Qinv}
\end{equation}
its eigenvalue multiplicities lead to the well-known structure
\begin{equation}
	\rev{\begin{aligned}
\QQ=\mathbf0 &: \ SO(3),\\
\QQ=s(\mathbf n\otimes\mathbf n-\II/3),\quad s\ne0 &: \ O(2),\\
q_i\ne q_j\quad\text{for all }i\ne j &: \ D_2 .
\end{aligned}}
	\label{eq:Qclasses}
\end{equation}
Here $O(2)$ denotes the appropriate embedded subgroup of proper
three-dimensional rotations preserving an unoriented axis. The quantity
\begin{equation}
	\beta_Q^2=1-\frac{6J_3^2}{J_2^3},\quad J_2>0,
\end{equation}
measures biaxiality but is undefined at the isotropic origin. This
representation does not resolve cubic, chiral, and other symmetry classes
which require different or higher-order descriptors. That limitation is a
principal motivation for Eq.~\eqref{eq:representation}.

The constraint in Eq.~\eqref{eq:Qinv} follows directly from the
characteristic polynomial, rather than from an additional constitutive
assumption. Since $\tr\QQ=0$, the Cayley--Hamilton identity is
\begin{equation}
 \QQ^3-\frac{J_2}{2}\QQ-\frac{J_3}{3}\II=0,
 \qquad
 \prod_{i<j}(q_i-q_j)^2
 =\frac12\big(J_2^3-6J_3^2\big)\ge0.
 \label{eq:Q-discriminant-derivation}
\end{equation}
The second identity is the discriminant of
$x^3-J_2x/2-J_3/3$. Equality corresponds to a repeated eigenvalue;
for a nonzero traceless tensor this is the uniaxial case. Three distinct
eigenvalues leave only the proper rotations preserving the three
principal axes, giving $D_2$. At the origin every rotation is admissible.
This also explains why biaxiality loses its definition as $J_2\to0$
without a prescribed approach direction~\citep{AllenderLonga2008}.

\subsection{Free energy and equilibrium phase transitions}
\label{sec:thermal}

Consider a small-strain representative for which a general local functional
can be written
\begin{equation}
	\cF[\bm\eta,\mathbf u;T,\lambda]
	=\int_\Omega\left\{
	f_0(\bm\eta;T,\lambda)
	+\frac12 K^{ij}_{ab}(\bm\eta)
	\partial_i\eta_a\,\partial_j\eta_b
	+W(\EE(\mathbf u),\bm\eta;T,\lambda)
	\right\}\dd V
	-\cW_{\mathrm{ext}},
	\label{eq:Fgeneral}
\end{equation}
where $\EE=\sym\nabla\mathbf u$ and $\lambda$ is an externally controlled
parameter (temperature, pressure, applied field, or loading). The local
coefficients are subject to the parent covariance condition
\begin{equation}
		f_0(\rho(g)\bm\eta;T,\lambda)=f_0(\bm\eta;T,\lambda),\qquad
		W(g\EE g^T,\rho(g)\bm\eta;T,\lambda)
		=W(\EE,\bm\eta;T,\lambda),
	\label{eq:invariance}
\end{equation}
with the corresponding transformed-gradient condition.

Here a control held fixed in a parent-invariance statement must
itself preserve the parent symmetry, as temperature or a scalar
coefficient does. An oriented applied field or stress instead
transforms jointly with the material variables; if it is held fixed,
the relevant invariance group is its residual subgroup. This avoids
attributing full-parent invariance to an explicitly biased specimen. Terms linear in
strain and terms quadratic in strain describe different physics:
\begin{equation}
	W(\EE,\bm\eta)
	=\frac12(\EE-\EE^*(\bm\eta)):
	\CCt(\bm\eta):
	(\EE-\EE^*(\bm\eta))+
	W_{\mathrm{higher}}.
	\label{eq:eigenstrain}
\end{equation}
The transformation strain $\EE^*$ can produce ferroelastic variants;
$\CCt(\bm\eta)$ controls incremental elasticity. In applications with
finite strain one instead constructs a frame-indifferent energy
$\Psi(\mathbf C,\bm\eta,\nabla\bm\eta)$ with
$\mathbf C=\FF^T\FF$ and computes the relevant push-forward tangent.

To identify the field equations, write $f$ for the complete local
integrand in Eq.~\eqref{eq:Fgeneral}, assume
$K^{ij}_{ab}=K^{ji}_{ba}$, and set
$\pi_{ai}=\partial f/\partial(\partial_i\eta_a)$.
Integration by parts gives the first variation
\begin{equation}
 \delta\cF={}\int_\Omega
 \left[(\partial_{\eta_a}f-\partial_i\pi_{ai})\delta\eta_a
       +\bm\sigma:\sym\nabla\delta\mathbf u\right]\dd V\qquad
 +\int_{\partial\Omega}n_i\pi_{ai}\delta\eta_a\,\dd S
 -\delta\cW_{\rm ext},\qquad
 \bm\sigma=\partial_{\EE}W.
 \label{eq:first-variation-derivation}
\end{equation}
Thus the natural structural boundary condition is $n_i\pi_{ai}=0$
when no structural surface power is prescribed. For a body force
$\mathbf b$ and an applied traction $\overline{\mathbf t}$, mechanical
stationarity gives $\nabla\cdot\bm\sigma+\mathbf b=0$ and
$\bm\sigma\mathbf n=\overline{\mathbf t}$ on the traction boundary.
These are the usual variational and microforce ingredients of gradient
theories~\citep{AllenCahn1979,CahnHilliard1958,FriedGurtin1999}.
In particular, a state-dependent gradient stiffness contributes
\begin{equation}
 \frac{\delta\cF}{\delta\eta_c}
 =\partial_{\eta_c}(f_0+W)
 +\frac12\partial_{\eta_c}K^{ij}_{ab}
       \partial_i\eta_a\partial_j\eta_b
 -\partial_i(K^{ij}_{cb}\partial_j\eta_b).
 \label{eq:variable-K-derivation}
\end{equation}
The derivative of $W$ is taken at fixed strain. If the displacement is
first minimized, the derivative of the reduced functional is evaluated
at that mechanical solution; its second derivative generally contains
an additional relaxation correction.

The mean-field equilibrium prediction is given by
\begin{equation}
	F_{\mathrm{MF}}(T,\lambda)
	=\inf_{\bm\eta,\mathbf u}\cF[\bm\eta,\mathbf u;T,\lambda].
	\label{eq:MF}
\end{equation}
Comparing the globally minimizing branches supplies mean-field
transition and coexistence points. Metastability requires retaining
the other local minima and their Hessians, which are not recorded
by the infimum alone. At nonzero
temperature, the equilibrium partition function of a specified finite,
regularized statistical model can be written schematically as
\begin{equation}
	Z_{L}(T,\lambda,h)
	=\int\mathcal D\bm\eta\,\mathcal D\mathbf u\,
	\exp\left[-\frac{\cF_L[\bm\eta,\mathbf u;T,\lambda]
		-h\cdot\int_{\Omega_L}\bm\eta\,\dd V}{\kb T}\right].
	\label{eq:Z}
\end{equation}

A spatial discretization or short-distance cutoff, together with the
mechanical ensemble and treatment of rigid displacement modes, is implicit
in this measure. The functional in the Boltzmann weight must be a
Hamiltonian at that specified coarse-graining scale. An already fully
fluctuation-renormalized thermodynamic potential cannot be used as a new
bare Hamiltonian without controlling the resulting double counting
of fluctuations~\citep{Fisher1998}.
The thermodynamic free-energy density is defined by
\begin{equation}
	f_{\mathrm{eq}}(T,\lambda,h)
	=-\lim_{L\to\infty}\frac{\kb T}{|\Omega_L|}
	\log Z_L(T,\lambda,h),
	\label{eq:thermolimit}
\end{equation}
when the limit exists. Genuine equilibrium phase transitions correspond
to singularities or nonanalyticities of the thermodynamic potential or its
appropriate derivatives; a sharp singularity generally emerges only in an
infinite-system limit. Spontaneous symmetry breaking is obtained, where
appropriate, using a prescribed order of limits, for instance
\begin{equation}
	\bm\eta_{\mathrm{sp}}
	=\lim_{h\to0^+}\lim_{L\to\infty}
	\frac{1}{|\Omega_L|}\left\langle\int_{\Omega_L}
	\bm\eta\,\dd V\right\rangle_{L,h}.
	\label{eq:SSB}
\end{equation}
The small conjugate field selects a branch and prevents a finite-volume
symmetric mixture from artificially averaging a broken-symmetry order
parameter to zero.
Changes in the stabilizer of the structural state and thermodynamic
phase transitions are distinct phenomena, and neither necessarily
implies the other. A change in structural symmetry does not, by
itself, establish a nonanalyticity in the thermodynamic potential.
Conversely, a thermodynamic phase transition may occur without
a change in the stabilizer of the structural state.
For the first implication, a smooth externally prescribed path
	$\bm\eta(t)=(t,0)$ for a vector representation has an enhanced stabilizer
	at $t=0$, whereas a smooth analytic energy evaluated on that prescribed
	path need not possess a thermodynamic singularity. The second implication
	fails for a liquid--gas first-order transition, whose two homogeneous
	phases can possess the same spatial rotational symmetry and differ by
	scalar density. Therefore the two criteria cannot be identified.

A complete characterization of an equilibrium branch therefore requires
more than its structural symmetry. In addition to the stabilizer of the
order parameter, the associated free-energy branch, coexistence and
stability conditions, and relevant material response must be considered.
These quantities collectively distinguish thermodynamically different
states that may share the same residual symmetry.
A thermodynamic phase and an orbit-type stratum are thus distinct
mathematical objects whose relationship is physically informative.

Let $I_1(\bm\eta),\ldots,I_r(\bm\eta)$ denote a symmetry-admissible
collection of scalar invariants. A smooth parent-invariant polynomial
potential can be written to specified order as
\begin{equation}
	f_0(\bm\eta;\lambda)
	=a_0(\lambda)+\sum_i a_i(\lambda)I_i(\bm\eta)
	+\sum_{ij}a_{ij}(\lambda)I_i(\bm\eta)I_j(\bm\eta)+\cdots.
	\label{eq:integrity}
\end{equation}
An arbitrarily selected finite list is not claimed to be a complete
integrity basis. Cubic invariants can permit first-order transitions,
while a vanishing cubic term often allows a continuous mean-field onset. Fluctuations, strain accommodation, or competing branches can change the
order of the actual transition.

\subsection{Dissipative evolution and stability}
\label{sec:dynamics}

For fixed mechanical and thermal controls, define the functional force
\begin{equation}
	\mathbf A_\eta=-\frac{\delta\cF}{\delta\bm\eta}
	=-\frac{\partial f}{\partial\bm\eta}
	+\partial_i\left(\frac{\partial f}{\partial(\partial_i\bm\eta)}\right).
	\label{eq:Aeta}
\end{equation}

For example, for constant scalar $\kappa$ and
$f=f_{\rm loc}+\kappa|\nabla\bm\eta|^2/2$, this reduces to
$\mathbf A_\eta=-\partial_{\bm\eta}f_{\rm loc}+\kappa\nabla^2\bm\eta$.
At fixed controls and mechanical equilibrium, the chain rule gives
$\dot\cF=-\int_\Omega\mathbf A_\eta\cdot\dot{\bm\eta}\,\dd V$
after the structural boundary term has vanished. This identifies the
force--rate pair used to construct the dissipative law.
Suppose the allowed rates lie in a linear constrained tangent space
$T_{\bm\eta}\cM$, with orthogonal projection $\PP_{\bm\eta}$.
For purely relaxational nonconserved kinetics consider
\begin{equation}
	\partial_t\bm\eta
	=\PP_{\bm\eta}\mathsf M_\eta\PP_{\bm\eta}\mathbf A_\eta,
	\qquad \mathsf M_\eta=\mathsf M_\eta^T\succeq0.
	\label{eq:Onsager}
\end{equation}

For a positive definite mobility, this law can equivalently be
obtained by minimizing the Rayleigh functional over admissible rates:
\begin{equation}
 \mathcal R[\mathbf v]=\int_\Omega
 \left[\frac12\mathbf v\cdot\mathsf M_T^{-1}\mathbf v
              -\mathbf A_\eta\cdot\mathbf v\right]\dd V,
 \quad \mathbf v\in T_{\bm\eta}\cM,
 \quad \mathsf M_T=
 (\PP_{\bm\eta}\mathsf M_\eta\PP_{\bm\eta})|_{T_{\bm\eta}\cM}.
 \label{eq:Rayleigh-derivation}
\end{equation}
Stationarity yields $\mathbf v=\mathsf M_T\PP_{\bm\eta}\mathbf A_\eta$.
The semidefinite case restricts rates to the range of $\mathsf M_T$.
Using the orthogonality of $\PP_{\bm\eta}$ in the chain rule then
gives Eq.~\eqref{eq:dissipation}.
The mobility must transform covariantly under the parent group. For
homogeneous thermal and mechanical controls with vanishing boundary
structural power,
\begin{equation}
	\frac{\dd\cF}{\dd t}
	=-\int_\Omega
	(\PP_{\bm\eta}\mathbf A_\eta)\cdot
	\mathsf M_\eta
	(\PP_{\bm\eta}\mathbf A_\eta)\,\dd V\leq0.
	\label{eq:dissipation}
\end{equation}
For $\QQ\in\mathrm{Sym}_0(3)$ the projection is
$\PP_0\mathbf A=\sym\mathbf A-\tr(\mathbf A)\II/3$.
If external strain, stress, field, or temperature is driven, the total
free energy can increase by external work; Eq.~\eqref{eq:dissipation}
must not be used to assert monotonic energy decrease for the driven
problem.

A conserved scalar density $n$ cannot obey an unconstrained Model-A-type
law. With chemical potential $\mu_n=\delta\cF/\delta n$, a standard
isothermal choice is
\begin{equation}
	\partial_t n=\nabla\cdot\left(M_n\nabla\mu_n\right),
	\qquad M_n\geq0,
	\label{eq:CH}
\end{equation}
under no-flux boundary conditions. The mass is preserved and
\begin{equation}
	\frac{\dd\cF}{\dd t}
	=-\int_\Omega M_n|\nabla\mu_n|^2\,\dd V\leq0.
\end{equation}
A theory with several slow modes may require a block kinetic operator,
reversible Poisson-bracket couplings, momentum transport, and thermal
transport. Parent symmetry constrains these blocks but does not fix their
magnitudes or the dynamic universality class. This follows the
Model-A/Model-B distinction and its generalizations
\citep{HohenbergHalperin1977}.

For a constant mobility in an unconstrained nonconserved order-parameter
space, a canonical stochastic closure is
\begin{align}
	&\partial_t\bm\eta=-\mathsf M_\eta
	\frac{\delta\cF}{\delta\bm\eta}+\bm\zeta,\label{eq:stochastic}\\
	&\langle\zeta_a(\mathbf X,t)\rangle=0,\qquad
	\langle\zeta_a(\mathbf X,t)\zeta_b(\mathbf X',t')\rangle
	=2\kb T(\mathsf M_\eta)_{ab}
	\delta(\mathbf X-\mathbf X')\delta(t-t').\label{eq:FDT}
\end{align}

The factor $2\kb T$ is fixed by detailed balance. On a finite
discretization with constant symmetric mobility, the probability
density satisfies the Fokker--Planck equation
\begin{equation}
 \partial_t\mathcal P
 =\nabla_{\bm\eta}\cdot\left\{
 \mathsf M_\eta\left[(\nabla_{\bm\eta}\cF)\mathcal P
                         +\kb T\nabla_{\bm\eta}\mathcal P\right]\right\}.
 \label{eq:Fokker-Planck-derivation}
\end{equation}
For $\mathcal P_{\rm eq}\propto\exp[-\cF/(\kb T)]$, the quantity
in square brackets vanishes identically. This establishes the stated
equilibrium measure for additive noise, rather than simply appending
a noise term to deterministic kinetics~\citep{Ohzeki2026}.
For state-dependent mobility, the stochastic calculus and corresponding
drift must be specified to obtain the intended equilibrium measure.
For a conserved scalar, noise must enter as a flux divergence:
\begin{align}
	\partial_t n&=\nabla\cdot(M_n\nabla\mu_n+\bm\xi),\\
	\langle\xi_i(\mathbf X,t)\xi_j(\mathbf X',t')\rangle
	&=2\kb T M_n\delta_{ij}
	\delta(\mathbf X-\mathbf X')\delta(t-t').
\end{align}
These expressions illustrate the required statistical extension.
Additional hydrodynamic or inertial variables demand corresponding
noise and reversible couplings.

Let $\bm\eta_*$ be a homogeneous stationary state and let
$\cH_\eta$ denote the Hessian of its local potential on the
admissible state space. With scalar mobility $\Gamma$ and isotropic
gradient coefficient $\kappa$, an eigenmode $\mathbf v_a$ of
$\cH_\eta$ with curvature $h_a$ has Fourier growth exponent
\begin{equation}
	\delta\bm\eta=\mathbf v_a e^{i\mathbf q\cdot\mathbf X+\nu_a(\mathbf q)t},
	\qquad
	\nu_a(\mathbf q)=-\Gamma(h_a+\kappa|\mathbf q|^2).
	\label{eq:relax}
\end{equation}
For $h_a>0$, $\tau_a(\mathbf q)=
[\Gamma(h_a+\kappa q^2)]^{-1}$;
for $h_a<0$, sufficiently long wavelengths are linearly unstable.
An orientational Goldstone direction of an unbiased continuous orbit
has zero homogeneous curvature, not the same positive curvature as
an amplitude mode. For conserved dynamics, the analogous simple scalar
linearization gives $\nu=-M_nq^2(h_n+\kappa q^2)$.

Indeed, the linearized nonconserved equation is
$\partial_t\delta\bm\eta=-\Gamma(\cH_\eta-\kappa\nabla^2)
\delta\bm\eta$; inserting the Fourier ansatz yields
Eq.~\eqref{eq:relax}. Here $\cH_\eta$ is evaluated under the specified
mechanical constraint. If strain is relaxed, the appropriate Hessian
is that of the mechanically reduced energy and may be nonlocal and
direction dependent. In that case $h_a+\kappa q^2$ must be replaced
by the eigenvalues of the complete Fourier Hessian.

For a deterministic equivariant ODE/PDE with uniqueness and exactly
parent-symmetric initial state and controls, the solution retains that
symmetry. \rev{For example, if $\bm\eta=0$ is stationary for all imposed controls,
then $\bm\eta(0)=0$ implies $\bm\eta(t)=0$ in a homogeneous
polynomial gradient flow, even after that stationary state becomes
linearly unstable. Parent symmetry by itself need not make the origin
stationary when the representation contains invariant scalar components.} Observed ordering requires fluctuations, a seed, boundaries,
external bias, or a limiting procedure selecting a branch.

\subsection{Critical scaling and fluctuations}
\label{sec:critical}

Near a continuous equilibrium transition, define the reduced distance
$t=(\lambda-\lambda_c)/\lambda_0$, with a nonzero fixed scale
$\lambda_0$. If an ordered branch possesses a local order parameter,
write the asymptotic laws
\begin{equation}
	\norm{\bm\eta_{\rm sp}}\sim B|t|^\beta,
	\qquad\xi\sim\xi_0|t|^{-\nu},
	\qquad\chi\sim\chi_0|t|^{-\gamma},
	\qquad\rev{\tau\sim\tau_0(\xi/\xi_0)^z},
	\label{eq:critical}
\end{equation}
where the last relation pertains to the relevant slow dynamic sector.
The exponents need not take their mean-field values. They depend on
spatial dimension, range of interactions, symmetry and number of active
components, conserved quantities, and other couplings. A symmetry
stabilizer does not uniquely determine an equilibrium or dynamic
universality class.

The distinction between symmetry and critical scaling is already
visible in the Gaussian approximation. For one amplitude mode with
curvature $h>0$, write the quadratic energy and source as
\begin{equation}
 \cF^{(2)}=\frac12\int\frac{\dd^dq}{(2\pi)^d}
 (h+\kappa q^2)|\widehat\varphi(\mathbf q)|^2
 -\int j\varphi\,\dd V.
 \label{eq:Gaussian-energy-derivation}
\end{equation}
Minimizing at fixed $j$ gives
$\widehat\varphi=\widehat j/(h+\kappa q^2)$, so
$\chi(0)=h^{-1}$ and $\xi=(\kappa/h)^{1/2}$. The same pole in
Eq.~\eqref{eq:relax} gives $\tau(q=0)=(\Gamma h)^{-1}
=\xi^2/(\Gamma\kappa)$, hence the Gaussian Model-A value $z=2$.
For the selected scalar branch of $a\varphi^2/2+c\varphi^4/4$,
$\varphi_*^2=-a/c$ below onset; if $a$ varies linearly with $t$,
this additionally gives $\beta=1/2$. These are approximation-specific
results~\citep{ChaikinLubensky1995,HohenbergHalperin1977,LiuTauber2016}.

An explicit route beyond mean field is to define the dimensionless
Euclidean statistical action $S=\cF/(\kb T)$ and introduce an infrared
regulator $R_k$. Its scale-dependent effective average action obeys
the standard functional renormalization-group equation
\begin{equation}
	\partial_k\Gamma_k[\bm\phi]
	=\frac12\Tr\left[
	(\Gamma_k^{(2)}[\bm\phi]+R_k)^{-1}
	\partial_k R_k\right],
	\label{eq:Wetterich}
\end{equation}
for a bosonic/classical field setting with a suitable regularization
\citep{Wetterich1993}.

For clarity, introduce
$\Delta S_k=\tfrac12\bm\phi\cdot R_k\bm\phi$ in the regulated
partition function, let $W_k[\mathbf J]=\log Z_k[\mathbf J]$, and define
$\Gamma_k[\bm\phi]=\mathbf J\cdot\bm\phi-W_k[\mathbf J]
-\Delta S_k[\bm\phi]$, with $\bm\phi=\delta W_k/\delta\mathbf J$.
The connected covariance is
$W_k^{(2)}=(\Gamma_k^{(2)}+R_k)^{-1}$.
Differentiation at fixed $\bm\phi$ cancels the mean-field part of
$\partial_k\Delta S_k$ and leaves
$\partial_k\Gamma_k=\tfrac12\Tr(W_k^{(2)}\partial_kR_k)$,
which is Eq.~\eqref{eq:Wetterich}. This construction specifies which
fluctuations have been integrated out. The group acts consistently at every coarse-graining
scale if $R_k$ preserves its symmetry. The running effective potential
and kinetic coefficients can renormalize the phase diagram and critical
exponents; Eq.~\eqref{eq:Wetterich} is a formal research route, not an
assertion that a specified truncation is automatically controlled.
A key question for the proposed symmetry-resolved framework is which
invariant representations remain active at a stable fixed point and
which operators are relevant, irrelevant, or dangerously irrelevant.
A proposed prediction of a universal exponent must survive this analysis
or a corresponding benchmark against known universality classes.

\section{Constitutive symmetry and acoustic signatures}
\label{sec:spectrum}
We now consider small-amplitude waves about a homogeneous equilibrium
state. We first hold the internal state fixed during the probe and derive
the directional selection rules for the elastic tangent. The final
subsection allows the internal state to relax in response to the wave.

\subsection{A parent-invariant constitutive map}
At a stress-free, homogeneous base state let the strain energy have
incremental tangent
\begin{equation}
	\CCt(\bm\eta)
	=\left.\frac{\partial^2W(\EE,\bm\eta)}
	{\partial\EE\,\partial\EE}\right|_{\EE=\EE_0(\bm\eta)}.
	\label{eq:tangent}
\end{equation}
Assume $\EE_0$ transforms covariantly under the parent group.
\rev{Then $\CCt$ is an equivariant map. Define its stabilizer within the
parent group by
\[
G_{\CCt(\bm\eta)}=\{g\in G_0:g\cdot\CCt(\bm\eta)=\CCt(\bm\eta)\}.
\]
The action on $\CCt$ is the natural fourth-order tensor action.
This definition distinguishes the inherited parent-group symmetry from
any additional symmetry of the elastic tensor outside $G_0$.}

Equivariance constrains the constitutive symmetry inherited from the structural state. For a parent-equivariant constitutive map $\bm\eta\mapsto\CCt(\bm\eta)$,
	\begin{equation}
		H_{\bm\eta}\subseteq G_{\CCt(\bm\eta)}.
		\label{eq:inherit}
	\end{equation}
	If the map is injective on the relevant states, then
	$H_{\bm\eta}=G_{\CCt(\bm\eta)}$, assuming the same group action and
	no independent structural variables.

The equivariance needed here follows by differentiating the energy
invariance twice. For arbitrary symmetric increments $\mathbf H$ and
$\mathbf K$,
\begin{equation}
 (g\mathbf H g^T):\CCt(\rho(g)\bm\eta):(g\mathbf K g^T)
 =\mathbf H:\CCt(\bm\eta):\mathbf K.
 \label{eq:tangent-covariance-derivation}
\end{equation}
Thus $C_{ijkl}(\rho(g)\bm\eta)=g_{ia}g_{jb}g_{kc}g_{ld}
C_{abcd}(\bm\eta)$, provided the evaluation strain also transforms
covariantly. The comparison is between stabilizers inside the same
parent group. An elastic tensor can possess additional symmetries
outside a finite structural parent~\citep{CowinMehrabadi1987,CowinMehrabadi1992}.
If $h\in H_{\bm\eta}$, equivariance yields
	$h\cdot\CCt(\bm\eta)=\CCt(\rho(h)\bm\eta)=\CCt(\bm\eta)$.
	Conversely, if $h$ preserves the tangent, equivariance implies
	$\CCt(\rho(h)\bm\eta)=\CCt(\bm\eta)$. Injectivity gives
	$\rho(h)\bm\eta=\bm\eta$.
\rev{Equality can fail when the constitutive map does not distinguish a
state from one of its parent-group transforms, for example when the
relevant coupling vanishes. Even when equality holds for the complete
tangent, an incomplete measurement can fail to distinguish structural
states; this is a separate loss of observational identifiability.}
The symmetry of a tangent evaluated under general finite deformation
must also account for the deformation and prestress.

\subsection{Christoffel/acoustic tensor and little-group selection}
\rev{For a classical local elastic solid with density $\rho_0$ and a frozen,
homogeneous tangent $C_{ijkl}$, evaluated under the specified thermal
constraint, define}
\begin{equation}
	A_{ik}(\hat{\mathbf n};\bm\eta)
	=C_{ijkl}(\bm\eta)\hat n_j\hat n_l,
	\qquad
	\AAA\mathbf p_m=\rho_0c_m^2\mathbf p_m,
	\qquad\omega_m=k c_m.
	\label{eq:acoustic}
\end{equation}

\rev{The thermal constraint can be expressed explicitly for a homogeneous
Helmholtz density $f(\EE,\bm\eta,T)$ per reference volume, with $\bm\eta$
frozen. Let $\mathfrak s=-\partial_T f$ be the entropy density,
$c_E=T(\partial_T\mathfrak s)_{\EE,\bm\eta}>0$, and
$\bm\beta_{\rm th}=(\partial_T\bm\sigma)_{\EE,\bm\eta}$.
The Maxwell relation gives
$\dd\mathfrak s=(c_E/T)\dd T-\bm\beta_{\rm th}:\dd\EE$.
Eliminating $\dd T$ at constant entropy then yields}
\begin{equation}
 \rev{\CCt^{S}=\CCt^{T}
       +\frac{T}{c_E}\bm\beta_{\rm th}\otimes\bm\beta_{\rm th}.}
 \label{eq:thermal-tangent-conversion}
\end{equation}
\rev{The superscripts $T$ and $S$ denote isothermal and isentropic
frozen-state tangents under the same mechanical constraint. The tensor
product acts as
$(\bm\beta_{\rm th}\otimes\bm\beta_{\rm th}):\mathbf H
=\bm\beta_{\rm th}(\bm\beta_{\rm th}:\mathbf H)$.
Thermal exchange during a wave period determines which constraint
approximates the measurement. Symmetry-consistent thermal coefficients
preserve the inherited group and its protected doublets. Applying
this conversion quantitatively requires thermal properties in
addition to the elastic and structural data.}

To obtain this eigenproblem, insert
$\mathbf u=\mathbf p\exp[i(k\hat{\mathbf n}\cdot\mathbf X-\omega t)]$
into $\rho_0\ddot u_i=\partial_j(C_{ijkl}E_{kl})$.
The minor symmetries of $C$ reduce the symmetric strain gradient to
$k^2C_{ijkl}\hat n_j\hat n_l p_k$, giving
$k^2\AAA\mathbf p=\rho_0\omega^2\mathbf p$.
Positive acoustic eigenvalues are required for propagating elastic
waves in this local nondissipative approximation~\citep{Shankar2017}.
In pre-stressed or finite-strain settings, this is replaced by the
appropriate incremental acoustic tensor. The relevant symmetry for a
fixed wave normal is the little group
\begin{equation}
	L_{\hat{\mathbf n}}(\bm\eta)
	=\{h\in H_{\bm\eta}:h\hat{\mathbf n}=\hat{\mathbf n}\}.
	\label{eq:littlegroup}
\end{equation}
The acoustic operator commutes with the representation of the little
group on polarization space. Irreducible representations of dimension
greater than one can enforce modal degeneracy; lower symmetry may
permit previously degenerate modes to separate. One must check the
representation, measurement direction, and perturbation: symmetry
breaking does not imply that every mode splits.

More explicitly, material covariance gives
$\AAA(h\hat{\mathbf n};\bm\eta)=h\AAA(\hat{\mathbf n};\bm\eta)h^T$
for $h\in H_{\bm\eta}$. At a fixed wave normal this becomes
$[\AAA,h]=0$. On a single irreducible real polarization subspace,
a commuting self-adjoint operator is scalar; equivalent copies may
mix, but the dimension of the irreducible factor still enforces
the corresponding degeneracy. For the uniaxial axial transverse
plane, rotations about the axis therefore force a scalar transverse
block. This is the operator statement behind the selection rule
used in Section~\ref{sec:models}.

\subsection{Degenerate perturbation theory and the first visible order}
Suppose at $\bm\eta=0$ the acoustic operator has an isolated
two-dimensional degenerate eigenvalue $a_0=\rho_0c_0^2$ for fixed
$\hat{\mathbf n}$. Let $P$ be its orthogonal eigenspace projector,
and approach the reference state along a fixed normalized direction
$\bm\eta=\varepsilon\mathbf v$ with $\varepsilon\downarrow0$.
If the acoustic operator is analytic along this path,
\begin{equation}
	\AAA(\varepsilon)=\AAA_0+\varepsilon\AAA_1(\mathbf v)
	+\varepsilon^2\AAA_2(\mathbf v)+\cdots.
	\label{eq:Aexp}
\end{equation}
At first order the two eigenvalue shifts are the eigenvalues of
$P\AAA_1P$ acting on the degenerate subspace. If it is proportional
to the identity, the first-order splitting vanishes. Higher orders
include both direct couplings and virtual mixing with other modes.

Let $P_\perp=\II-P$ and define the reduced resolvent on the
complement by
\begin{equation}
 R_0=P_\perp\left[(a_0\II-\AAA_0)|_{\operatorname{range}P_\perp}
             \right]^{-1}P_\perp.
 \label{eq:reduced-resolvent-derivation}
\end{equation}
The inverse is only on that complement, where spectral isolation
makes it well defined. Eliminating the complementary polarization
to the required order gives the effective doublet operator
\begin{equation}
 \AAA_{\rm dbl}=a_0P+\varepsilon B_1+\varepsilon^2B_2+O(\varepsilon^3),
 \quad B_1=P\AAA_1P,\quad
 B_2=P\AAA_2P+P\AAA_1R_0\AAA_1P.
 \label{eq:doublet-second-order-derivation}
\end{equation}
The mixing term in $B_2$ is the usual second-order correction
from perturbation theory~\citep{Chernyshev2004}. If $B_1$ is scalar
and $B_2$ is not, then $p=2$. In an orthonormal doublet basis the
gap coefficient at that order is
\begin{equation}
 D=\left[(B_{2,11}-B_{2,22})^2+4B_{2,12}^2\right]^{1/2}.
 \label{eq:doublet-gap-coefficient}
\end{equation}
Only the traceless part of the first nonscalar effective block
produces splitting; scalar shifts move both modes together.
Denote the first nonzero splitting order by $p\in\{1,2,\ldots\}$.
Then, when that order exists,
\begin{equation}
	|a_+(\varepsilon)-a_-(\varepsilon)|
	=D(\hat{\mathbf n},\mathbf v)\varepsilon^p
	+o(\varepsilon^p),\qquad D>0.
	\label{eq:ap}
\end{equation}
Since $\omega= k\sqrt{a/\rho_0}$,
\begin{equation}
	\Delta\omega
	=\frac{kD(\hat{\mathbf n},\mathbf v)}{2\rho_0c_0}
	\varepsilon^p+o(k\varepsilon^p).
	\label{eq:omegap}
\end{equation}

The conversion to a frequency gap can also be made without an
independent Taylor expansion of each mode:
\begin{equation}
 |\omega_+-\omega_-|
 =\frac{k|a_+-a_-|}{\rho_0(c_++c_-)}.
 \label{eq:exact-gap-conversion}
\end{equation}
Since $c_\pm\to c_0>0$, Eq.~\eqref{eq:omegap} follows.
The positive nonzero reference speed is essential: the same
exponent conversion need not hold for a mode softening to zero.
A perturbation symmetry may forbid the leading matrix element.
Accidental crossings can disappear under a generic perturbation,
whereas protected degeneracies remain only when the relevant little
group and its representation still enforce them. \rev{The value of $p$ depends on the constitutive coupling, the selected
state-space path, and the observation direction. Symmetry can forbid a
coefficient, but an allowed coefficient can also vanish through parameter
choice or cancellation; the stabilizer alone therefore does not fix $p$.}

\subsection{Conditional spectral critical-exponent relation}
The spectral--critical reduction is restricted to a single selected ordered branch near a continuous transition, subject to the following conditions:
	(i) the measured system is sufficiently homogeneous or domain-selected
	for a nonzero $\bm\eta_{\rm sp}$ to characterize the probe volume;
	(ii) the selected branch lies on a fixed state-space ray,
	$\bm\eta_{\rm sp}=r(t)\mathbf v$, with $\norm{\mathbf v}=1$ and
	$r(t)=B|t|^\beta+o(|t|^\beta)$, $B>0$;
	(iii) the relevant spectral observable is analytic in the amplitude
	with first nonzero order $p$ and nonzero coefficient;
	(iv) singular background elastic self-energies, critical attenuation,
	and probe-induced dynamic order-parameter relaxation do not supply a
	more relevant contribution to the measured gap; and
	(v) the experimental direction does not lie on a protected zero.

\rev{At fixed nonzero probe wave number $k$, under these assumptions and
with a positive limiting reference speed $c_0$, the directional spectral
gap obeys}
	\begin{equation}
		\Delta\omega(\hat{\mathbf n},t)
			=\mathcal A(\hat{\mathbf n})|t|^{p\beta}
			+o(|t|^{p\beta}),\qquad\mathcal A(\hat{\mathbf n})>0.
		\label{eq:pbeta}
	\end{equation}

By hypothesis, the spectral gap has the asymptotic representation
	$\Delta\omega=A(\hat{\mathbf n})\norm{\bm\eta_{\rm sp}}^p
	+o(\norm{\bm\eta_{\rm sp}}^p)$ with $A>0$. Substitution of the
	ordered-branch scaling $\norm{\bm\eta_{\rm sp}}=
	B|t|^\beta+o(|t|^\beta)$ gives
	$\Delta\omega=AB^p|t|^{p\beta}+o(|t|^{p\beta})$.
Equation~\eqref{eq:pbeta} is a conditional composition of two scaling
	laws, not a new identity obeyed by all critical spectra. A different
	RG scaling dimension for the probe operator, nonanalytic dynamic
	self-energies, macroscopic multidomain cancellation, a leading
	critical contribution unrelated to the order parameter, or a protected
	zero can invalidate it. The exponent $p$ need not be universal and
	$\beta$ alone is not fixed by the broken-symmetry subgroup.

For a curved branch, merely specifying
$\bm\eta_{\rm sp}=B|t|^\beta\mathbf v+o(|t|^\beta)$ is not
enough when $p>1$: a subleading transverse component can activate
a lower-degree observation channel and dominate the nominal
$|t|^{p\beta}$ term. One must instead verify the same degree-$p$
expansion along the actual branch, or show that all branch departures
contribute $o(r^p)$ to the gap. The distinction between an analytic
constitutive readout and a fluctuation-renormalized critical observable
also has to be checked~\citep{SchorggSchwabl1992}.

For first-order transitions, an order-parameter jump may generate
an observable jump,
\begin{equation}
	[\Delta\omega]_{\rm coex}
	=\Delta\omega(\bm\eta_+)-\Delta\omega(\bm\eta_-),
	\label{eq:jump}
\end{equation}
but only if the selected measurement is sensitive to the difference.
The jump should be evaluated at coexistence, not confused with a
spinodal or kinetic switching threshold.

If the wave probe also drives the structural variable, linearized
mechanics and internal-state dynamics generically take the form
\begin{equation}
	\begin{pmatrix}
		\mathsf K_{uu}(\omega)&\mathsf K_{u\eta}(\omega)\\
		\mathsf K_{\eta u}(\omega)&\mathsf K_{\eta\eta}(\omega)
	\end{pmatrix}
	\begin{pmatrix}\delta\mathbf u\\\delta\bm\eta\end{pmatrix}=0.
\end{equation}
When the structural block is invertible, eliminating $\delta\bm\eta$
gives
\begin{equation}
	\mathsf K_{\rm eff}(\omega)
	=\mathsf K_{uu}(\omega)
	-\mathsf K_{u\eta}(\omega)
	\mathsf K_{\eta\eta}(\omega)^{-1}
	\mathsf K_{\eta u}(\omega).
	\label{eq:Schur}
\end{equation}
Thus a slow structural relaxation can generate dispersion and
attenuation in a dynamically relaxed tangent. The frozen-state wave
approximation requires a comparison between probe and structural
time scales; $\omega\tau_\eta\gg1$ is a common sufficient condition
when coupling is present. If the linear cross block vanishes at the
chosen stress-free base state, frozen and linear relaxed responses may
coincide even without that inequality.

The relaxation correction can be derived directly from a quadratic
local energy. Let $\mathbf e$ be incremental strain coordinates and
$\mathbf z$ independent structural coordinates, and write
\begin{equation}
 f^{(2)}=\tfrac12\mathbf e^T C_\infty\mathbf e
       +\mathbf e^T B\mathbf z+\tfrac12\mathbf z^TH\mathbf z,
 \qquad \dot{\mathbf z}=-M(H\mathbf z+B^T\mathbf e).
 \label{eq:coupled-quadratic-derivation}
\end{equation}
Here $H\succ0$ and $M\succ0$ on the active structural subspace.
With convention $e^{-i\omega t}$, elimination gives
\begin{equation}
 \mathbf z=-(H-i\omega M^{-1})^{-1}B^T\mathbf e,\qquad
 C_{\rm eff}(\omega)=C_\infty-B(H-i\omega M^{-1})^{-1}B^T,\qquad
 C_{\rm eff}(0)=C_\infty-BH^{-1}B^T.
 \label{eq:dynamic-tangent-derivation}
\end{equation}
The static expression is the Schur complement of the full energy
Hessian. For $L_\omega=H-i\omega M^{-1}$ and $\omega>0$,
\begin{equation}
 -\operatorname{Im}C_{\rm eff}
 =\omega B L_\omega^{-\dagger}M^{-1}L_\omega^{-1}B^T\succeq0.
 \label{eq:dynamic-passivity-derivation}
\end{equation}
Thus eliminating a passive relaxing variable produces a passive
complex modulus, with the stated sign convention~\citep{Soprunyuk2026}.
For a single variable $H=a$, $M=\Gamma_\eta$, and coupling
$B=-g$, Eq.~\eqref{eq:dynamic-tangent-derivation} reduces to
Eq.~\eqref{eq:bench-dynamic-modulus}. Stability of the complete
quadratic energy requires $C_\infty-BH^{-1}B^T\succ0$ as well as
$H\succ0$; symmetry alone supplies neither inequality.

\section{Tensor and ferroelastic realizations}
\label{sec:models}
The two models below separate distinct sources of acoustic selectivity.
For an isotropic parent, the mechanical couplings determine whether
uniaxial ordering produces a linear or quadratic splitting. For a cubic
parent, the little group of the propagation direction determines which
transverse degeneracy is protected after tetragonal ordering.

\subsection{Uniaxial ordering in a parent-isotropic solid}
\label{sec:Qtensor}

Using the tensor representation of Section~\ref{sec:geometry}, we recover the constitutive model of Ref.~\cite{SourceManuscript} by specializing
to $G_0=SO(3)$ and $\bm\eta=\QQ\in\mathrm{Sym}_0(3)$. At small strain,
\rev{consider the following potential}
\begin{equation}
	\Psi(\EE,\QQ,\nabla\QQ)
	={}\frac{\lambda_L}{2}(\tr\EE)^2
	+\mu\EE:\EE
	+\gamma_1(\tr\EE)(\EE:\QQ)
	+\gamma_2\tr(\QQ\EE^2)
	+\frac a2\tr\QQ^2-\frac b3\tr\QQ^3
	+\frac c4(\tr\QQ^2)^2
	+\frac{\kappa_Q}{2}|\nabla\QQ|^2.
	\label{eq:Qcoupling}
\end{equation}
Here $\lambda_L$ is a Lam\'e parameter (distinguished from the control
parameter $\lambda$), $\mu>0$, $3\lambda_L+2\mu>0$, and
$\gamma_1,\gamma_2$ are calibrated mechanical--structure couplings.
The isotropic inequalities ensure a positive quadratic strain energy
at $\QQ=0$. The coupled tangent must remain stable on the states probed. This model has zero stress at
$\EE=0$ for any frozen $\QQ$, but the tangent depends on $\QQ$.
The first strain derivative, at fixed $\QQ$, is
\begin{equation}
 \bm\sigma=\lambda_L(\tr\EE)\II+2\mu\EE
 +\gamma_1[(\EE:\QQ)\II+(\tr\EE)\QQ]
 +\gamma_2(\QQ\EE+\EE\QQ).
 \label{eq:Q-stress-derivation}
\end{equation}
This follows from $\delta\tr(\QQ\EE^2)
=(\QQ\EE+\EE\QQ):\delta\EE$, without assuming that $\EE$ and
$\QQ$ commute. Taking a further variation in a symmetric direction
$\mathbf H$ gives
\begin{equation}
		\CCt(\QQ):\mathbf H={}
		\lambda_L(\tr\mathbf H)\II+2\mu\mathbf H
		+\gamma_1\big[(\mathbf H:\QQ)\II+(\tr\mathbf H)\QQ\big]
		+\gamma_2(\QQ\mathbf H+\mathbf H\QQ).
	\label{eq:Qelastic}
\end{equation}
The structural perturbation in components is
\begin{equation}
		\delta C_{ijkl}={}\gamma_1(\delta_{ij}Q_{kl}+Q_{ij}\delta_{kl})
		+\frac{\gamma_2}{2}
		(Q_{ik}\delta_{jl}+Q_{il}\delta_{jk}
		+\delta_{ik}Q_{jl}+\delta_{il}Q_{jk}).
	\label{eq:Qelasticindices}
\end{equation}

A simple sufficient stability range follows by bounding the coupling
energy for every symmetric $\mathbf H$:
\begin{equation}
 W_{\rm el}(\mathbf H,\QQ)\ge
 \left[m_0-(\sqrt3|\gamma_1|+|\gamma_2|)\norm{\QQ}\right]
 \norm{\mathbf H}^2,
 m_0=\min\left\{\mu,\frac{3\lambda_L+2\mu}{2}\right\}>0.
 \label{eq:Q-stability-bound}
\end{equation}
Here $|\tr\mathbf H|\le\sqrt3\norm{\mathbf H}$ and
$|\tr(\QQ\mathbf H^2)|\le\norm{\QQ}\norm{\mathbf H}^2$.
Positivity of the bracket is sufficient, although not necessary,
for a positive strain Hessian and hence for strong ellipticity.
The two conditions are not generally equivalent~\citep{BardenhagenTriantafyllidis1994}.
\rev{Equation~\eqref{eq:Qcoupling} is used here as a local small-strain expansion on an admissible neighborhood with a stable elastic tangent. The linear-in-$\QQ$ elastic couplings do not ensure a globally bounded energy over unrestricted $(\EE,\QQ)$. Applying the global minimization in Eq.~\eqref{eq:MF} to this realization would therefore require a specified restricted admissible domain or a stabilizing constitutive completion. The local bound above does not provide that completion.}
For constant isotropic structural mobility, the same potential gives
the explicitly projected evolution
\begin{equation}
 \dot{\QQ}=-\Gamma_Q\left[
 (a+cJ_2)\QQ-b(\QQ^2-J_2\II/3)-\kappa_Q\nabla^2\QQ
 +\gamma_1(\tr\EE)\dev\EE+\gamma_2\dev(\EE^2)\right].
 \label{eq:Q-projected-evolution}
\end{equation}
All terms are symmetric and traceless, so the evolution preserves
the defining state space without an independent trace multiplier.

The structural part of Eq.~\eqref{eq:Qcoupling} for a homogeneous, stress-free state is
\begin{equation}
	f_Q(\QQ)=\frac a2J_2-\frac b3J_3+\frac c4J_2^2,
	\qquad c>0.
	\label{eq:Qlandau}
\end{equation}
For $\QQ=s(\mathbf n\otimes\mathbf n-\II/3)$,
$J_2=2s^2/3$ and $J_3=2s^3/9$, so
\begin{equation}
	f_Q(s)=\frac a3s^2-\frac{2b}{27}s^3+\frac c9s^4,
	\qquad
	f_Q'(s)=\frac{2s}{9}(3a-bs+2cs^2).
	\label{eq:Qbranch}
\end{equation}
For $b>0$, the uniaxial branch obeys
\begin{equation}
	s_\pm=\frac{b\pm\sqrt{b^2-24ac}}{4c},\qquad
	a_t=\frac{b^2}{27c},\qquad
	s_t=\frac{b}{3c}.
	\label{eq:Qcoexistence}
\end{equation}

To obtain coexistence, use stationarity
$a=(bs-2cs^2)/3$ to eliminate $a$ from the energy:
\begin{equation}
 f_Q(s)|_{f_Q'=0}=\frac{s^3}{27}(b-3cs),
 \qquad
 f_Q''(s)|_{f_Q'=0}=\frac{2s}{9}(4cs-b).
 \label{eq:Q-coexistence-derivation}
\end{equation}
Equality with $f_Q(0)=0$ gives $s_t=b/(3c)$ and then
$a_t=b^2/(27c)$. The nonzero pair appears at
$a=b^2/(24c)$, while the parent curvature changes sign at $a=0$.
Neither threshold is the coexistence point. For $b>0$, minimizing
at fixed $J_2$ maximizes $J_3$. The bound in
Eq.~\eqref{eq:Qinv} then selects a prolate uniaxial tensor, so this
uniaxial reduction identifies the bulk minimum of the stated
quartic model, rather than an arbitrary tensor restriction
~\citep{Pajak2018}.
The values $a_t$ and $s_t$ in Eq.~\eqref{eq:Qcoexistence} are the
coexistence control and order-parameter jump for the specified
uniaxial branch, not universal numbers. For
$b=0$ and $a<0$, the quartic model fixes $J_2=-a/c$ but does not
select uniaxiality over biaxiality without another mechanism. Therefore
assigning a uniaxial continuous branch in that case is an explicit
selection assumption.

Let $D_{ij}=C_{ijkk}$ and $V_{ik}=C_{ijkj}$ denote the dilatation and
Voigt stiffness contractions.
The isotropic parts are
$\mathbf D_0=(3\lambda_L+2\mu)\II$ and
$\mathbf V_0=(\lambda_L+4\mu)\II$.
Tracing the perturbation over $(k,l)$ and $(j,l)$, respectively,
uses $\tr\QQ=0$ and removes all isotropic contributions after taking
the deviator. This gives the two independent coefficients below. Equation~\eqref{eq:Qelasticindices}
implies
\begin{align}
	\dev\mathbf D&=(3\gamma_1+2\gamma_2)\QQ=:a_D\QQ,\\
	\dev\mathbf V&=(2\gamma_1+\tfrac52\gamma_2)\QQ=:a_V\QQ.
	\label{eq:contractions}
\end{align}
The mapping $(\gamma_1,\gamma_2)\mapsto(a_D,a_V)$ has determinant
$7/2$. Hence, for any fixed $(\gamma_1,\gamma_2)\ne(0,0)$,
\begin{equation}
		\QQ=\frac{a_D\dev\mathbf D+a_V\dev\mathbf V}
		{a_D^2+a_V^2}
	\label{eq:recon}
\end{equation}
and the calibrated complete tangent distinguishes $\QQ$ exactly.
Consequently $H_{\QQ}=G_{\CCt(\QQ)}$ as a proper rotational
stabilizer in this restricted model. If both couplings vanish, the
complete tangent is isotropic and cannot reveal a nonzero $\QQ$.
A particular incomplete set of wave speeds need not reconstruct the
full tangent even when Eq.~\eqref{eq:recon} holds.

Put $\QQ=s(\mathbf N\otimes\mathbf N-\II/3)$, and let $\vartheta$
be the angle between wave normal $\hat{\mathbf n}$ and the structural
axis $\mathbf N$. For sufficiently small $s$, with a nonzero
longitudinal--transverse separation in the isotropic reference, the
two quasi-transverse modes obey
\begin{equation}
		|c_{T1}^2-c_{T2}^2|
		=\frac{|\gamma_2s|}{2\rho_0}\sin^2\vartheta
		+O(s^2).
	\label{eq:Qgap}
\end{equation}
The $\gamma_1$ coupling drops out of the first-order transverse
projection and affects the splitting through higher-order mixing.
The axial zero at $\vartheta=0$ is protected by the uniaxial
rotational little group. At fixed non-axial observation angle, if
$\gamma_2\ne0$ the gap has $p=1$ along the selected uniaxial path.
If $\gamma_2=0$ and $\gamma_1\ne0$, the leading oblique splitting
from the source model is instead
\begin{equation}
	|c_{T1}^2-c_{T2}^2|
	=\frac{\gamma_1^2s^2}{\rho_0(\lambda_L+\mu)}
	\sin^2\vartheta\cos^2\vartheta
	+O(|s|^3),
	\label{eq:QgapSecond}
\end{equation}
assuming the positive-energy isotropic reference specified above.
\rev{This gives $p=2$ at generic oblique angles. In this
$\gamma_1$-only model, both the axial and equatorial directions have
zero transverse splitting: the axial zero is symmetry protected,
whereas the equatorial zero follows from the selected coupling.
Thus an angular zero must be interpreted together with the constitutive
law, rather than attributed to structural symmetry alone.}

If a continuous ordered branch has $|s|\sim |t|^\beta$ and the
regularity and observability assumptions underlying Eq.~\eqref{eq:pbeta} hold, the observable
exponent is $\beta$ for the nonzero $\gamma_2$ channel and $2\beta$
for the leading $\gamma_1$-only oblique channel. In the mean-field
selected $b=0$ uniaxial branch, $\beta=1/2$. This choice is not
necessarily a generic full-tensor equilibrium phase and must be
stabilized by the material's selection mechanism.
The cases above concern a specified coupling channel and observation
angle. In particular, a vanishing coefficient caused by calibration
or by an angular zero must be identified before assigning a fitted
temperature exponent.

\subsection{Cubic-to-tetragonal ferroelasticity}
\label{sec:ferroelastic}

Consider a parent cubic crystal with point group $G_0=O_h$.
A cubic-to-tetragonal distortion is represented at minimal order by
a two-dimensional $E_g$ multiplet
\begin{equation}
	\bm\eta=(\eta_1,\eta_2),\qquad
	r^2=\eta_1^2+\eta_2^2,\quad
	\eta_1=r\cos\phi,\ \eta_2=r\sin\phi.
	\label{eq:Eg}
\end{equation}
Fix the basis explicitly by assigning the multiplet the same
transformation law as $(-e_3,e_2)$, with the strain components defined
in Eq.~\eqref{eq:strainEg}. Its diagonal tensor representative is
\begin{equation}
 \mathbf D_E(\bm\eta)=\operatorname{diag}\left(
 -\frac{\eta_1}{\sqrt6}+\frac{\eta_2}{\sqrt2},
 -\frac{\eta_1}{\sqrt6}-\frac{\eta_2}{\sqrt2},
 \frac{2\eta_1}{\sqrt6}\right).
 \label{eq:Eg-basis-derivation}
\end{equation}
Cubic rotations permute its diagonal entries. Consequently
$\tr\mathbf D_E^2=r^2$ and
$\sqrt6\tr\mathbf D_E^3=\eta_1^3-3\eta_1\eta_2^2
=r^3\cos3\phi$ are invariants~\citep{Catalan2012}.
This fixes the angular convention used in the following polynomial;
the unrotated $(e_2,e_3)$ basis instead produces a sine form of the
same cubic invariant. A minimal Landau density is
\begin{equation}
		f_{E_g}(r,\phi)
		=\frac a2r^2+\frac b3r^3\cos3\phi+\frac c4r^4,
		\qquad c>0.
	\label{eq:EgF}
\end{equation}
For $b>0$, the three preferred tetragonal variants have
$\cos3\phi=-1$. Their symmetry groups are conjugate tetragonal
subgroups $D_{4h}\subset O_h$ (up to extra couplings and specimen
conditions); a generic nonzero $E_g$ direction may have lower
symmetry. The cubic parent-to-tetragonal transition can therefore be
represented through a single invariant potential, without selecting
three separate elastic laws. Aizu's classification of ferroelastic
species provides the established group-theoretic context
\citep{Aizu1969,Aizu1970}.

Along a preferred variant, Eq.~\eqref{eq:EgF} reduces to
\begin{equation}
	f_{\rm var}(r)=\frac a2r^2-\frac b3r^3+\frac c4r^4,
	\qquad
	\frac{\dd f_{\rm var}}{\dd r}
	=r(a-br+cr^2).
	\label{eq:EgBranch}
\end{equation}
The stationary radii are
\begin{equation}
	r=0,\qquad
	r_\pm=\frac{b\pm\sqrt{b^2-4ac}}{2c}.
	\label{eq:EgR}
\end{equation}
For $b>0$ the two equal-energy minima occur at
\begin{equation}
	a_t=\frac{2b^2}{9c},\qquad
		r_t=\frac{2b}{3c}.
	\label{eq:EgCoexist}
\end{equation}

For $r>0$, angular stationarity gives
$\partial_\phi f_{E_g}=-br^3\sin3\phi=0$ and
$\partial_{\phi\phi}f_{E_g}=-3br^3\cos3\phi$.
Thus the preferred rays for $b>0$ have positive angular curvature.
On those rays the radial curvature at a stationary point is
$f_{\rm var}''=r(2cr-b)$; $r_+$ is locally stable radially and
$r_-$ is unstable whenever the roots are distinct and positive.
At stationarity the energy is $br^3/6-cr^4/4$, whose equality with
zero yields Eq.~\eqref{eq:EgCoexist}. Reversing the sign of $b$
interchanges elongational and compressional preferred rays and
replaces $b$ by $|b|$ in the preferred-ray radial polynomial.
The nonzero stationary pair first appears at $a=b^2/(4c)$, whereas
\rev{the cubic parent state loses quadratic local stability at}
$a=0$. Coexistence, emergence of metastable stationary branches,
and loss of metastability are distinct thresholds. At $b=0$ the
quartic model has accidental continuous angular degeneracy at fixed
$r$. Higher-order cubic invariants, elastic compatibility, and
boundaries are needed to determine the selected variants. These are
mean-field conclusions, subject to fluctuation and elastic
renormalization.

Define symmetry-adapted small-strain combinations
\begin{equation}
	e_1=\frac{E_{xx}+E_{yy}+E_{zz}}{\sqrt3},\quad
	e_2=\frac{E_{xx}-E_{yy}}{\sqrt2},\quad
	e_3=\frac{E_{xx}+E_{yy}-2E_{zz}}{\sqrt6}.
	\label{eq:strainEg}
\end{equation}
The doublet $(e_2,e_3)$ transforms as $E_g$, up to basis conventions.
If $\bm\eta$ denotes an internal optical/structural distortion distinct
from the macroscopic strain, the leading coupling can be
\begin{equation}
	W_{E_g}=\frac{K_E}{2}(e_2^2+e_3^2)
	-g(\eta_2 e_2-\eta_1 e_3)
	+W_{\rm other}(\EE,\bm\eta),
	\qquad K_E>0.
	\label{eq:Egstrain}
\end{equation}
This is the invariant scalar product of the two $E_g$ doublets
$(\eta_1,\eta_2)$ and $(-e_3,e_2)$.
If the strain is fully relaxed in a homogeneous stress-free
idealization and $W_{\rm other}$ does not couple these sectors,
\begin{equation}
	(e_2,e_3)=\frac{g}{K_E}(\eta_2,-\eta_1),\qquad
	f_{E_g}^{\rm relaxed}
	=f_{E_g}-\frac{g^2}{2K_E}r^2.
	\label{eq:strainrelaxed}
\end{equation}
\rev{Thus $a_{\rm eff}=a-g^2/K_E$ in this homogeneous stress-free idealization.}
This result follows by completing the square:
\begin{equation}
 \frac{K_E}{2}\left[\left(e_2-\frac{g\eta_2}{K_E}\right)^2
             +\left(e_3+\frac{g\eta_1}{K_E}\right)^2\right]
 -\frac{g^2}{2K_E}r^2.
 \label{eq:Eg-square-completion}
\end{equation}
The bare and relaxed values of $a$ must therefore be distinguished
when comparing the branch thresholds. If $a$ in
Eq.~\eqref{eq:EgF} is already a stress-free fitted coefficient,
the same relaxation correction must not be subtracted again.
Finite-body compatibility, external constraints, domains, and
nonlocal elastic interactions obstruct independent local strain
minimization. If instead the strain is itself chosen as the primary
order parameter, Eq.~\eqref{eq:Egstrain} must not introduce a
redundant, independently evolving copy of the same degree of freedom.

A parent-invariant higher-order elastic energy can make the tangent
phase dependent,
\begin{equation}
	W_{\rm elastic}(\EE,\bm\eta)
	=\frac12\EE:\CCt_0:\EE
	+\sum_{a=1}^2\eta_a\,\mathcal B_a(\EE,\EE)
	+O(r^2\norm{\EE}^2,\norm{\EE}^3),
	\label{eq:EgElastic}
\end{equation}
where the bilinear elastic forms $\mathcal B_a$ must transform as a
covariant $E_g$ doublet. This notation encodes allowed symmetry
couplings but does not assume their coefficients or positive
ellipticity a priori.

The distinction between the two couplings is explicit: the term
linear in strain in Eq.~\eqref{eq:Egstrain} changes the equilibrium
strain but has zero second strain derivative at fixed $\bm\eta$.
It cannot by itself generate a phase-dependent frozen stiffness.
For a concrete symmetry-allowed example of Eq.~\eqref{eq:EgElastic},
let $d_i$ be the diagonal entries of $\mathbf D_E$ and add
\begin{equation}
 W_{\rm an}=2\gamma_E
 (d_xE_{yz}^2+d_yE_{xz}^2+d_zE_{xy}^2).
 \label{eq:Eg-explicit-elastic-coupling}
\end{equation}
This expression is invariant under cubic signed coordinate
permutations. Since the tensorial shear energy is
$2C_{ijij}E_{ij}^2$, it gives
$\delta(C_{44},C_{55},C_{66})=\gamma_E(d_x,d_y,d_z)$.
For a $z$-tetragonal ray $\eta_2=0$, one obtains
$C_{66}-C_{44}=\sqrt{3/2}\gamma_E\eta_1$.
Thus a calibrated nonzero $\gamma_E$ realizes $p=1$ explicitly;
the abstract expansion in Eq.~\eqref{eq:EgCouplingPower} also
allows models in which that term is absent~\citep{Berryman2012}.

For a tetragonal variant whose unique axis is $z$, consider wave
propagation along $x$. Standard tetragonal stiffness notation gives
\begin{equation}
	c_{T,y}^2=\frac{C_{66}}{\rho_0},\qquad
	c_{T,z}^2=\frac{C_{44}}{\rho_0},\qquad
	\Delta c_T^2\big|_{[100]}
		=\frac{|C_{66}-C_{44}|}{\rho_0}.
	\label{eq:tetragap}
\end{equation}
For the cubic parent $C_{66}=C_{44}$, so these two transverse
branches are degenerate along $[100]$. In the tetragonal ordered
state they generally split. Along the tetragonal $[001]$ axis,
by contrast, fourfold rotation protects the two transverse
polarizations and
\begin{equation}
	\Delta c_T^2\big|_{[001]}=0.
	\label{eq:tetrprotected}
\end{equation}

\rev{The tetragonal acoustic matrix is
$\operatorname{diag}(C_{11},C_{66},C_{44})$ along $[100]$.
Along $[001]$, it is $\operatorname{diag}(C_{44},C_{44},C_{33})$.}
These diagonal matrices give both statements without identifying
a general cubic propagation direction with a degenerate axis.
The important claim is an angular selection rule, not a claim
that every propagation direction is degenerate in a cubic crystal.
At a small selected $E_g$ amplitude $r$, suppose the calibrated
constitutive map has
\begin{equation}
	C_{66}-C_{44}=B_E r^p+o(r^p),\qquad B_E\ne0.
	\label{eq:EgCouplingPower}
\end{equation}
Then the $[100]$ squared-speed gap has leading order
$|B_E|r^p/\rho_0$, whereas the $[001]$ protected gap is zero.
Near a continuous critical branch with $r\sim|t|^\beta$, the
$[100]$ gap follows $|t|^{p\beta}$ only under the hypotheses in
Section~\ref{sec:spectrum}. For a first-order branch, the signal
can exhibit a finite coexistence discontinuity. Measuring both
crystallographic directions distinguishes a predicted lifted
degeneracy from a symmetry-protected one.

\section{Numerical benchmarks}
\label{sec:benchmarks}

\rev{The benchmarks below test distinct parts of the framework using
separately specified models, where the homogeneous strain marginal tests
invariant free-energy reconstruction, the $E_g$ lattice tests coexistence,
the compatible strain model tests domain organization, and the fixed-cell
bond network tests acoustic selection rules. The Ising calculations test
prescribed linear and quadratic spectral readouts, with and without
driving. These calculations provide complementary consistency checks.}

\subsection{Symmetry strata and free-energy reconstruction}
\label{sec:benchmark-strata}
In the basis of Eq.~\eqref{eq:strainEg}, write the diagonal deviatoric
strain as $s\mathbf B\mathbf q$, where $\mathbf q=(e_2,e_3)/s$ and
\begin{equation}
 \mathbf B=
 \begin{pmatrix}
  1/\sqrt 2 & 1/\sqrt 6\\
  -1/\sqrt 2 & 1/\sqrt 6\\
  0 & -2/\sqrt 6
 \end{pmatrix}.
 \label{eq:bench-basis}
\end{equation}

\begin{figure}[pos=!htbp]
 \centering
 \includegraphics[width=\linewidth]{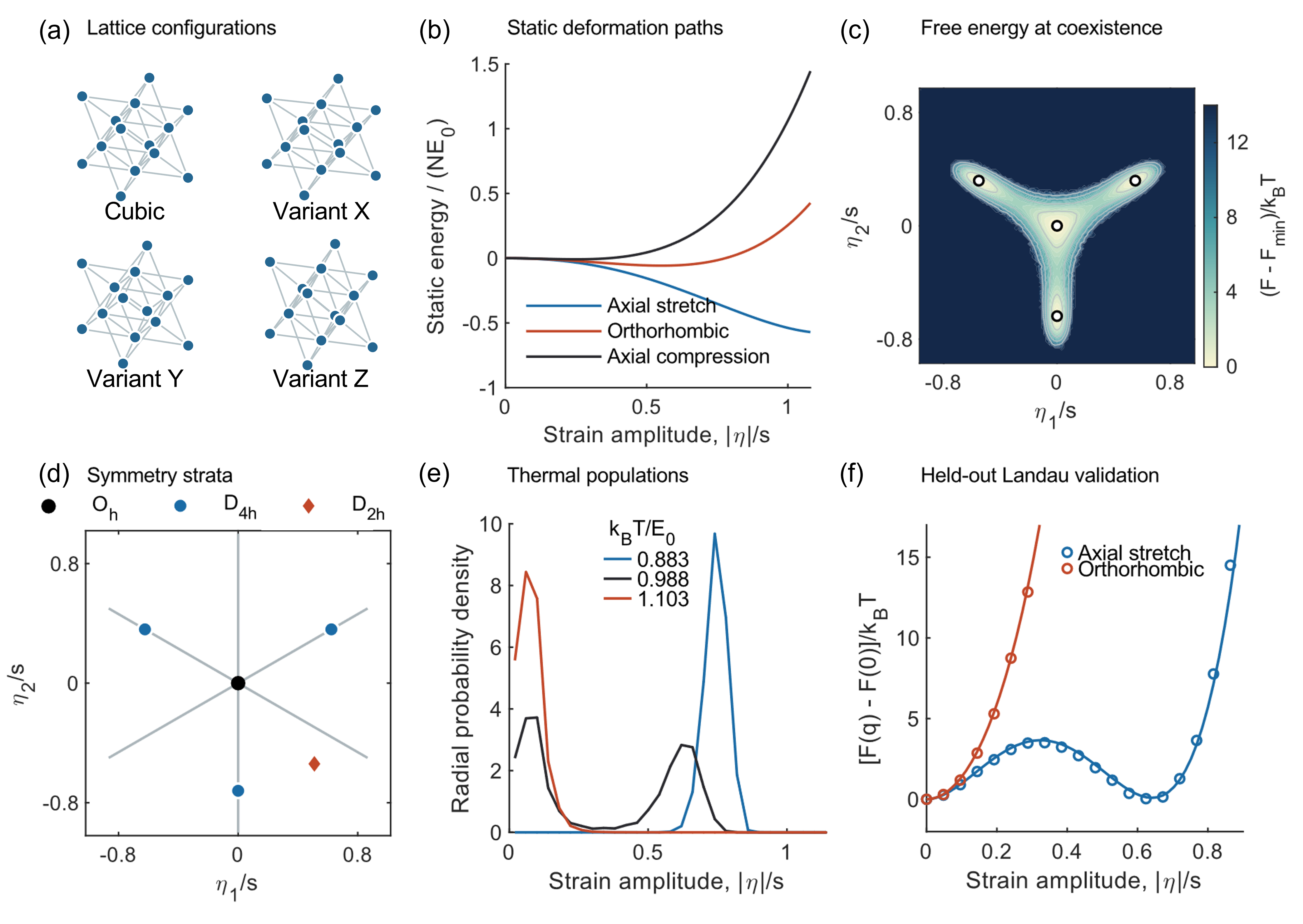}
 \caption{\textbf{Symmetry-stratified states in a cubic-to-tetragonal
 lattice benchmark.} \textbf{(a)} Undeformed face-centred cubic
 reference and the three volume-preserving elongational variants.
 \textbf{(b)} Static bond energies along the indicated deformation
 paths. \textbf{(c)} WHAM free energy at $t_*=0.9881$ with the
 equal-depth minima marked. \textbf{(d)} The origin, the tetragonal
 lines, and the generic directions have stabilizers $O_h$, $D_{4h}$,
 and $D_{2h}$, respectively. Blue points mark the elongational orbit.
 \textbf{(e)} Radial probability densities below, at, and above
 $t_*$. \textbf{(f)} Sixth-order invariant fit (lines) and an
 independent WHAM estimate along excluded directions (circles).}
 \label{fig1}
\end{figure}

For a proper-strain realization in the canonical basis of
Section~\ref{sec:ferroelastic}, the components are related by
$(q_1,q_2)=(\eta_2,-\eta_1)/s$. This orthogonal basis change leaves
the quadratic norm unchanged and converts the cosine cubic invariant
into $3q_1^2q_2-q_2^3$. It must be applied before comparing angular
positions across the two conventions.
For the homogeneous lattice benchmark in Fig.~\ref{fig1}, the
deformation gradient is $\operatorname{diag}(\exp[s(\mathbf B\mathbf q)_j])$
with $s=0.10$. \rev{Thus the plotted order parameter is obtained from the logarithmic
strain of the deformed face-centred cubic reference. This finite-deformation
construction illustrates the symmetry classification of a homogeneous
lattice; it is distinct from the infinitesimal-strain constitutive
realizations in Section~\ref{sec:models}.} A
permutation of the three diagonal strains acts on the doublet as the
cubic group projected onto $E_g$. The origin has stabilizer $O_h$.
Two equal principal strains give $D_{4h}$ on each of the three lines
away from the origin. Three unequal principal strains give $D_{2h}$.
The three elongational rays form one orbit. This classification
concerns the diagonal strain state represented by the doublet.

The static energy sums over the three axial second-neighbour
bonds counted per site, with reduced extensions $d_j=(\mathbf B\mathbf q)_j$,
\begin{equation}
 \frac{U(\mathbf q)}{NE_0}=\sum_{j=1}^3
 \left(-0.375d_j^2-\frac{2.4\sqrt6}{3}d_j^3+1.2d_j^4+0.12d_j^6\right).
 \label{eq:bench-static-bonds}
\end{equation}
The displacements have polarization-dependent stiffness $K_j$
on a periodic harmonic nearest-neighbour graph.
For $N=512$ sites and $t=k_{\mathrm B}T/E_0$, integration over the
$N-1$ nonzero displacement modes per polarization gives the
shape-dependent free energy
\begin{equation}
 \frac{F_t(\mathbf q)}{N E_0}
 =\frac{U(\mathbf q)}{N E_0}
 +\frac{t}{2}\left(1-\frac{1}{N}\right)
   \sum_{j=1}^3 \ln \frac{K_j(\mathbf q)}{K_0}+\mathrm{const.},
 \qquad
 \frac{K_j}{K_0}=\exp\!\left(\frac{1.15|\mathbf q|^2}{3}\right)
                  (1+0.14d_j^2).
 \label{eq:bench-bond-free}
\end{equation}

The logarithmic term follows from a Gaussian integral. If
$\ell_\alpha>0$ are the $N-1$ nonzero eigenvalues of the common
periodic graph Laplacian, the mode energy is
$\tfrac12K_j\ell_\alpha x_{j\alpha}^2$. Hence
\begin{equation}
 Z_{\rm vib}(\mathbf q)\propto
 \prod_{j=1}^3\prod_{\alpha=1}^{N-1}
 (\beta_TK_j(\mathbf q)\ell_\alpha)^{-1/2},
 \qquad \beta_T=(\kb T)^{-1}.
 \label{eq:harmonic-logdet-derivation}
\end{equation}
Taking $-\kb T\log Z_{\rm vib}$ and dividing by $NE_0$ gives
the coefficient $t(1-1/N)/2$ in Eq.~\eqref{eq:bench-bond-free}.
The translational zero modes are fixed, and graph and momentum
factors independent of $\mathbf q$ enter the constant. This simple
product requires the specified factorization into $K_j(\mathbf q)$
times a common graph operator~\citep{Monacelli2025}.
The static bond energy includes the cubic invariant
$I_3=\sum_j d_j^3=(3q_1^2q_2-q_2^3)/\sqrt 6$, the basis-specific
form of the angular term in Eq.~\eqref{eq:EgF}. Figure~\ref{fig1}b
compares three deformations of this bond model.

Independent draws from 81 biased windows of the analytically
integrated strain marginal were combined by weighted histogram analysis (WHAM) to obtain $F_t=-k_{\mathrm B}T\ln P_t+\mathrm{const.}$
on a flat two-dimensional strain grid. The equal-depth temperature
$t_*=0.9881$ gives one cubic and three tetragonal minima, with a
barrier of $3.54\,k_{\mathrm B}T_*$ along a preferred variant ray.
Equal depth refers to the minima of the two-dimensional map.
The radial probability in Fig.~\ref{fig1}e also includes the
polar measure $r\,\mathrm{d}r\,\mathrm{d}\phi$, with $r=|\mathbf q|$,
and the multiplicity of the tetragonal wells. An invariant fit
using $1,r^2,I_3,r^4,r^2 I_3,r^6,I_3^2$ excludes the symmetry
orbits of the two test directions and their angular neighbourhoods.
Its root-mean-square error against a separate umbrella sample on
those directions is $0.30\,k_{\mathrm B}T_*$. \rev{This tests sampling, reconstruction, and angular prediction against
the specified strain marginal. Because the reference marginal is
analytically available, it is a controlled reconstruction benchmark,
rather than an independent test of the material Hamiltonian.}

Table~\ref{tab:free-energy-validation} quantifies both the strain-marginal
reconstruction and the invariant fit. We use the total-variation distance
$\tfrac12\sum_\ell|P_\ell^{\rm WHAM}-P_\ell^{\rm ref}|$ on the common
discrete grid to compare normalized probabilities. The angular test
uses an independently sampled reference and excludes the complete
symmetry-related neighbourhoods of the test rays from fitting. The
probability error and the energy-fit error measure complementary
parts of this reconstruction.

\begin{table}[pos=!htbp]
 \centering
 \caption{Sampling and free-energy reconstruction for the homogeneous
 lattice benchmark in Fig.~\ref{fig1}. The reference marginal follows
 from the integrated harmonic model.}
 \label{tab:free-energy-validation}
 \small
 \renewcommand{\arraystretch}{1.15}
 \begin{tabular}{@{}>{\raggedright\arraybackslash}p{.53\linewidth}
                    >{\raggedright\arraybackslash}p{.42\linewidth}@{}}
 \toprule
 Quantity & Value or protocol\\
 \midrule
 Lattice sites and logarithmic-strain scale & $N=512$, $s=0.10$\\
 Reduced temperatures $t=\kb T/E_0$ & $0.8831,\ 0.9881,\ 1.1031$\\
 Umbrella sampling at each temperature & 81 windows, 14,000 independent draws per window\\
 Independent reconstruction check & A separate sample at $t_*$\\
 Equal-depth temperature $t_*$ & $0.9881$\\
 Tetragonal-minimum and barrier radii at $t_*$ & $0.6375,\ 0.3214$\\
 Preferred-ray barrier $\,\Delta F/(\kb T_*)$ & $3.54$\\
 Total-variation distance, all four reconstructions & $0.0080\text{--}0.0171$\\
 Invariant fit versus independent excluded-ray WHAM & RMSE $0.299\,\kb T_*$\\
 Invariant fit versus exact marginal on test rays & RMSE $0.240\,\kb T_*$\\
 Independent WHAM versus exact marginal on test rays & RMSE $0.123\,\kb T_*$\\
 \bottomrule
 \end{tabular}
\end{table}

\FloatBarrier

\subsection{Coexistence and a smooth symmetry-stratum crossing}
\label{sec:benchmark-coexistence}
The finite-volume statistical test of
Section~\ref{sec:thermal} uses a continuous $E_g$ field
$\mathbf q_i=r_i(\cos\phi_i,\sin\phi_i)$ on a periodic cubic lattice.
Here the doublet uses the rotated cosine convention of
Section~\ref{sec:ferroelastic}; its angular origin is a preferred
elongational ray for the negative $b$ used below. It differs from
the fixed $(e_2,e_3)$ coordinates of Section~\ref{sec:benchmark-strata}. Its
Hamiltonian is
\begin{equation}
 \mathcal H=
 \sum_i\left(\frac{a}{2}r_i^2+\frac{b}{3}r_i^3\cos3\phi_i
             +\frac{c}{4}r_i^4\right)
 +\frac{J}{2}\sum_{\langle i,j\rangle}|\mathbf q_i-\mathbf q_j|^2 .
 \label{eq:bench-eg-lattice}
\end{equation}

\begin{figure}[pos=!htbp]
 \centering
 \includegraphics[width=\linewidth]{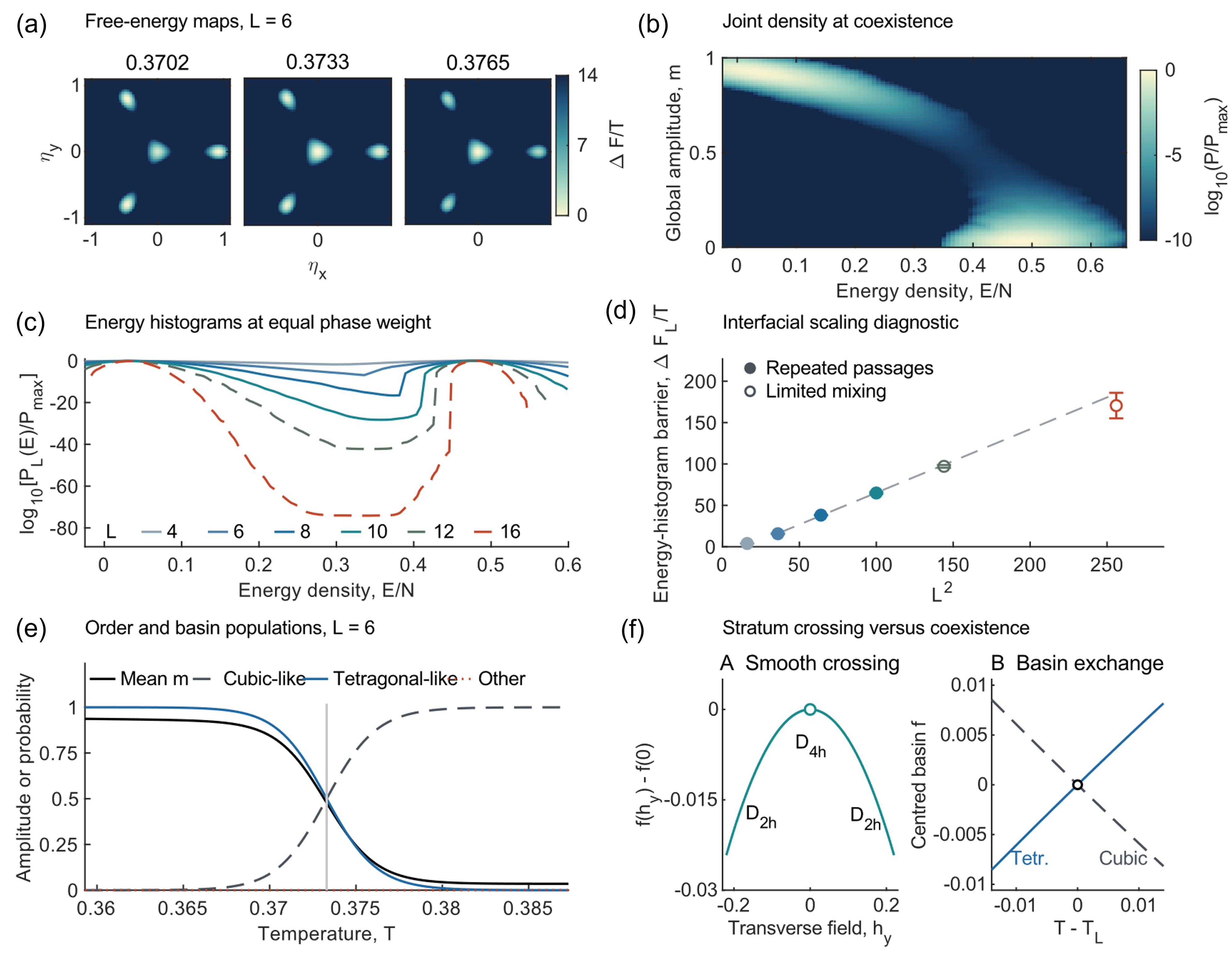}
 \caption{\textbf{Thermodynamic coexistence and symmetry-stratum
 crossing.} \textbf{(a)} Free-energy maps of the global $E_g$
 doublet for $L=6$ at ordered-basin weights $0.9$, $0.5$, and $0.1$.
 \textbf{(b)} Joint distribution of energy density and global
 amplitude at equal basin weight. \textbf{(c)} Energy histograms
 at each size's equal-weight temperature. \textbf{(d)} Barrier
 from Eq.~\eqref{eq:bench-hist-barrier} against $L^2$. Bars show
 the standard deviation across three umbrella runs. Dashed
 histograms and open markers denote limited mixing. The dashed
 line in (d) uses filled symbols with $L\ge6$. \textbf{(e)}
 Mean amplitude and threshold-based basin populations.
 \textbf{(f)} Analytic field-driven stratum crossing in the
 harmonic model and exchange of the two restricted free-energy
 branches in the anharmonic model. The latter curves are centred
 on their common average.}
 \label{fig2}
\end{figure}

We use $(a,b,c,J)=(19.10,-60,40,0.55)$ and sample by Metropolis updates with replica exchanges between
umbrella windows in $L^{-3}\sum_i r_i^2$ and reweight to
equal integrated basin weight at $T_L$, using $0.48$ as the threshold
in the window coordinate. The ordered weight
includes all three variants.
\rev{The reference sampling temperature is $0.374$. Each of the
three independent umbrella ladders discards 4500 sweeps and records
3200 measurements at intervals of three sweeps, with neighbouring
window exchanges attempted each sweep. Temperature reweighting then
locates $T_L$ for each lattice size.}
Figure~\ref{fig2}a,b shows the
distribution of the global average
$\overline{\mathbf q}=L^{-3}\sum_i\mathbf q_i$ and its amplitude,
as well as the coexistence of the low- and high-amplitude basins.
The energy-histogram barrier is
\begin{equation}
 \frac{\Delta F_L}{k_{\mathrm B}T_L}
 =\ln\frac{\sqrt{P_L(E_1)P_L(E_2)}}{P_L(E_{\rm v})},
 \label{eq:bench-hist-barrier}
\end{equation}
where $E_1,E_2$ are the two peak positions and $E_{\rm v}$ is the
intervening valley. For $L=4,6,8,10$, where repeated passages were
confirmed in the umbrella or auxiliary fixed-bias runs, the peaks
remain separated by $0.44$--$0.45$ energy units per site and the
barrier rises from $3.92$ to $64.93$ in units of $k_{\mathrm B}T_L$.
These are finite-size first-order signatures. The $L=12,16$ data
have limited mixing. For a periodic box with two planar interfaces of tension
$\sigma$, the leading area contribution would be $2\sigma L^2$.
The $L^2$ line is a finite-range guide. Figure~\ref{fig2}e
labels samples with $|\overline{\mathbf q}|<0.18$ as cubic-like
and the remaining samples within $10^\circ$ of a preferred ray
as tetragonal-like. The response estimators available from the same
histograms are $c_V=L^3\operatorname{Var}(E/L^3)/(k_{\mathrm B}T^2)$
and $\chi_m=L^3\operatorname{Var}(|\overline{\mathbf q}|)/(k_{\mathrm B}T)$.
\rev{Here $\chi_m$ measures fluctuations of the radial amplitude.
The linear susceptibility to a vector field conjugate to
$\overline{\mathbf q}$ is instead the component covariance
$L^3[\langle\overline q_\alpha\overline q_\beta\rangle
-\langle\overline q_\alpha\rangle\langle\overline q_\beta\rangle]/(\kb T)$.
The observable and its conjugate source must be specified when
interpreting these fluctuation estimators as response functions.}

The two paths in Fig.~\ref{fig2}f isolate the distinction in
Section~\ref{sec:thermal}. Path A uses a convex harmonic lattice
with local energy $|\mathbf q_i|^2/2$ and field term
$-\mathbf h\cdot\mathbf q_i$, where $\mathbf h=(0.65,h_y)$.
Shifting all $\mathbf q_i$ by $\mathbf h$ leaves the neighbour
differences invariant. Consequently the free-energy density and
the mean satisfy
\begin{equation}
 f(h_y)-f(0)=-\frac{h_y^2}{2},\qquad
 \langle\overline{\mathbf q}\rangle=\mathbf h .
 \label{eq:bench-smooth-cross}
\end{equation}
The mean passes through $D_{2h}$, $D_{4h}$, and $D_{2h}$ as
$h_y$ crosses zero. The free energy is analytic at every size
and in the thermodynamic limit. Path B compares the free energies
restricted to the two basins of Eq.~\eqref{eq:bench-eg-lattice}.
Their exchange identifies the equal-weight condition while the
finite-volume partition function remains smooth.

Table~\ref{tab:coexistence-sizes} reports the finite-size estimates
together with the passage diagnostics. The energy-density peak
separation varies little over the simulated sizes, while the
barrier grows substantially. The independent fixed-bias checks
at $L=8,10$ support the corresponding umbrella estimates. At $L=12$
the two checks give barriers of approximately $96.5$ and $61.2$,
so that the small spread of the umbrella estimates understates the
sampling ambiguity. The filled-symbol trend in Fig.~\ref{fig2}d
therefore uses the diagnostic-supported sizes $L=6,8,10$.

\begin{table}[pos=!htbp]
 \centering
 \caption{Finite-size coexistence diagnostics for Fig.~\ref{fig2}.
 The barrier central value uses the pooled umbrella reconstruction.
 Its uncertainty is the sample standard deviation across three
 independent umbrella ladders.}
 \label{tab:coexistence-sizes}
 \small
 \setlength{\tabcolsep}{4pt}
 \renewcommand{\arraystretch}{1.18}
 \begin{tabular}{@{}r r r r >{\raggedright\arraybackslash}p{.35\linewidth}@{}}
 \toprule
 $L$ & $T_L$ & $\Delta F_L/(\kb T_L)$ & Peak gap per site & Sampling diagnostic\\
 \midrule
4 & 0.37535 & $3.92\pm0.19$ & 0.441 & Repeated umbrella passages\\
6 & 0.37330 & $15.70\pm0.55$ & 0.444 & Repeated umbrella passages\\
8 & 0.37348 & $38.16\pm0.39$ & 0.452 & Fixed-bias check passed\\
10 & 0.37275 & $64.93\pm1.15$ & 0.446 & Fixed-bias check passed\\
12 & 0.37197 & $97.11\pm1.68$ & 0.450 & Inconsistent fixed-bias barriers\\
16 & 0.37194 & $170.52\pm15.43$ & 0.449 & Exchange bottleneck\\
 \bottomrule
 \end{tabular}
 \par\smallskip
 \parbox{\linewidth}{\footnotesize The peak gap is the separation of the energy-density maxima
 in model energy units. The $L=12,16$ rows are exploratory
 estimates with limited mixing. The present size range supports
 a finite-range area trend for the diagnostic-supported runs.}
\end{table}

\FloatBarrier

\subsection{Compatible three-dimensional microstructure}
\label{sec:benchmark-compatibility}
To extend the proper-strain case in
Section~\ref{sec:ferroelastic} to spatially varying states, the
periodic phase-field model takes $\mathbf q$ to be the $E_g$ part
of the strain itself, expressed in units of $s$. The dimensionless energy and local density are
\begin{equation}
 F[\mathbf q]=\int_\Omega\left[f(\mathbf q)+\frac{\kappa}{2}
               |\nabla\mathbf q|^2\right]\mathrm dV+F_{\rm compat},\qquad
 f(\mathbf q)=\frac{a}{2}|\mathbf q|^2
    +\frac{b}{3}(3q_1^2q_2-q_2^3)+\frac{c}{4}|\mathbf q|^4 ,
 \label{eq:bench-phasefield}
\end{equation}
with $(a,b,c)=(-0.50,-1.50,2)$.
\begin{figure}[pos=!htbp]
 \centering
 \includegraphics[width=\linewidth]{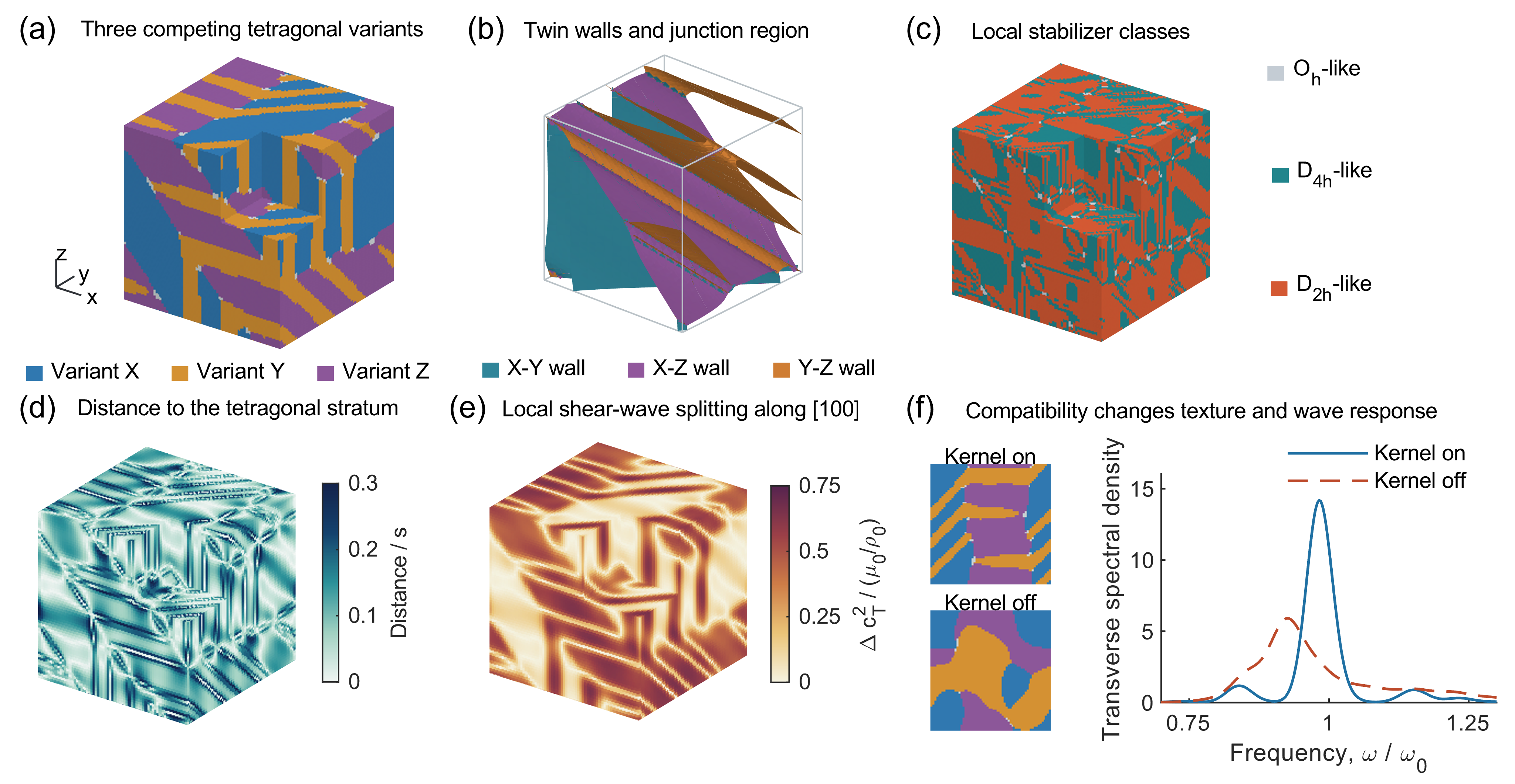}
 \caption{\textbf{Three-dimensional microstructure under nonlocal
 elastic compatibility.} \textbf{(a)} Cutaway coloured by the
 nearest tetragonal variant. \textbf{(b)} Pairwise walls and
 junction region in a selected subvolume. \textbf{(c)}
 Finite-tolerance stabilizer classes of the $E_g$ projection.
 \textbf{(d)} Distance to the closure of the tetragonal stratum.
 \textbf{(e)} Frozen local squared-speed splitting along $[100]$
 from Eq.~\eqref{eq:bench-distance-shear}. \textbf{(f)}
 Matched mid-plane cuts and transverse plane-wave spectral
 measures with and without compatibility. The spectra use the
 same frozen constitutive law for both microstructures.}
 \label{fig3}
\end{figure}
At $\mathbf k\ne0$,
let $\mathbf n=\mathbf k/|\mathbf k|$ and
$\widehat{\bm\varepsilon}/s=\operatorname{sym}(\mathbf n\otimes\mathbf v)$.
This is the Fourier representation of the Saint-Venant condition
for compatible linear strain. Set
\begin{equation}
 A_{\alpha i}=B_{i\alpha}n_i,\quad
 \mathbf Q=\mu\mathbf I+(\lambda+\mu)\mathbf n\otimes\mathbf n,\quad
 \mathbf G=\mathbf A\mathbf Q^{-1}\mathbf A^{\mathsf T},\quad
 \mathbf P=\mathbf G\mathbf G^+ .
 \label{eq:bench-compat-matrices}
\end{equation}
Minimizing the auxiliary isotropic elastic energy subject to
$\mathbf A\mathbf v=\widehat{\mathbf q}$ gives
$\mathbf v=\mathbf Q^{-1}\mathbf A^{\mathsf T}\mathbf G^+\widehat{\mathbf q}$.
Removing the auxiliary $E_g$ penalty, already represented in the
local potential, leaves the compatibility kernel
\begin{equation}
 \mathbf U(\mathbf n)=\mathbf G^+-2\mu\mathbf P,\qquad
 \widehat{\mathbf q}\in\operatorname{range}\mathbf A,\qquad
 F_{\rm compat}=\frac{|\Omega|}{2}\sum_{\mathbf k\ne0}
 \widehat{\mathbf q}^{\,\dagger}\mathbf U(\mathbf n)\widehat{\mathbf q},
 \label{eq:bench-compat-kernel}
\end{equation}
where the Fourier coefficients are normalized by $|\Omega|^{-1}$.
The pseudoinverse handles the rank-one modes along cubic axes.
The homogeneous mode is clamped in the matched comparison.
This construction follows the general strain-compatibility
approach for ferroelastics \citep{LookmanCompatibility2003}.

The elimination can be checked with a multiplier $\bm\Lambda$.
For the auxiliary energy $\tfrac12\mathbf v^\dagger\mathbf Q\mathbf v$
and constraint $\mathbf A\mathbf v=\widehat{\mathbf q}$,
stationarity gives $\mathbf Q\mathbf v=\mathbf A^T\bm\Lambda$
and $\mathbf G\bm\Lambda=\widehat{\mathbf q}$.
The minimum energy is therefore
$\tfrac12\widehat{\mathbf q}^{\,\dagger}\mathbf G^+
\widehat{\mathbf q}$. Orthogonality of the two diagonal strain
basis tensors makes their auxiliary energy $\mu|\widehat{\mathbf q}|^2$;
subtracting it gives Eq.~\eqref{eq:bench-compat-kernel}.
This remaining energy is nonnegative on $\operatorname{range}\mathbf A$.
Components outside that range are incompatible and must be excluded,
rather than assigned a finite energy by a pseudoinverse. The projector
$\mathbf P$ in the evolution enforces this restriction~\citep{Rasmussen2001}.

We evolved the two components by projected semi-implicit gradient
flow with $\lambda=\mu=4$ and gradient coefficient $\kappa=0.85$.
\rev{The primary comparison uses 240 steps of size $0.78$, unit
structural mobility, and stabilization coefficient 2. The gradient
and compatibility terms are implicit, and the local potential
derivative is explicit with the matching stabilization term.
Both grids resolve the same physical box and initial field.
The mean doublet is clamped to zero, and the continuous-derivative
Nyquist planes are set to zero in both comparisons.}
On the $128^3$ grid, $83.8\%$ of candidate compatible walls have
normals within $15^\circ$ of their ideal pair-specific
$\{110\}$ directions, compared with $10.8\%$ in the local control
initialized with the same field. The $256^3$ compatible result is
$82.8\%$. Both compatible runs give a normalized Fourier
compatibility residual below $1.6\times10^{-7}$. Let
$\mathbf v_j$ be unit vectors along the three variant rays. The
distance from the $E_g$ projection to the closure of the
tetragonal stratum and the frozen local $[100]$ squared-speed
gap used in Fig.~\ref{fig3} are
\begin{equation}
 d_{4h}/s=\min_j\left|\mathbf q-(\mathbf v_j\cdot\mathbf q)\mathbf v_j\right|,
 \qquad
 \Delta c_T^2([100])=\frac{\gamma}{\rho_0}|d_z-d_y|,
 \quad d_i=(\mathbf B\mathbf q)_i .
 \label{eq:bench-distance-shear}
\end{equation}
The generic $D_{2h}$ stratum is dense in the doublet plane,
so its distance is identically zero. The speed gap uses the illustrative
frozen constitutive map
$(C_{yzyz},C_{xzxz},C_{xyxy})=\mu_0+\gamma(d_x,d_y,d_z)$,
with $\mu_0=4$, $\gamma=2.3$, and $\rho_0=1$.
The classes in Fig.~\ref{fig3}c apply tolerances $|\mathbf q|<0.16$
and $d_{4h}/s<0.075$ in this projected state space.

For the spectrum in Fig.~\ref{fig3}f, the same frozen law defines
the spatial wave operator
$\mathbf K=\rho_0^{-1}\mathcal D^{*}\CCt(\mathbf x)\mathcal D$,
where $\mathcal D$ is the periodic discrete symmetric gradient.
The plotted density is the spectral measure of $\sqrt{\mathbf K}$
projected on the transverse plane-wave inputs. At the selected
$[100]$ wavevector on the $64^3$ wave grid, its normalized width
is $0.0995$ with compatibility and $0.1404$ in the local control.
These widths describe the mode distribution in a finite periodic
body.

Table~\ref{tab:compatibility-controls} compares the compatible and local
evolutions under the same clamped mean strain, physical box of side 64,
and smooth initial field. Refining the compatible grid changes the
selected-wall fraction by about one percentage point, with very small
field discrepancy after comparison on the same $64^3$ grid. The frozen
wave calculation uses the identical strain-dependent stiffness map
for both microstructures. Its normalized spectral width is reduced
by approximately $29.2\%$ in the compatible case. \rev{This comparison connects domain arrangement to the plane-wave spectral
measure under the chosen frozen constitutive map. Here the wave stiffness
is prescribed as an observation law, separately from the auxiliary elastic
energy used to construct the compatibility kernel. The comparison therefore
tests this specified microstructure-to-spectrum map. It does not establish
that the wave operator is the Hessian of the microstructure-generating
functional.}

\begin{table}[pos=!htbp]
 \centering
 \caption{Compatibility, discretization and frozen-wave checks for
 Fig.~\ref{fig3}. All cases use the same physical domain.
 Percentages refer to the candidate pairwise walls.}
 \label{tab:compatibility-controls}
 \small
 \setlength{\tabcolsep}{4pt}
 \renewcommand{\arraystretch}{1.18}
 \begin{tabular}{@{}>{\raggedright\arraybackslash}p{.43\linewidth}r r r@{}}
 \toprule
 Quantity & Compatible $128^3$ & Compatible $256^3$ & Local $128^3$\\
 \midrule
 Walls within $15^\circ$ of pair-specific $\{110\}$ normals
 & $83.8\%$ & $82.8\%$ & $10.8\%$\\
 Normalized Saint-Venant residual
 & $1.45\times10^{-7}$ & $1.57\times10^{-7}$ & --\\
 Relative $L^2$ field discrepancy on a common $64^3$ grid
 & \multicolumn{2}{c}{$8.17\times10^{-7}$} & --\\
 Normalized frozen-wave spectral width on the $64^3$ wave grid
 & $0.0995$ & -- & $0.1404$\\
 \bottomrule
 \end{tabular}
 \par\smallskip
 \parbox{\linewidth}{\footnotesize The residual is
 $\|P_\perp\widehat{\bm\varepsilon}P_\perp\|/
 \|\widehat{\bm\varepsilon}\|$, evaluated over nonzero Fourier modes
 with $P_\perp=\II-\mathbf n\otimes\mathbf n$.
 Spectral widths are standard deviations of the projected frequencies
 before visualization smoothing, divided by the homogeneous reference frequency.
 An em dash marks an unevaluated quantity.}
\end{table}

\FloatBarrier

\subsection{Polarization-resolved finite-temperature lattice spectra}
\label{sec:benchmark-phonon}
We test the directional rules in
Eqs.~\eqref{eq:tetragap} and~\eqref{eq:tetrprotected} with classical
molecular dynamics on a monatomic face-centred cubic bond network.
The homogeneous cell is assigned to the cubic or $z$-tetragonal
branch of Fig.~\ref{fig1}. Each trajectory evolves at fixed cell
shape. The bond energy is
$k_b\delta_b^2/2+\alpha\delta_b^4/4$ with $\alpha=120$,
$\delta_b=\mathbf n_b\cdot(\mathbf u_j-\mathbf u_i)$,
and $\mathbf n_b=\mathbf R_b/|\mathbf R_b|$ along the deformed
reference bond. The six face-diagonal bonds have
$k_b=4.5+2.10d_{\perp b}$, where $d_{\perp b}$ is the reduced
strain normal to the bond plane. The three axial bonds have
$k_b=4.5$. The harmonic frequency-squared matrix in unit mass is
\begin{equation}
 \mathbf D_0(\mathbf k)=2\sum_b k_b[1-\cos(\mathbf k\cdot\mathbf R_b)]
                   \,\mathbf n_b\otimes\mathbf n_b,\qquad
 \mathbf A_0(\mathbf n)=\lim_{k\to0}\mathbf D_0(k\mathbf n)/k^2 .
 \label{eq:bench-md-harmonic}
\end{equation}
The sum uses one bond per undirected direction per site. Dashed
dispersion curves use $\mathbf D_0$ and the directional maps use
$\mathbf A_0$. The standard simulation uses 2048 atoms in an
$8^3$ conventional-cell box. Each branch is equilibrated before
a microcanonical production run.
\rev{The original temperature series uses 4200 BAOAB Langevin
steps with unit friction, followed by velocity-Verlet production
at $\Delta t=0.012$ with 8192 frames spaced by $0.072$ model time
units. The updated $T=0.967$ point uses 24 independent states
equilibrated for 33334 steps at $\Delta t=0.003$, with unit
friction. Each production record then has 32768 frames spaced
by $0.072$, for duration $2359.296$. After removing centre-of-mass
motion, the same equilibrated state is copied to production at
$\Delta t=0.012,0.006,0.003$. The main near-transition point uses
the finest step. Mass and $\kb$ are one.}
The common path in the reduced
reciprocal coordinates of the conventional cell is
$X_x$--$\Gamma$--$X_z$--$W_z$--$L$--$\Gamma$--$K$.

Table~\ref{tab:phonon-controls} tests the two directional selection rules
over the sampled temperatures and records the principal resolution
checks at $T=0.967$. The tetragonal states give a positive
$[100]$ peak difference, while the $[001]$ estimates fluctuate near
zero. Above the assigned branch-selection temperature, both axial
differences fluctuate near zero. At $T=0.967$ the ratio of the
$[100]$ gap to the linewidth sum is approximately $0.41$.
\rev{A matched-state and matched-duration calculation now compares
24 paired replicas at three integration steps. The $[100]$ gap
difference between $\Delta t=0.006$ and $0.003$ is
$-0.0030$, with a paired 95\% bootstrap interval
$[-0.0093,0.0036]$. This interval lies within the preset
absolute tolerance $[-0.01,0.01]$ in model angular-frequency units.
The near-transition estimate therefore satisfies this finite-precision
step criterion. The earlier six-replica, short-record comparison is
superseded by this controlled calculation.}
\begin{table}[pos=!htbp]
 \centering
 \caption{Polarization-resolved MD peak differences and numerical
 controls for Fig.~\ref{fig4}, at reduced wavevector $0.5$.
 The main data use 2048 atoms. Frequencies are in model angular-frequency
 units and quoted errors are one trajectory-jackknife standard error.}
 \label{tab:phonon-controls}
 \small
 \setlength{\tabcolsep}{4pt}
 \renewcommand{\arraystretch}{1.17}
 \begin{tabular}{@{}>{\raggedright\arraybackslash}p{.28\linewidth}r r r r@{}}
 \toprule
 $T$ or control & Trajectories & Gap $[100]$ & Gap $[001]$ & FWHM sum $[100]$\\
 \midrule
$0.700$ & 5 & $0.186\pm0.011$ & $-0.007\pm0.006$ & $0.265\pm0.020$\\
$0.850$ & 5 & $0.140\pm0.012$ & $-0.018\pm0.009$ & $0.232\pm0.022$\\
$0.967$ & 24 & $0.121\pm0.002$ & $-0.003\pm0.003$ & $0.296\pm0.007$\\
$1.006$ & 5 & $-0.007\pm0.009$ & $0.002\pm0.009$ & $0.285\pm0.023$\\
$1.110$ & 5 & $0.004\pm0.007$ & $-0.012\pm0.006$ & $0.297\pm0.018$\\
$1.250$ & 5 & $-0.003\pm0.012$ & $0.004\pm0.013$ & $0.267\pm0.026$\\
 \midrule
 \multicolumn{5}{@{}l}{Controls at $T=0.967$}\\
 Bare harmonic reference & -- & $0.444$ & $0$ & --\\
 Original 6912-atom check & 3 & $0.100\pm0.024$ & $0.006\pm0.003$ & $0.308$\\
Matched states, $\Delta t=0.012$ & 24 & $0.118\pm0.002$ & $-0.000\pm0.002$ & $0.298$\\
Matched states, $\Delta t=0.006$ & 24 & $0.118\pm0.002$ & $0.004\pm0.002$ & $0.300$\\
Matched states, $\Delta t=0.003$ & 24 & $0.121\pm0.002$ & $-0.003\pm0.003$ & $0.296$\\
Half refined record & 24 & $0.119$ & $-0.002$ & $0.299$\\
 \bottomrule
 \end{tabular}
 \par\smallskip
 \parbox{\linewidth}{\footnotesize The signed gaps are
 $\omega_y-\omega_z$ along $[100]$ and $\omega_x-\omega_y$ along
 $[001]$. The tabulated FWHM values refer to the Lorentzian before
 window convolution. The original temperature points and enlarged-box
 check retain their original protocol. The refined $T=0.967$ data
 have spacing $0.00266$ and a single-mode Hann reference FWHM
 of $0.00384$. The matched-state controls all use duration
 $2359.296$ and the same sampling interval, with trajectory-jackknife
 errors. The half refined record has duration $1179.648$.
 An em dash marks an inapplicable quantity.}
\end{table}

\begin{figure}[pos=!htbp]
 \centering
 \includegraphics[width=\linewidth]{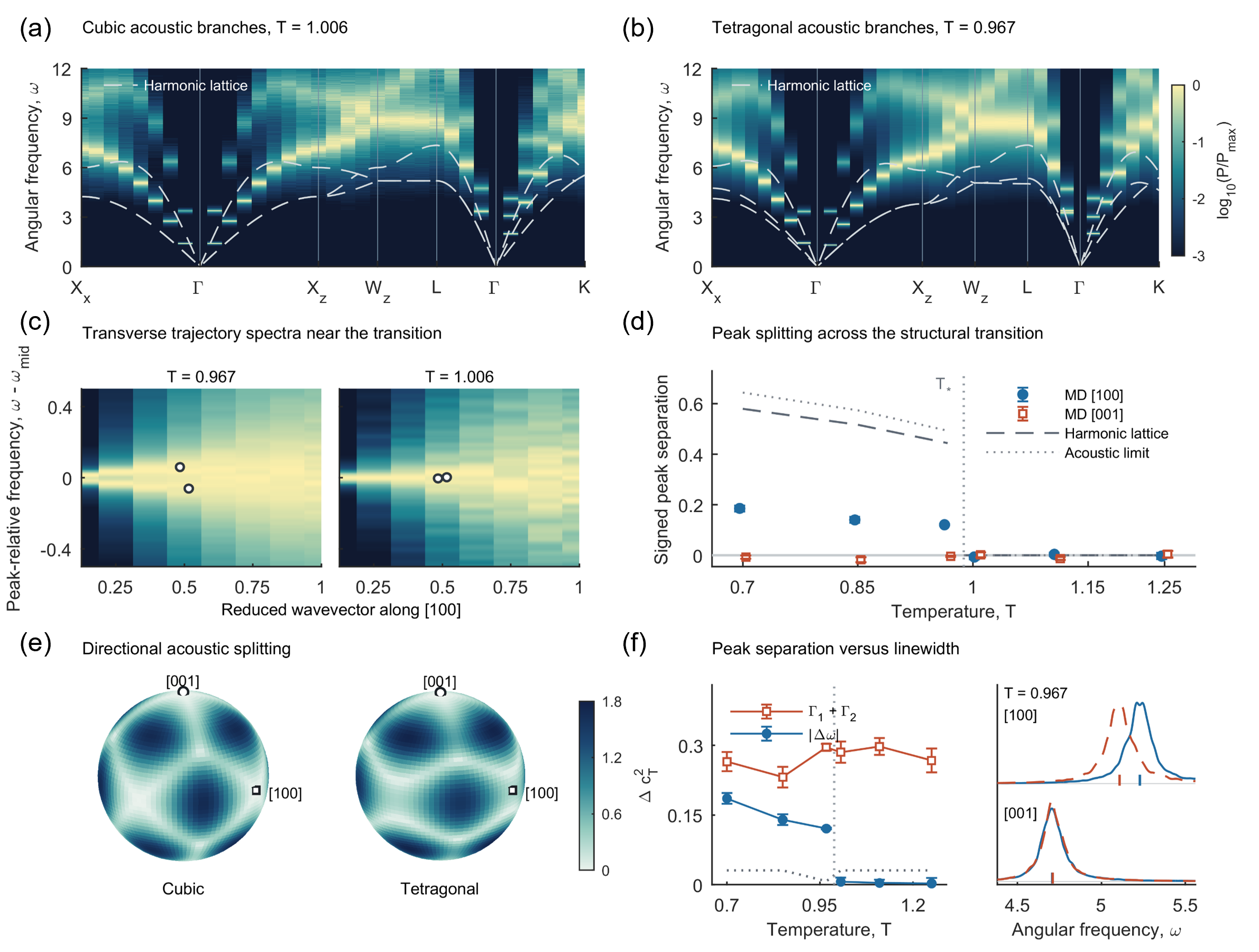}
 \caption{\textbf{Protected and lifted transverse degeneracies
 in a finite-temperature lattice model.} \textbf{(a,b)}
 Velocity power on a common reduced reciprocal-space path for
 cubic and $z$-tetragonal cells. Dashed curves are the harmonic
 eigenfrequencies. \textbf{(c)} Transverse power along $[100]$,
 centred on the fitted peak midpoint at each wavevector. Circles
 indicate the two fits at reduced wavevector $0.5$, horizontally
 offset for visibility. \textbf{(d)} Signed MD peak differences
 and the harmonic finite-wavevector and long-wave references.
 \textbf{(e)} Directional squared-speed gap of the two
 quasi-transverse acoustic branches. \textbf{(f)} Absolute
 $[100]$ gap and sum of fitted linewidths, with the summed
 record-specific Hann-window width dotted. The $T=0.967$ point
 and tetragonal spectra use the refined trajectories described in
 Table~\ref{tab:phonon-controls}. The right-hand profiles show
 independent polarizations at $T=0.967$ with fitted centres
 marked at the baselines.}
 \label{fig4}
\end{figure}

The MD observable is a polarization-resolved velocity power
spectrum \citep{ThomasSED2010}. For a unit polarization $\mathbf e$,
reference sites $\mathbf R_i$, and a Hann taper $w(t)$, its
positive-frequency contribution is proportional to
\begin{equation}
 S_{\mathbf e}(\mathbf k,\omega)
 \ \propto\ 
 \left\langle\left|\int w(t)e^{i\omega t}
      \sum_i e^{-i\mathbf k\cdot\mathbf R_i}
      \,\mathbf e\cdot\dot{\mathbf u}_i(t)\,\mathrm dt
 \right|^2\right\rangle .
 \label{eq:bench-md-power}
\end{equation}
The estimate combines both temporal frequency signs and averages
independent trajectories. The original record spacing in frequency
is $0.01065$. At $T=0.967$ the refined records give $0.00266$,
with a single-mode Hann reference width of $0.00384$.
For peak initialization, smoothing uses the larger of two bins
and $0.02$ model angular-frequency units, so that a longer record
retains a comparable physical smoothing scale. This smoothing
initializes the fit to the unsmoothed power.
Peak positions and linewidths are fitted independently to the
polarizations using Lorentzians convolved with the window response.

On the cubic branch the axial little group is $C_{4v}$ for
both directions. On the $z$-tetragonal branch, the $[001]$ little
group retains its two-dimensional transverse representation.
Along $[100]$ it reduces to $C_{2v}$, whose transverse
polarizations belong to distinct representations. These symmetry
constraints also apply to the ensemble-averaged velocity correlations
within each branch.
Figure~\ref{fig4} summarizes the resulting polarization-resolved
spectral signatures. The cubic and $z$-tetragonal dispersions in
Fig.~\ref{fig4}a,b distinguish the symmetry-protected transverse
doublet from the branch-dependent splitting along the common
reciprocal-space path. The polarization-resolved $[100]$ profiles
in Fig.~\ref{fig4}c resolve the two fitted peak centres, whose
signed difference is compared with harmonic references in
Fig.~\ref{fig4}d. The directional squared-speed gap in
Fig.~\ref{fig4}e provides the complementary long-wavelength
symmetry diagnostic. Finally, Fig.~\ref{fig4}f compares the
finite-temperature $[100]$ peak separation with the combined
linewidth and the window-resolution reference, distinguishing
a nonzero polarization-resolved splitting from two peaks that
would necessarily be resolved in a polarization-summed spectrum.
At $T=0.967$ and reduced wavevector $0.5$, the signed MD peak
difference along $[100]$ is $0.121\pm0.002$, whereas the $[001]$ difference is $-0.003\pm0.003$. The uncertainty is a
one-standard-error trajectory jackknife. Along $[100]$ the sum
of fitted linewidths is $0.296\pm0.007$. Thus the split channels
overlap in the polarization-summed spectrum. The bare
finite-wavevector splitting at the same state is $0.444$.
An enlarged box of $6912$ atoms gives $0.100\pm0.024$ for
the $[100]$ MD gap at the same reduced wavevector. \rev{These comparisons separate the symmetry constraint, the undamped
reference, and the anharmonic fitted peak positions.}

\FloatBarrier

\subsection{Independent static exponents and held-out spectral scaling}
\label{sec:benchmark-ising}
\begin{figure}[pos=!htbp]
 \centering
 \includegraphics[width=\linewidth]{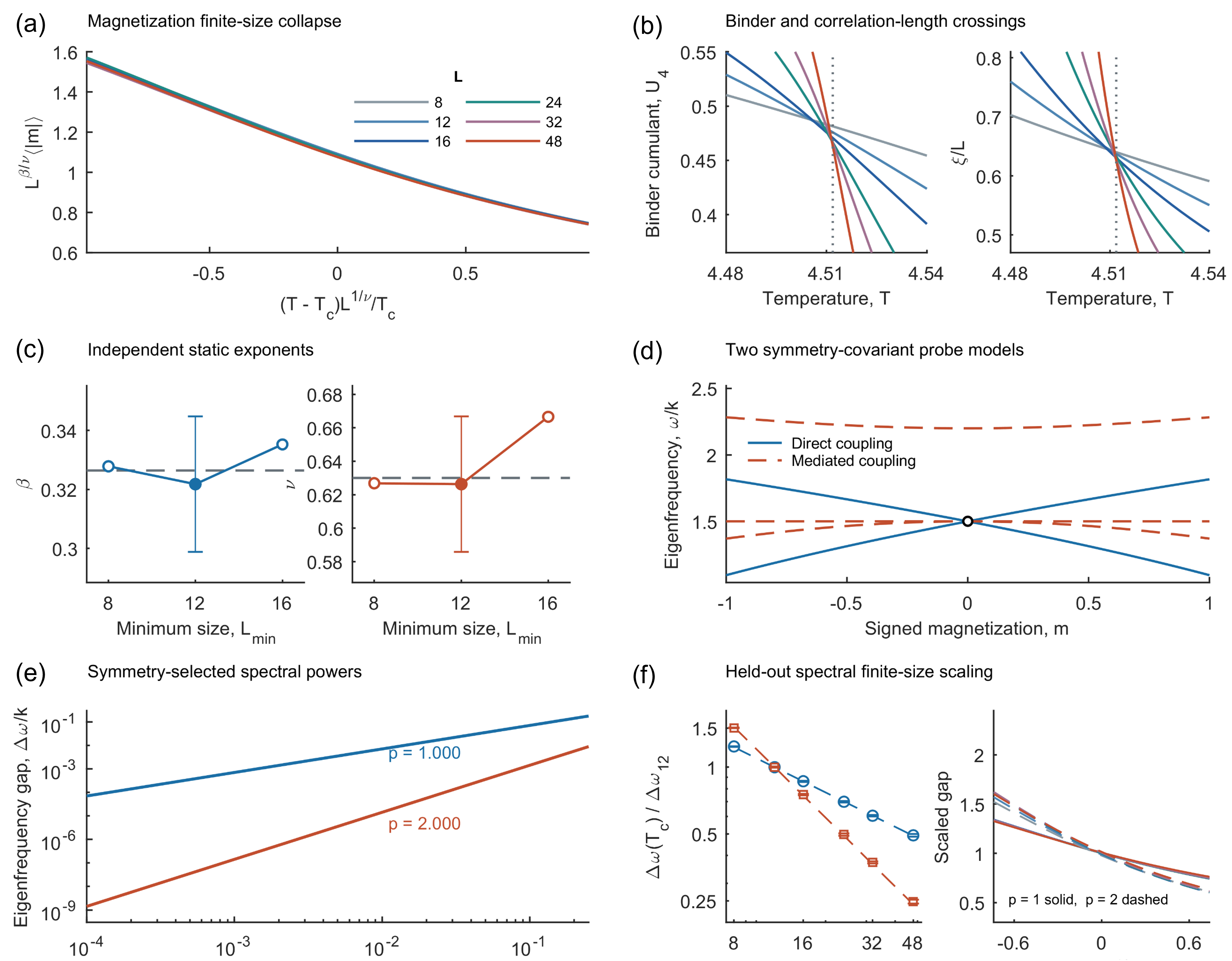}
 \caption{\textbf{Critical universality and symmetry-selected
 spectral scaling.} \textbf{(a)} Magnetization collapse using
 parameters from the static training chains. \textbf{(b)}
 Binder cumulant and correlation-length ratio, with the common
 fitted critical temperature dotted. \textbf{(c)} Static
 exponents across lower size cutoffs. The filled symbols and
 bars mark the primary estimates and bootstrap standard errors.
 \textbf{(d)} Eigenbranches of the direct and mediated probes.
 \textbf{(e)} Their transverse gaps at fixed $k$ versus
 magnetization amplitude. \textbf{(f)} Critical size dependence
 and spectral collapse on held-out chains. The dashed guides
 on the left use the static estimate of $\beta/\nu$ multiplied
 by $p=1$ or $2$.}
 \label{fig5}
\end{figure}

To test Eq.~\eqref{eq:pbeta} with non-mean-field fluctuations, we
sampled the periodic nearest-neighbour Ising model
$\mathcal H=-J\sum_{\langle i,j\rangle}s_i s_j$ in three dimensions
using Wolff updates on sizes $L=8,12,16,24,32,48$ with $J=k_{\mathrm B}=1$.
Half of the independent chains were used to determine the static
parameters, using eight of the sixteen chains at each size.
\rev{The main chains are sampled at $T_0=4.5115$ with one stored
observation every two Wolff cluster flips. Single-histogram
reweighting supplies the nearby temperatures used in the fits.}
With $m=L^{-3}\sum_i s_i$, the Binder cumulant and
second-moment length are
\begin{equation}
 U_4=1-\frac{\langle m^4\rangle}{3\langle m^2\rangle^2},\qquad
 \xi_L=\frac{1}{2\sin(\pi/L)}
     \left[\frac{S(0)}{S(k_{\min})}-1\right]^{1/2},
 \quad S(0)=L^3\langle m^2\rangle .
 \label{eq:bench-ising-binder}
\end{equation}
The slopes of $U_4$ and $\xi_L/L$ near the common crossing
scale as $L^{1/\nu}$, up to corrections. Together with
$\langle|m|\rangle_{T_c}\propto L^{-\beta/\nu}$, the fit on
$L\ge12$ gives $T_c=4.51190\pm0.00054$,
$\nu=0.626\pm0.041$, and $\beta=0.322\pm0.023$.
The fits include a leading correction proportional to $L^{-0.83}$.
The uncertainties use whole-chain resampling. Figure~\ref{fig5}c
compares different lower size cutoffs and established Ising
reference values \citep{HasenbuschIsing2010}.

\rev{The two auxiliary probes in Fig.~\ref{fig5}d are prescribed algebraic
readouts of the sampled magnetization and share a transverse doublet at
$m=0$. Their dimensionless stiffness matrices are}
\begin{equation}
 \frac{\mathbf D_{\rm A}}{k^2}=
 \begin{pmatrix}v_T^2&gm\\gm&v_T^2\end{pmatrix},
 \qquad
 \frac{\mathbf D_{\rm B}}{k^2}=
 \begin{pmatrix}
  v_T^2&0&gm\\
  0&v_T^2&0\\
  gm&0&v_L^2
 \end{pmatrix},\qquad v_L>v_T .
 \label{eq:bench-ising-probes}
\end{equation}
\rev{The numerical probes use $v_T=1.50$, $v_L=2.20$, and $g=1.05$.
Their positivity conditions are $|gm|<v_T^2$ for A and
$g^2m^2<v_T^2v_L^2$ for B. These conditions hold throughout the
sampled Ising range $|m|\le1$, so both readouts have real positive
frequencies at every sampled magnetization.}

Both obey $\mathbf D(-m)=\mathbf S\mathbf D(m)\mathbf S^{-1}$,
with $\mathbf S=\operatorname{diag}(1,-1)$ and
$\operatorname{diag}(1,1,-1)$, respectively. \rev{The equal transverse reference stiffness is imposed in both probes;
spin-inversion symmetry alone does not enforce this degeneracy.
In B, the same parity of the transverse states rules out terms odd
in $m$ in the transverse block.} Diagonalization gives
\begin{equation}
 \frac{\Delta\omega_{\rm A}}{k}=\frac{|g|}{v_T}|m|+O(|m|^3),
 \qquad
 \frac{\Delta\omega_{\rm B}}{k}=
 \frac{g^2m^2}{2v_T(v_L^2-v_T^2)}+O(m^4).
 \label{eq:bench-ising-probe-powers}
\end{equation}

For A the two squared frequencies are $k^2(v_T^2\pm gm)$.
For B, the uncoupled transverse eigenvalue is $v_T^2$, and the
lower eigenvalue of the coupled block is
\begin{equation}
 \lambda_-(m)=\frac{v_T^2+v_L^2
 -\sqrt{(v_L^2-v_T^2)^2+4g^2m^2}}{2}
 =v_T^2-\frac{g^2m^2}{v_L^2-v_T^2}+O(m^4).
 \label{eq:probe-B-eigenvalue-derivation}
\end{equation}
Taking the square roots gives Eq.~\eqref{eq:bench-ising-probe-powers}
within the positive-stiffness range. If the critical magnetization
density obeys $P_L(m)=L^{\beta/\nu}\mathcal P_*(mL^{\beta/\nu})$
up to scaling corrections, substitution of
$x=mL^{\beta/\nu}$ into the moment integral gives
$\langle|m|^p\rangle=L^{-p\beta/\nu}\int|x|^p\mathcal P_*(x)\,\dd x$.
This is the moment-level origin of the two gap exponents; it does
not replace the average gap by the gap at the averaged signed
magnetization~\citep{Xu2019}.
For the held-out chains, we average the positive gap of each
configuration at fixed $k$. Finite-size scaling then predicts
$\langle\Delta\omega\rangle/k\propto L^{-p\beta/\nu}$ at $T_c$,
with a collapse coordinate $(T-T_c)L^{1/\nu}/T_c$.
The two independently fitted size exponents are $0.492\pm0.065$
and $1.012\pm0.101$, consistent with the static predictions
$0.514$ and $1.028$. For the second channel, the measured
$\langle m^2\rangle/\langle|m|\rangle^2$ is $1.22$--$1.24$,
so taking the mean before evaluating the probe would change the
leading amplitude. The test concerns controlled eigenfrequency
gaps evaluated on fluctuating configurations. As an ordered-side
check, independent $L=32$ runs over $4.29\le T\le4.45$ give
effective temperature exponents of $0.304$ for $|m|$ and
$0.308$ and $0.587$ for the two gaps. These finite-window slopes
complement the critical size scaling.

Table~\ref{tab:critical-validation} makes the independence of the
static calibration and spectral test explicit. The static chains
supply $\beta/\nu$ and its uncertainty, while the other chains
supply the two spectral size exponents. The differences between
the measured size exponents and $p\beta/\nu$ are evaluated within
the joint whole-chain bootstrap, including the uncertainty of
$T_c$. Both differences are compatible with zero at the quoted
precision. \rev{This comparison tests the moment-level prediction in
Eq.~\eqref{eq:bench-ising-probe-powers} for two specified readouts.}

\begin{table}[pos=!htbp]
 \centering
 \caption{Independent static calibration and held-out spectral
 scaling in Fig.~\ref{fig5}. The primary fits use $L\ge12$,
 a leading $L^{-0.83}$ correction, and 500 whole-chain bootstrap
 resamples. Errors are bootstrap standard errors.}
 \label{tab:critical-validation}
 \small
 \setlength{\tabcolsep}{4pt}
 \renewcommand{\arraystretch}{1.18}
 \begin{tabular}{@{}>{\raggedright\arraybackslash}p{.37\linewidth}r r r@{}}
 \toprule
 Quantity & Static estimate & Held-out estimate & Difference\\
 \midrule
 $T_c$ & $4.51190\pm0.00054$ & -- & --\\
 $\nu$ & $0.626\pm0.041$ & -- & --\\
 $\beta$ & $0.322\pm0.023$ & -- & --\\
 $\beta/\nu$ & $0.514\pm0.012$ & -- & --\\
 Gap size exponent, $p=1$ & $0.514\pm0.012$ & $0.492\pm0.065$ & $-0.022\pm0.063$\\
 Gap size exponent, $p=2$ & $1.028\pm0.023$ & $1.012\pm0.101$ & $-0.015\pm0.097$\\
 \bottomrule
 \end{tabular}
 \par\smallskip
 \parbox{\linewidth}{\footnotesize For the gap rows, the static column
 is the prediction $p\beta/\nu$, and the difference is the
 held-out estimate minus this prediction. Covariance is retained
 when computing the difference error, and differences are evaluated
 before rounding. Eight independent chains
 per size supply each of the static and held-out groups.
 An em dash marks a quantity outside that group's assigned role.}
\end{table}

At criticality, $\langle m^2\rangle/\langle|m|\rangle^2$ spans
$1.22$--$1.24$ over the sampled sizes. Consequently, replacing
the second moment by the square of the mean amplitude would lower
the leading quadratic-response amplitude by approximately
$18$--$19\%$. Retaining the configuration-level gap preserves this
fluctuation contribution.

\FloatBarrier

\subsection{Relaxing probes and nonequilibrium critical dynamics}
\label{sec:benchmark-quench}
\begin{figure}[pos=!htbp]
 \centering
 \includegraphics[width=\linewidth]{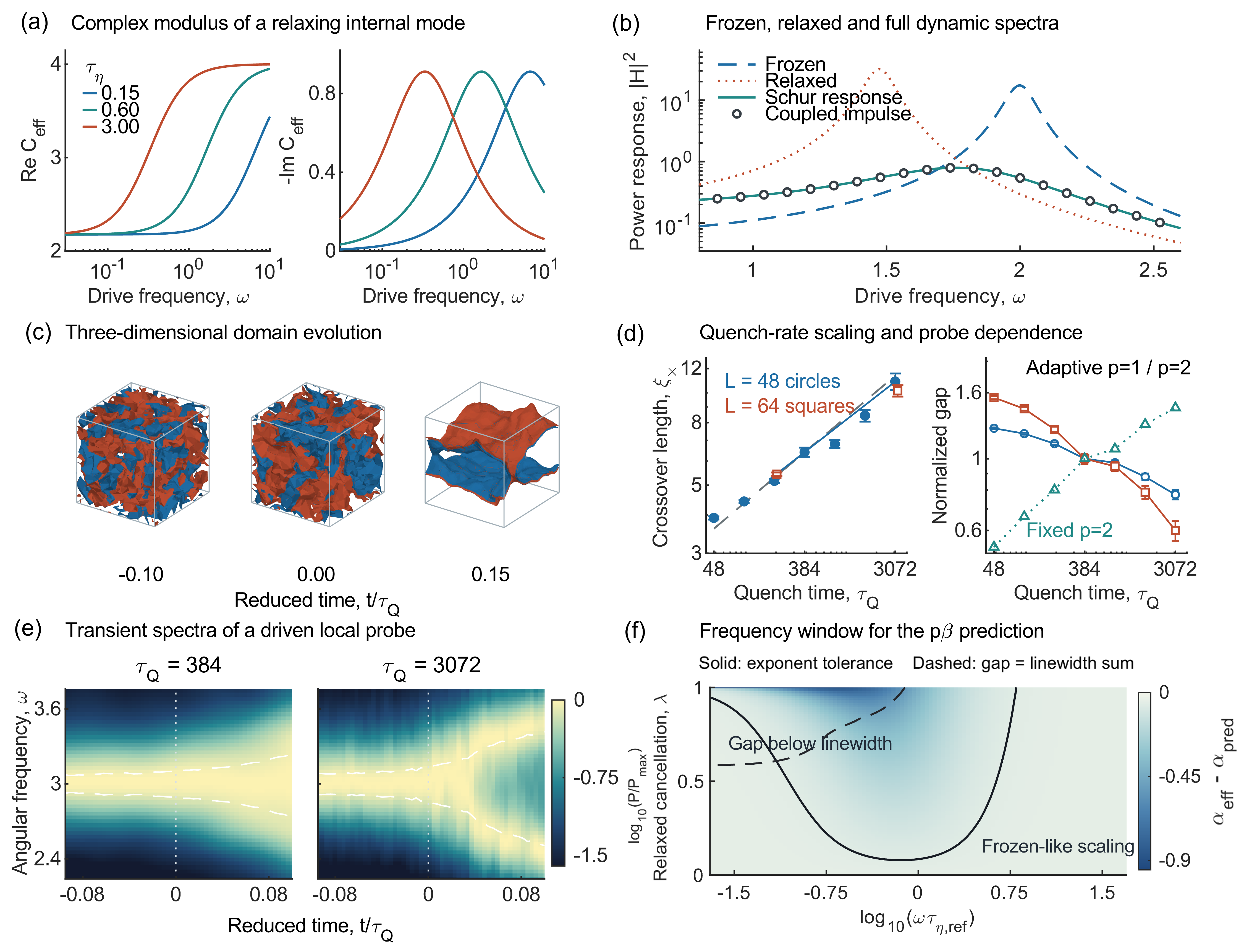}
 \caption{\textbf{Nonequilibrium critical dynamics and limits
 of the frozen response.} \textbf{(a)} Storage and loss moduli
 for three internal relaxation times. \textbf{(b)} Frozen,
 relaxed and coupled spectra, with numerical impulse results
 overlaid. \textbf{(c)} Oppositely signed coarse-grained domains
 during a three-dimensional Ising quench. \textbf{(d)}
 Crossover length at $T_c$ and normalized frozen-probe gaps
 for adaptive and fixed spatial windows. The dashed length
 guide is the Model-A Kibble--Zurek reference. \textbf{(e)}
 Short-time spectra of the driven probe velocities. White
 dashed curves show the frozen reference. \textbf{(f)}
 Exponent deviation for the mediated probe under relaxation.
 The solid contour is the prescribed $0.06$ exponent tolerance.
 The dashed contour equates the smallest storage-modulus gap proxy
 across the fitted rate interval with the prescribed linewidth sum.}
 \label{fig6}
\end{figure}

For a scalar mechanical coordinate coupled to one relaxing
internal variable, Eq.~\eqref{eq:Schur} takes the form
\begin{equation}
 C_{\rm eff}(\omega)=C_\infty-\frac{g^2}{a-i\omega/\Gamma_\eta},
 \qquad
 \tau_\eta=(a\Gamma_\eta)^{-1},\qquad
 a>0,\quad aC_\infty>g^2.
 \label{eq:bench-dynamic-modulus}
\end{equation}
Here the time convention is $e^{-i\omega t}$, so the loss
modulus is $-\operatorname{Im} C_{\rm eff}$. Figures~\ref{fig6}a,b
use $C_\infty=4$, $a=1$, $g=1.35$, and mechanical damping
$\gamma_{\rm m}=0.12$. At low frequency
the modulus tends to $C_\infty-g^2/a$ and at high frequency to
$C_\infty$. For unit mass and wave number with mechanical
damping $\gamma_{\rm m}$, the response
$H(\omega)=[C_{\rm eff}(\omega)-\omega^2-i\gamma_{\rm m}\omega]^{-1}$
agrees with the Fourier transform of the integrated coupled
impulse response in Fig.~\ref{fig6}b. The frequency changes the
effective response while the symmetry of the underlying model
remains fixed.

The kinetic Ising calculations use local nonconserved heat-bath
updates with $T(t)=T_c(1-t/\tau_Q)$.
\rev{For this driving schedule the fixed numerical value is
$T_c=4.511524$, which lies inside the uncertainty interval of the
independent static fit. The seven rates are
$\tau_Q=48,96,192,384,768,1536,3072$ sweeps. One sweep updates
every spin once, providing the time unit for this local dynamics.}
We averaged 20 independent
quenches at each of seven rates on $48^3$ sites and checked two
rates on $64^3$ sites. At $t=0$ we define $\xi_\times$ by the first
crossing of $35\%$ of the initial nonzero-shell value of
$r^{1+\eta_{\rm I}}C(r)$, where $C(r)$ is the shell-averaged spin
correlation and $\eta_{\rm I}=0.03627$ is the Ising anomalous dimension
\cite{HasenbuschIsing2010}. For $\tau_Q\ge192$ sweeps the fitted exponent of
$\xi_\times\sim\tau_Q^{\alpha_\xi}$ is $0.255\pm0.019$.
Changing the radial threshold yields $0.237$--$0.263$.
Using the static estimates above and the literature Model-A
$z=2.0245$ gives the Kibble--Zurek reference
$\nu/(1+z\nu)=0.276$ \citep{HasenbuschDynamic2020,ChandranKZ2012}.

For a linear ramp, let $\epsilon_T=(T-T_c)/T_c=-t/\tau_Q$.
Equating the equilibrium relaxation time
$\tau_0|\widehat\epsilon_T|^{-z\nu}$ to the remaining driving time
$|\widehat\epsilon_T/\dot\epsilon_T|=\tau_Q|\widehat\epsilon_T|$
gives
\begin{equation}
 |\widehat\epsilon_T|\sim(\tau_0/\tau_Q)^{1/(1+z\nu)},\qquad
 \widehat\xi\sim\tau_Q^{\nu/(1+z\nu)},\qquad
 \widehat r\sim\tau_Q^{-\beta/(1+z\nu)}.
 \label{eq:KZ-derivation}
\end{equation}
A degree-$p$ frozen readout then has the conditional rate exponent
$p\beta/(1+z\nu)$ when its spatial coarse-graining length remains
a fixed multiple of $\widehat\xi$ and the dynamic scaling regime
is reached. A fixed microscopic window probes a different moment
and need not obey this same rate law~\citep{Deutschlander2015}.

The same two frozen probes as in Eq.~\eqref{eq:bench-ising-probes}
are applied to locally averaged magnetization. For a Gaussian
width of $0.55\xi_\times$ in lattice units,
$\langle\Delta\omega\rangle\sim\tau_Q^{-\alpha}$ gives
$\alpha=0.123\pm0.010$ and $0.242\pm0.022$ for $p=1,2$.
The conditional estimates $p\beta/(1+z\nu)$ are $0.142$
and $0.284$. A fixed Gaussian width of $3$ lattice units gives $-0.204\pm0.010$
for the second probe. The sign change reflects the different
coarse-grained observable. For the transient spectra in
Fig.~\ref{fig6}e, the sampled block magnetization $m_B(t)$ drives
two damped oscillators with independent white-noise forcing with instantaneous squared frequencies
$4[2.25\pm1.05m_B(t)]$. The plotted power uses a finite-time
Hann-window transform of the velocities from the driven dynamics
over $40$ sweeps. The probe is a one-way readout of the spin dynamics.

For the two-channel parameter scan in Fig.~\ref{fig6}f, the
stiffer channel relaxes by $\lambda\delta C/(1-i\omega\tau_\eta)$,
where $\delta C>0$ is the frozen stiffness contrast and
$0\le\lambda\le1$. The storage-modulus contrast is therefore
\begin{equation}
 \Delta C'(\omega)=\delta C\left[1-
        \frac{\lambda}{1+(\omega\tau_\eta)^2}\right],\qquad
 \tau_\eta\propto \xi_\times^z.
 \label{eq:bench-dynamic-contrast}
\end{equation}
\rev{Taking the square-root difference of the two storage moduli
defines a gap proxy. This proxy is not, in general, the separation of
the peaks or complex poles of the full dissipative response.} The scan compares its fitted quench-rate
exponent with $2\beta/(1+z\nu)$ and its magnitude with a prescribed
linewidth sum.
\rev{The contours mark an exponent-tolerance criterion and a prescribed
detectability criterion for this storage-modulus proxy. Establishing
resolved peaks in a material requires evaluation of the complete
dissipative response and the measurement window.}

Table~\ref{tab:quench-validation} compares the measured rate exponents
with the conditional Kibble--Zurek predictions using the static
calibration and fixed literature value $z=2.0245$. The adaptive-window
gap exponents lie below the central reference values over this finite
rate interval. The fixed-window exponents have the opposite sign,
showing that the spatial averaging scale changes the measured
rate dependence.

\begin{table}[pos=!htbp]
 \centering
 \caption{Nonequilibrium rate exponents in Fig.~\ref{fig6}. All fits
 use $\tau_Q\ge192$ sweeps in the primary $48^3$ runs, with 20
 independent quenches per rate. Measured errors are trajectory-bootstrap
 standard errors.}
 \label{tab:quench-validation}
 \small
 \setlength{\tabcolsep}{4pt}
 \renewcommand{\arraystretch}{1.18}
 \begin{tabular}{@{}>{\raggedright\arraybackslash}p{.53\linewidth}r r@{}}
 \toprule
 Observable and averaging protocol & Measured exponent & Conditional reference\\
 \midrule
 $\xi_\times\propto\tau_Q^{\alpha_\xi}$, $35\%$ crossing
 & $0.255\pm0.019$ & $0.276\pm0.008$\\
 Gap, $p=1$, Gaussian width $0.55\xi_\times$
 & $0.123\pm0.010$ & $0.142\pm0.005$\\
 Gap, $p=2$, Gaussian width $0.55\xi_\times$
 & $0.242\pm0.022$ & $0.284\pm0.011$\\
 Gap, $p=1$, fixed Gaussian width 3
 & $-0.107\pm0.006$ & --\\
 Gap, $p=2$, fixed Gaussian width 3
 & $-0.204\pm0.010$ & --\\
 \bottomrule
 \end{tabular}
 \par\smallskip
 \parbox{\linewidth}{\footnotesize Gap exponents use
 $\langle\Delta\omega\rangle\propto\tau_Q^{-\alpha}$.
 The reference uncertainties propagate the static bootstrap samples
 with $z$ fixed. A fixed microscopic window has a distinct
 coarse-graining protocol. The em dashes mark cases outside the
 stated reference scaling hypothesis.}
\end{table}

Changing the correlation threshold from $35\%$ to $25\%$ or $50\%$
gives length exponents between $0.237$ and $0.263$. This spread
quantifies an observable-definition sensitivity in addition to the
bootstrap error. Independently, the numerical impulse response
reproduces the scalar transfer spectrum with relative error
$1.44\times10^{-5}$, checking the mechanical elimination used in
Fig.~\ref{fig6}a,b.

\FloatBarrier

\section{Discussion}
\label{sec:discussion}
\subsection{Experimental identification of the structural state}
\label{sec:inverse}

Let $\mathbf y=\mathcal O(\bm\eta,\bm\nu)$ collect measurable wave
speeds, modal polarization projectors, stiffness contractions,
thermographic observables, or diffraction-sensitive structural data.
The vector $\bm\nu$ denotes nuisance parameters such as density,
prestress, calibration coefficients, and spatial texture. Near a
regular state,
\begin{equation}
	\delta\mathbf y
	=J_\eta\,\delta\bm\eta+J_\nu\,\delta\bm\nu,
	\qquad J_\eta=\frac{\partial\mathcal O}{\partial\bm\eta}.
	\label{eq:ident}
\end{equation}
If nuisance parameters are independently fixed, full column rank
of $J_\eta$ is sufficient for first-order local distinguishability,
though not global injectivity. If nuisance parameters vary, let
$\Sigma$ be the noise covariance and define
\begin{equation}
	\widetilde J_\eta=\Sigma^{-1/2}J_\eta,\quad
	\widetilde J_\nu=\Sigma^{-1/2}J_\nu,\quad
	P_\nu=\widetilde J_\nu\widetilde J_\nu^+.
\end{equation}
The nuisance-projected information is
\begin{equation}
	\cI_{\eta\mid\nu}
		=\widetilde J_\eta^T(\II-P_\nu)
		\widetilde J_\eta.
	\label{eq:schurFisher}
\end{equation}

To derive the projection, consider a Gaussian measurement model
with parameter-independent $\Sigma\succ0$. For a trial structural
increment, nuisance optimization minimizes
$\norm{\widetilde J_\eta\delta\bm\eta+
\widetilde J_\nu\delta\bm\nu}^2$.
The fitted nuisance increment removes the component in
$\operatorname{range}\widetilde J_\nu$, leaving
$(\II-P_\nu)\widetilde J_\eta\delta\bm\eta$.
The squared residual is the quadratic form in
Eq.~\eqref{eq:schurFisher}. If the nuisance block has full column rank,
the same result is the ordinary Schur complement
\begin{equation}
 \cI_{\eta\mid\nu}=\cI_{\eta\eta}
 -\cI_{\eta\nu}\cI_{\nu\nu}^{-1}\cI_{\nu\eta},
 \quad \cI_{ab}=J_a^T\Sigma^{-1}J_b.
 \label{eq:Fisher-Schur-derivation}
\end{equation}
The projector form remains valid with redundant nuisance columns.
Its interpretation as Fisher information uses the stated Gaussian
noise model; more general likelihoods require their own information
metric. Known state constraints and unobservable orientation gauges
must be removed before applying a full-rank test~\citep{GuptaChouhan2026}.
Positive definiteness is a regular local identifiability condition;
small eigenvalues imply noise amplification. At exact modal
degeneracy the polarization basis is not unique. Use invariant
projectors and subspace distances instead of arbitrary eigenvector
labels.

A practical calibration for cubic-to-tetragonal transformation is to
track $(e_2,e_3)$ or diffraction-derived variants concurrently with
$C_{44}$, $C_{66}$, and the angular shear gap. At a first-order
transition, compare coexistence predictions with measured hysteresis,
not merely a fitted discontinuity in a normalized phase fraction.
At a continuous transition, test the spectral exponent against
structurally measured $\beta$, rather than extracting both $p$ and
$\beta$ from one gap curve.

\subsection{\rev{Assessment against NiTi measurements}}
\label{sec:niti-validation}
\begin{figure}[pos=!htbp]
 \centering
 \includegraphics[width=\linewidth]{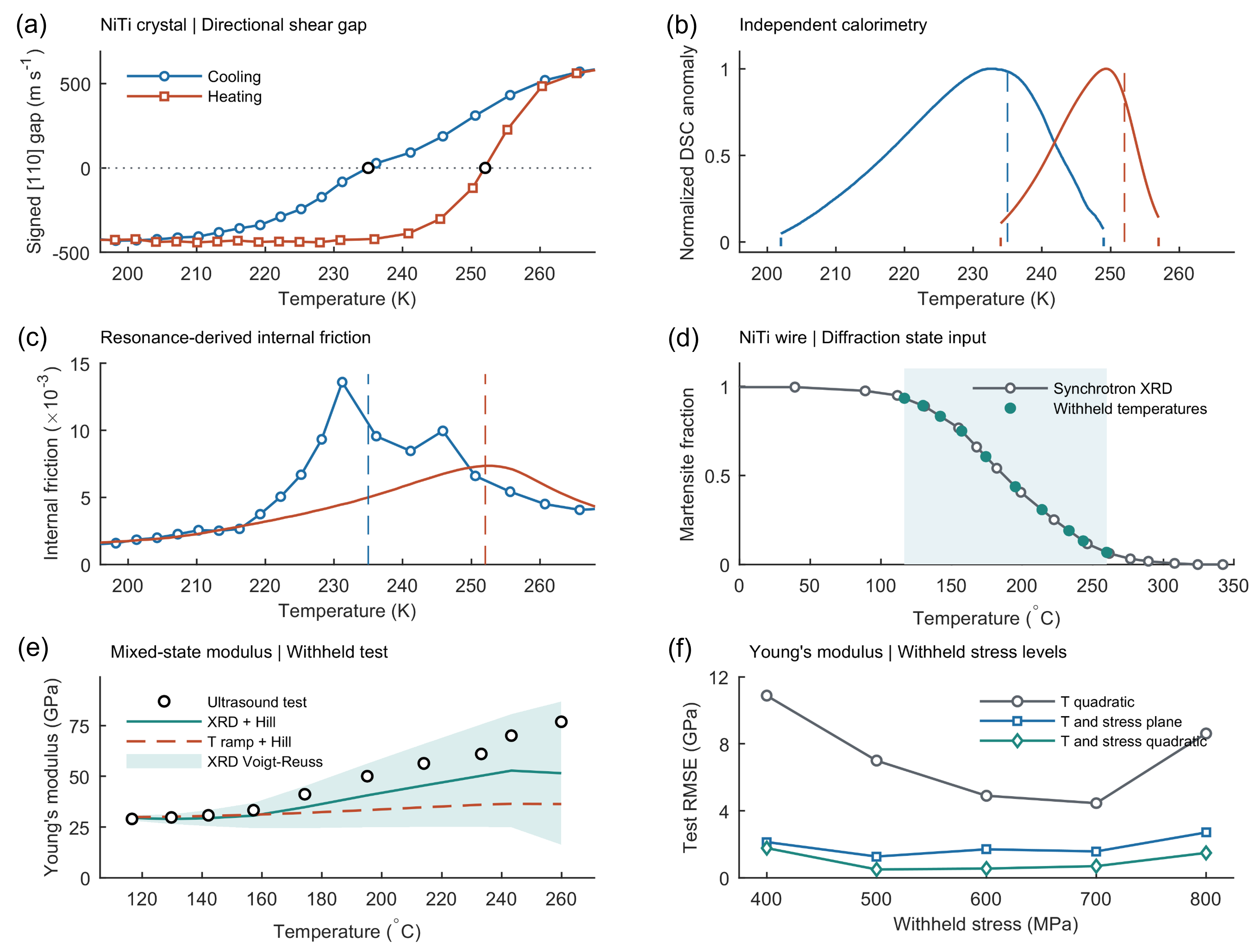}
 \caption{\textbf{Effective elastic degeneracy and state-conditioned
 response in NiTi.} \textbf{(a)} Signed $[110]$ transverse speed
 difference inferred from the two fitted shear constants, with
 interpolated zeros shown as open circles. \textbf{(b)} Same-specimen
 DSC anomalies normalized separately on cooling and heating. Short
 bars indicate the reported transformation intervals and dashed
 lines indicate the zero crossings in (a). \textbf{(c)} Internal
 friction obtained from the single-crystal resonance measurements.
 \textbf{(d)} Diffraction-derived martensite fraction from a matching
 wire experiment. Coloured markers give the fractions used at the
 withheld temperatures. \textbf{(e)} Withheld wire moduli and
 predictions using the pure-phase calibration. Shading spans the
 scalar Voigt and Reuss predictions based on diffraction.
 \textbf{(f)} Wire-modulus prediction error by withheld stress level
 for three regression baselines.}
 \label{fig7}
\end{figure}

\rev{Figure~\ref{fig7} assesses three aspects of the state-variable
description using NiTi data. The single-crystal data compare
an effective elastic degeneracy with a calorimetric transformation marker.
Two separate wire analyses assess the predictive value of phase-fraction
and loading information. These comparisons test the value of additional
state information within the stated analysis models; they do not directly
test the cubic-to-tetragonal constitutive realization or the continuous
critical relation in Eq.~\eqref{eq:pbeta}.}

We use the cooling and heating branches of the single-crystal
resonant-ultrasound data of Bodn\'arov\'a et al.~\citep{BodnarovaNiTi2025}.
For a cubic effective stiffness, the $[100]$ transverse subspace is
a doublet with squared speed $C_{44}/\rho_0$. For propagation along
$[110]$, the Christoffel eigenvectors polarized along $[001]$ and
$[1\bar10]$ have eigenvalues $C_{44}$ and $C'$, respectively, in
units of stiffness. Thus the signed velocity difference and
Zener anisotropy are
\begin{equation}
 \delta c_{[110]}=\frac{\sqrt{C_{44}}-\sqrt{C'}}{\sqrt{\rho_0}},
 \qquad C'=\frac{C_{11}-C_{12}}{2},\qquad A=\frac{C_{44}}{C'} .
 \label{eq:niti-christoffel}
\end{equation}
The equality $C_{44}=C'$ is a tuned degeneracy relative to the
cubic class. At this equality the entire fitted cubic tensor has
the isotropic Lam\'e form with $\lambda=C_{12}$ and $\mu=C_{44}$.
The stabilizer of this effective tangent can therefore exceed
the symmetry identified by a structural probe.

Interpolation of $C_{44}-C'$ gives crossings at $234.99$ and
$252.02\,\mathrm K$ on cooling and heating, respectively
(Fig.~\ref{fig7}a). Interpolating $A-1$ gives $234.75$ and
$251.63\,\mathrm K$. \rev{The difference is a consequence of interpolating two different
nonlinear functions of the sampled moduli. At an exactly resolved
temperature with $C'>0$, the conditions $C_{44}-C'=0$ and $A-1=0$
are equivalent.} The measured brackets are
$231.20$--$236.15\,\mathrm K$ and $250.19$--$255.25\,\mathrm K$.
The crossings lie within the independent, same-specimen calorimetric
intervals of $202$--$249\,\mathrm K$ on cooling and
$234$--$257\,\mathrm K$ on heating (Fig.~\ref{fig7}b).
The inferred shear speeds use the fitted elastic constants, and the
longitudinal pulse-echo data contribute to their elastic inversion.
The internal friction in Fig.~\ref{fig7}c comes from the resonances.
The calorimetry independently locates the thermal transformation
interval, while identifying local variant symmetry and phase fraction
requires structural data.

Table~\ref{tab:niti-markers} places the two elastic crossing definitions
alongside the independently measured calorimetric markers. The
cooling--heating separation is $17.03\,\mathrm K$ for the
modulus-difference interpolation and $16.88\,\mathrm K$ for the
ratio interpolation. Their shared branch dependence and their
locations inside the calorimetric transformation intervals \rev{support
the role of thermal history in the effective response.}

\begin{table}[pos=!htbp]
 \centering
 \caption{Elastic degeneracy and calorimetric markers from the
 single-crystal NiTi archive used in Fig.~\ref{fig7}a--c
 \citep{BodnarovaNiTi2025}. Temperatures are in kelvin.}
 \label{tab:niti-markers}
 \small
 \setlength{\tabcolsep}{4pt}
 \renewcommand{\arraystretch}{1.18}
 \begin{tabular}{@{}>{\raggedright\arraybackslash}p{.57\linewidth}r r@{}}
 \toprule
 Quantity & Cooling & Heating\\
 \midrule
 Elastic measurement points & 38 & 36\\
 Zero of linearly interpolated $C_{44}-C'$ & 234.99 & 252.02\\
 Zero of linearly interpolated $A-1$ & 234.75 & 251.63\\
 Measured crossing bracket & 231.20--236.15 & 250.19--255.25\\
 Reported DSC transformation interval & 202--249 & 234--257\\
 DSC anomaly peak & 232.11 & 249.35\\
 \bottomrule
 \end{tabular}
 \par\smallskip
 \parbox{\linewidth}{\footnotesize Both elastic zeros follow
 interpolation of archived measurements. The modulus-difference
 crossing is used for the markers in Fig.~\ref{fig7}a,b.
 The DSC peaks use the branch-normalized anomaly with the
 stated baseline treatment. Brackets indicate adjacent measured
 temperatures enclosing the crossing.}
\end{table}

For any single-valued prediction $f(T)$ of $A$, pointwise completion
of the square in the errors on the cooling and heating branches gives
the uniformly temperature-weighted bound
\begin{equation}
 \mathrm{RMSE}_{\mathrm{pair}}(f)
 \ge
 \left[\frac{\displaystyle\int_{T_1}^{T_2}
   [A_{\mathrm{cool}}(T)-A_{\mathrm{heat}}(T)]^2\,\mathrm dT}
   {4(T_2-T_1)}\right]^{1/2}=0.0981 .
 \label{eq:niti-history-bound}
\end{equation}
Here $T_1=165.085\,\mathrm K$ and $T_2=294.9872\,\mathrm K$
bound the common interval of the two piecewise-linear interpolants.
Equality holds for the midpoint of the branches.

Explicitly, set $A_c=A_{\rm cool}$, $A_h=A_{\rm heat}$ and
$\overline A=(A_c+A_h)/2$. Pointwise,
\begin{equation}
 \frac{(f-A_c)^2+(f-A_h)^2}{2}
 =(f-\overline A)^2+\frac{(A_c-A_h)^2}{4}.
 \label{eq:history-bound-derivation}
\end{equation}
Integrating over the common interval and dropping the first,
nonnegative term proves Eq.~\eqref{eq:niti-history-bound} for the
specified equal-branch, uniform-temperature weighting. This bound
quantifies the predictive information carried by thermal history
for these interpolated data.
\rev{Its numerical value is conditional on the specified interpolation
and temperature weighting. It is a deterministic bound for the two
measured paths, with measurement uncertainty outside the present
bound calculation.}

For the wire experiment of Sedl\'ak et al.~\citep{SedlakNiTi2025},
we fit linear martensite and austenite endmember Young's moduli
$E_M(T)$ and $E_A(T)$ to six reference-phase points each.
\rev{The martensite calibration uses the six heating points at or below
$100\,^\circ\mathrm C$, and the austenite calibration uses the six
cooling points at or above $100\,^\circ\mathrm C$. The phase fractions
are linearly interpolated to the withheld measurement temperatures.}
Ten mixed-state heating measurements are withheld. The martensite
fraction $\varphi(T)$ comes from diffraction on a matching wire
under a matched thermomechanical protocol~\citep{SittnerDiffraction2018}.
The scalar mixture closures are
\begin{equation}
 E_V = \varphi E_M + (1-\varphi)E_A,\qquad
 E_R = \big[\varphi/E_M+(1-\varphi)/E_A\big]^{-1},\qquad
 E_H = (E_V+E_R)/2 .
 \label{eq:niti-mixture}
\end{equation}

These scalar closures follow from distinct loading idealizations.
Under a common axial strain, averaging phase stresses gives $E_V$;
under a common axial stress, averaging phase strains gives $E_R$.
The Hill expression is their arithmetic mean~\citep{PaggiWriggers2012}.
For textured, constrained, or anisotropic transforming wires, these
scalar relations are illustrative approximations: rigorous elastic
homogenization uses tensor stiffness and compliance together with
the appropriate boundary conditions.
With the same endmember calibration and Hill closure, the withheld
RMSE is $11.45\,\mathrm{GPa}$ with diffraction and
$20.56\,\mathrm{GPa}$ with a fraction that decreases linearly from
one at $100\,^\circ\mathrm C$ to zero at $320\,^\circ\mathrm C$
(Fig.~\ref{fig7}d,e). The diffraction-conditioned Voigt and Reuss
RMSE values are $7.72$ and $30.02\,\mathrm{GPa}$. Extrapolating
the endmember stiffnesses with a positive exponential instead of a
linear law changes the ranking of the closures.

In a separate martensite-only test, we withhold one entire stress
group at a time across five groups spanning $400$--$800\,\mathrm{MPa}$
with 40 total observations. The pooled RMSE of the three-parameter
temperature quadratic is $7.59\,\mathrm{GPa}$. The three-parameter
temperature--stress plane gives $2.12\,\mathrm{GPa}$ and the
six-parameter quadratic surface gives $1.19\,\mathrm{GPa}$
(Fig.~\ref{fig7}f). The comparison at equal parameter count
quantifies the predictive information supplied by stress for these
conditions. The scalar closures and regressions are classical
baselines.

Table~\ref{tab:niti-prediction} summarizes two complementary withheld-data
tests. At fixed endmember calibration and Hill closure, replacing
the prescribed temperature-ramp fraction by diffraction-derived
fraction reduces the mixed-state RMSE by $44.3\%$ for linear
endmember continuation. The improvement persists with positive
exponential continuation of the martensite endmember, while the ranking of the mixture closures
changes. For the stress test, adding stress to a model with three
coefficients reduces the pooled RMSE by $72.1\%$. These comparisons
\rev{quantify the value of structural and loading information within
specified classical models.}

\begin{table}[pos=!htbp]
 \centering
 \caption{Withheld NiTi wire-modulus prediction errors for the data
 in Fig.~\ref{fig7}d--f \citep{SedlakNiTi2025,SittnerDiffraction2018}.
 RMSE values are in gigapascals.}
 \label{tab:niti-prediction}
 \small
 \setlength{\tabcolsep}{4pt}
 \renewcommand{\arraystretch}{1.16}
 \begin{tabular}{@{}>{\raggedright\arraybackslash}p{.50\linewidth}
                    >{\raggedright\arraybackslash}p{.26\linewidth}r@{}}
 \toprule
 Model or closure & State input & RMSE\\
 \midrule
 \multicolumn{3}{@{}l}{Mixed-state heating, linear endmember continuation, 10 withheld points}\\
 Hill & Temperature-ramp fraction & 20.56\\
 Hill & Diffraction fraction & 11.45\\
 Voigt & Diffraction fraction & 7.72\\
 Reuss & Diffraction fraction & 30.02\\
 \midrule
 \multicolumn{3}{@{}l}{Sensitivity check with positive exponential martensite continuation}\\
 Hill & Temperature-ramp fraction & 12.22\\
 Hill & Diffraction fraction & 1.76\\
 Voigt & Diffraction fraction & 9.36\\
 Reuss & Diffraction fraction & 10.84\\
 \midrule
 \multicolumn{3}{@{}l}{Entire-stress-group holdout, 40 observations in five folds}\\
 Temperature quadratic, 3 coefficients & $1,T,T^2$ & 7.59\\
 Temperature--stress plane, 3 coefficients & $1,T,\sigma$ & 2.12\\
 Quadratic surface, 6 coefficients & $1,T,\sigma,T^2,T\sigma,\sigma^2$ & 1.19\\
 \bottomrule
 \end{tabular}
 \par\smallskip
 \parbox{\linewidth}{\footnotesize Mixed-state tests share six
 martensite and six austenite calibration points and the same ten
 mixed-state test temperatures, spanning $116.65$--$259.91\,^\circ\mathrm C$.
 The exponential sensitivity check fits the logarithm of the martensite
 calibration moduli versus temperature. The austenite continuation is
 linear in both blocks. The diffraction data come from a matching wire experiment.
 The stress test withholds complete groups at
 $400,500,600,700,800\,\mathrm{MPa}$, preserving all observations
 at a given stress in the same fold. Its errors pool all
 withheld predictions, including the two endpoint extrapolation folds.
 The input column in the regression block lists the basis functions,
 with $\sigma$ denoting tensile stress.}
\end{table}
\FloatBarrier

\rev{The strongest closed-form results concern the frozen, homogeneous constitutive models and their direction-dependent acoustic doublets. For comparison with a physical transforming solid, one must additionally specify the thermal constraint of the measured tangent, calibrate the structural couplings, resolve domain averaging, and compare the probe period with internal relaxation times. The numerical readout models and NiTi reanalyses illuminate these requirements without completing all of them for a single material.}

A finite-dimensional local order parameter and its stabilizer cannot
completely classify topologically ordered quantum phases or every
defect-mediated transition. The $2$D XY/BKT mechanism, for example,
requires vortex physics and long-distance correlation behavior
\citep{Berezinskii1971,KosterlitzThouless1973}.
When the physical system admits meaningful topological or nonlocal
descriptors, a larger phase-state record may be written
\begin{equation}
	\cS_{\mathrm{phase}}
	=\big([H_{\bm\eta}],\cT_{\mathrm{top}},
	\cF_{\mathrm{therm}},\cD_{\mathrm{slow}},
	\cO_{\mathrm{meas}}\big),
	\label{eq:largestate}
\end{equation}
where $\cT_{\mathrm{top}}$ is a system-specific topological or
nonlocal descriptor, not a universally existing scalar invariant.
This expression is a classification program rather than a proven
one-to-one parametrization of all phases.

No single internal variable (scalar, vector, or tensor) is guaranteed
to identify all phases. Neither the group-subgroup relation nor the
form of a local Landau polynomial alone fixes microscopic energy
scales, the RG fixed point, elastic constants, nucleation rates,
or dissipation. The spectrum additionally depends on the full
constitutive operator. As a result, the most defensible target is a
common covariance/thermodynamics/observation architecture,
with explicit hypotheses under which particular cross-material
laws are valid.

\section{Conclusions}
\label{sec:conclusion}

\rev{We have formulated a symmetry-stratified description of structural
phase transitions that links an internal-state representation to an
invariant free energy, incremental elasticity, and directional acoustic
observables. The framework distinguishes orbit motion, changes of
structural symmetry, and thermodynamic transitions.

The central mechanical results are the inheritance of the structural
stabilizer by a parent-equivariant frozen tangent and the identification
of the first nonzero acoustic splitting order by degenerate perturbation
theory. For a selected continuous branch, fixed wave number, and positive
reference wave speed, these results imply $\Delta\omega\sim|t|^{p\beta}$
only when the stated regularity and observability conditions hold.
The tensor model provides an explicit reconstruction from calibrated
stiffness contractions and distinguishes linear and quadratic splitting
channels. The cubic-to-tetragonal realization contrasts a lifted
$[100]$ doublet with a protected $[001]$ doublet.

The numerical benchmarks test complementary parts of this construction.
They address invariant free-energy reconstruction, finite-size coexistence,
compatible domain patterns, and polarization-resolved spectra at fixed
cell shape. The Ising-based probes verify moment-level scaling for
prescribed readouts and demonstrate sensitivity to coarse-graining and
relaxation.

The NiTi data illustrate the distinction between effective
elastic degeneracy and structural symmetry. They quantify thermal-history
dependence, improved phase-fraction predictions within the specified Hill
comparisons, and the value of stress in the reported regression baselines.
 A material-specific test should combine independently resolved
structural states with directional, polarization-resolved spectra and
a calibrated tangent under matched loading and thermal conditions.
This defines a concrete route from the present symmetry constraints
to experimentally testable constitutive predictions.}

\section*{Declaration of Competing Interest}
The authors declare that they have no known competing financial interests or personal relationships that could have appeared to influence the work reported in this paper.
\section*{Acknowledgments}
This work was supported by the Adolf Martens Fellowship (Grant No. BAM-AMF-2025-1).
\section*{Data availability}
\rev{Data will be made available on request.}

\printcredits

\appendix
\section{\rev{Acoustic splitting in the structural-tensor model}}
\label{app:Qacoustic}
For completeness, the following calculation makes explicit why one
mechanical coupling can be spectrally silent at first order. Let
$\mathbf m=\hat{\mathbf n}$ and
$q_m=\mathbf m\cdot\QQ\mathbf m$. Contracting
Eq.~\eqref{eq:Qelasticindices} with $m_jm_l$ yields
\begin{equation}
		\delta\AAA={}\gamma_1\big[\mathbf m\otimes(\QQ\mathbf m)
		+(\QQ\mathbf m)\otimes\mathbf m\big]
		+\frac{\gamma_2}{2}\big[\QQ+q_m\II
		+(\QQ\mathbf m)\otimes\mathbf m
		+\mathbf m\otimes(\QQ\mathbf m)\big].
	\label{eq:Qacousticfull}
\end{equation}
Let $P_T=\II-\mathbf m\otimes\mathbf m$ be the unperturbed shear
projector. Since $P_T\mathbf m=0$,
\begin{equation}
	P_T\delta\AAA P_T
	=\frac{\gamma_2}{2}
	\big[P_T\QQ P_T+q_mP_T\big].
	\label{eq:PTQ}
\end{equation}
Thus the $\gamma_1$ term is exactly absent from the first-order
shear block. For uniaxial
$\QQ=s(\mathbf N\otimes\mathbf N-\II/3)$,
the orthonormal transverse basis
\begin{equation}
	\mathbf e_1=\frac{P_T\mathbf N}{\sin\vartheta},
	\qquad\mathbf e_2=\frac{\mathbf m\times\mathbf N}{\sin\vartheta}
	\qquad(\sin\vartheta\ne0)
\end{equation}
diagonalizes the block:
\begin{equation}
 (P_T\delta\AAA P_T)_{\{\mathbf e_1,\mathbf e_2\}}
 =\frac{\gamma_2s}{2}
 \begin{pmatrix}1/3&0\\0&\cos^2\vartheta-2/3\end{pmatrix}.
 \label{eq:Q-transverse-block-derivation}
\end{equation}
The difference of the two shear corrections is
$\gamma_2s\sin^2\vartheta/2$, proving
Eq.~\eqref{eq:Qgap} after division by density. At $\vartheta=0$
the chosen basis is undefined but the symmetric eigenvalue pair
remains protected by axial rotational symmetry.

For $\gamma_2=0$, the leading shear splitting originates instead
from second-order coupling between the in-plane transverse mode
and the longitudinal mode. The unperturbed longitudinal and transverse acoustic eigenvalues are
$a_L=\lambda_L+2\mu$ and $a_T=\mu$, respectively. In the basis
$(\mathbf e_1,\mathbf e_2,\mathbf m)$, the only transverse--longitudinal
matrix element is
\begin{equation}
 \mathbf e_1\cdot\delta\AAA\mathbf m
 =\gamma_1\mathbf e_1\cdot\QQ\mathbf m
 =\gamma_1s\sin\vartheta\cos\vartheta,
 \qquad \mathbf e_2\cdot\delta\AAA\mathbf m=0.
 \label{eq:Q-mixing-element-derivation}
\end{equation}
Since the source model is linear in $\QQ$ at the tangent level,
there is no direct $\AAA_2$ term. The reduced resolvent is
$R_0=-\mathbf m\otimes\mathbf m/(\lambda_L+\mu)$, so
\begin{equation}
 \delta a_{T1}^{(2)}
 =-\frac{\gamma_1^2s^2\sin^2\vartheta\cos^2\vartheta}
             {\lambda_L+\mu},\qquad
 \delta a_{T2}^{(2)}=0.
 \label{eq:Q-mixing-shift-derivation}
\end{equation}
Taking the magnitude of their difference and dividing by density
gives Eq.~\eqref{eq:QgapSecond}; the positive-energy isotropic
reference used here has $\lambda_L+\mu>0$.
The mechanism is the same virtual mixing term in
Eq.~\eqref{eq:doublet-second-order-derivation}~\citep{Chernyshev2004}. Its angular zeros differ from the
generic first-order $\gamma_2$ channel, illustrating that $p$
contains actual constitutive information.

\section{Free-energy comparison for the two-component cubic invariant}
The cubic-to-tetragonal preferred-variant polynomial is
$f(r)=ar^2/2-br^3/3+cr^4/4$.
Along a nonzero stationary branch,
$a=br-cr^2$. Inserting this identity into $f(r)$ gives
\begin{equation}
	f(r)\big|_{f'=0}
	=\frac b6r^3-\frac c4r^4.
\end{equation}
Equality with the parent branch $f(0)=0$ requires
$r_t=2b/(3c)$. Inserting into $a=br-cr^2$ produces
$a_t=2b^2/(9c)$. The nonzero roots exist for
$a\leq b^2/(4c)$, while the parent-state curvature is
$f''(0)=a$. In this elementary homogeneous setting
$a=b^2/(4c)$ is an ordered-branch saddle-node threshold,
$a=0$ is the parent-state spinodal, and $a=a_t$ is
equilibrium coexistence.

\section{Mathematical notation}
\label{app:notation}

Table~\ref{tab:notation} lists the principal symbols used in the
derivations, numerical benchmarks, and experimental analysis. The
last column identifies the relevant context when a symbol has several
meanings. The benchmarks in Figs.~\ref{fig1}--\ref{fig6} use
dimensionless model units. The NiTi temperatures and moduli in
Fig.~\ref{fig7} retain their experimental units.

\begingroup
\small
\renewcommand{\arraystretch}{1.1}
\begin{longtable}{@{}>{\raggedright\arraybackslash}p{.225\textwidth}
                     >{\raggedright\arraybackslash}p{.56\textwidth}
                     >{\raggedright\arraybackslash}p{.152\textwidth}@{}}
\caption{Principal mathematical symbols and their meanings.}
\label{tab:notation}\\
\toprule
\textbf{Symbol} & \textbf{Meaning} & \textbf{Context}\\
\midrule
\endfirsthead
\multicolumn{3}{@{}l}{\small\textbf{Table~\thetable} (continued)}\\
\toprule
\textbf{Symbol} & \textbf{Meaning} & \textbf{Context}\\
\midrule
\endhead
\midrule
\multicolumn{3}{r@{}}{\small Continued on the next page}\\
\endfoot
\bottomrule
\endlastfoot

\multicolumn{3}{@{}l}{\textbf{Structural state and symmetry}}\\*
$G_0,\ g$ & Parent symmetry group and one of its elements. & Sec.~\ref{sec:geometry}\\
$\cV,\ V_a,\ \rho(g)$ & Structural state space, its representation subspaces, and the orthogonal action of $g$. & Eq.~\eqref{eq:representation}\\
$\bm\eta,\ \bm\eta_a$ & Structural order parameter and its $a$th component or multiplet. & Sec.~\ref{sec:geometry}\\
$H_{\bm\eta},\ H_{\rm loaded}$ & Stabilizer of the structural state and residual symmetry with mechanical and other state constraints included. & Sec.~\ref{sec:geometry}\\
$\cO_{\bm\eta},\ \cV/G_0$ & Orbit of a state and space of all parent-group orbits. & Sec.~\ref{sec:geometry}\\
$[H],\ \cV_{(H)}$ & Conjugacy class of a subgroup and stratum of states with stabilizers in that class. & Eq.~\eqref{eq:orbits}\\
$\QQ,\ \mathrm{Sym}_0(3)$ & Structural tensor and space of symmetric traceless $3\times3$ tensors. & Eq.~\eqref{eq:Qspace}\\
$q_i,\ J_2,\ J_3,\ \beta_Q$ & Eigenvalues of $\QQ$, trace invariants $\tr\QQ^2$ and $\tr\QQ^3$, and biaxiality measure. & Sec.~\ref{sec:geometry}\\
$s,\ \mathbf N$ & Signed uniaxial tensor amplitude and unit ordering axis. & Sec.~\ref{sec:Qtensor}\\
$O_h,\ D_{4h},\ D_{2h},\ E_g$ & Cubic, tetragonal and orthorhombic point groups, and the even two-dimensional cubic representation. & Sec.~\ref{sec:ferroelastic}, Fig.~\ref{fig1}\\
$\bm\xi$ & Other symmetry-bearing internal variables included in the loaded-state stabilizer. & Eq.~\eqref{eq:loaded}\\
$r,\ \phi,\ e_1,e_2,e_3$ & Amplitude and angle of the $E_g$ state, followed by the dilatational strain and the two $E_g$ strain components. & Sec.~\ref{sec:ferroelastic}\\

\addlinespace
\multicolumn{3}{@{}l}{\textbf{Free energy, kinetics, and critical behavior}}\\*
$\Omega,\ \Omega_L,\ \mathbf X$ & Reference body, finite domain of linear size $L$, and material position. & Sec.~\ref{sec:framework}\\
$\mathbf u,\ \EE,\ \EE^*,\ \EE_0$ & Displacement, infinitesimal strain, transformation strain, and base-state strain. & Secs.~\ref{sec:thermal}, \ref{sec:spectrum}\\
$\cS_{\rm phase}$ & Extended phase record that includes system-specific topological and dynamical descriptors; equilibrium branch data are described in Sec.~\ref{sec:thermal}. & Eq.~\eqref{eq:largestate}\\
$\FF,\ \mathbf C$ & Deformation gradient and right Cauchy--Green tensor $\FF^T\FF$. & Sec.~\ref{sec:thermal}\\
$\cF,\ f_0,\ W,\ \cW_{\rm ext}$ & Free-energy functional, local structural density, strain energy density, and external work. & Eq.~\eqref{eq:Fgeneral}\\
$K^{ij}_{ab},\ \kappa,\ \kappa_Q$ & General gradient-energy coefficients and scalar gradient coefficients for the corresponding state fields. & Eqs.~\eqref{eq:Fgeneral}, \eqref{eq:relax}, \eqref{eq:Qcoupling}\\
$T,\ \lambda,\ \kb,\ h$ & Temperature, external control parameter, Boltzmann constant, and conjugate field selecting a structural branch. & Sec.~\ref{sec:thermal}\\
$F_{\rm MF},\ Z_L,\ f_{\rm eq}$ & Minimized mean-field free energy, finite-volume partition function, and thermodynamic-limit equilibrium free-energy density. & Sec.~\ref{sec:thermal}\\
$\bm\eta_{\rm sp},\ I_i$ & Spontaneous order parameter after branch selection, and scalar invariants of the structural state. & Sec.~\ref{sec:thermal}\\
$a,b,c,\ a_t,\ s_t,r_t$ & Quadratic, cubic and quartic Landau coefficients, the coexistence value of $a$, and ordered amplitudes at coexistence. & Sec.~\ref{sec:models}\\
$s_\pm,\ r_\pm,\ a_{\rm eff}$ & Nonzero stationary branch amplitudes and the $E_g$ quadratic coefficient after homogeneous strain relaxation. & Sec.~\ref{sec:models}\\
$\mathbf A_\eta,\ \PP_{\bm\eta}$ & Free-energy driving force and projector onto admissible structural rates. & Sec.~\ref{sec:dynamics}\\
$\mathsf M_\eta,\ \Gamma$ & Order-parameter mobility operator and scalar mobility in the linear stability model. & Sec.~\ref{sec:dynamics}\\
$n,\ \mu_n,\ M_n$ & Conserved scalar density, its chemical potential, and its mobility. & Eq.~\eqref{eq:CH}\\
$\bm\zeta,\ \bm\xi$ & Additive structural noise and conserved stochastic flux in the kinetic equations. & Sec.~\ref{sec:dynamics}\\
$\cH_\eta,\ h_a,\ \mathbf v_a$ & Structural Hessian at the stationary state, a curvature eigenvalue, and its eigenmode. & Eq.~\eqref{eq:relax}\\
$\mathbf q,\ \nu_a(\mathbf q),\ \tau_a$ & Fourier wavevector, mode growth rate, and relaxation time in the structural stability calculation. & Eq.~\eqref{eq:relax}\\
$t$ & Time in an evolution equation or a time-dependent trajectory. & Sec.~\ref{sec:dynamics}, Fig.~\ref{fig6}\\
$t,\ \lambda_c,\ \lambda_0$ & Reduced critical distance $t=(\lambda-\lambda_c)/\lambda_0$, critical control value, and fixed control scale. & Eq.~\eqref{eq:critical}\\
$\xi,\ \chi,\ \tau$ & Correlation length, susceptibility, and characteristic relaxation time. & Eq.~\eqref{eq:critical}\\
$\beta,\ \nu,\ \gamma,\ z$ & Order-parameter, correlation-length, susceptibility, and dynamic critical exponents. & Eq.~\eqref{eq:critical}\\
$B,\ \xi_0,\ \chi_0,\ \tau_0$ & Amplitudes of the critical laws for structural order, correlation length, susceptibility, and relaxation time. & Eq.~\eqref{eq:critical}\\
$k,\ R_k,\ \Gamma_k,\ \bm\phi$ & Infrared coarse-graining scale, regulator, effective average action, and its field argument. The superscript $(2)$ denotes the field Hessian. & Eq.~\eqref{eq:Wetterich}\\

\addlinespace
\multicolumn{3}{@{}l}{\textbf{Elasticity, acoustic modes, and identifiability}}\\*
$\CCt,\ C_{ijkl},\ \CCt_0$ & Fourth-order tangent stiffness, its Cartesian components, and its reference value. & Secs.~\ref{sec:spectrum}, \ref{sec:models}\\
$\CCt^T,\CCt^S,\ \mathfrak s,\ c_E,\ \bm\beta_{\rm th}$ & \rev{Isothermal and isentropic frozen tangents, entropy density, fixed-strain heat capacity per reference volume, and thermal stress coefficient.} & Eq.~\eqref{eq:thermal-tangent-conversion}\\

$G_{\CCt(\bm\eta)}$ & Subgroup preserving the complete frozen tangent. & Eq.~\eqref{eq:inherit}\\
$\rho_0,\ \mathbf k,\ k,\ \hat{\mathbf n}$ & Mass density, wavevector, wave number $k=|\mathbf k|$, and unit propagation direction. & Sec.~\ref{sec:spectrum}\\
$\AAA,\ L_{\hat{\mathbf n}}$ & Christoffel tensor in stiffness units and little group of the propagation direction. & Eqs.~\eqref{eq:acoustic}, \eqref{eq:littlegroup}\\
$\mathbf p_m,\ c_m,\ \omega_m$ & Polarization, phase speed, and angular frequency of mode $m$. & Eq.~\eqref{eq:acoustic}\\
$\varepsilon,\ \mathbf v$ & Amplitude and fixed normalized direction of the structural perturbation $\bm\eta=\varepsilon\mathbf v$. & Eq.~\eqref{eq:Aexp}\\
$P,\ P_T,\ a_0,\ a_\pm$ & Projectors onto a degenerate eigenspace and the reference transverse subspace, the reference acoustic eigenvalue, and the two perturbed eigenvalues. & Sec.~\ref{sec:spectrum}, App.~\ref{app:Qacoustic}\\
$D,\ \mathcal A,\ p$ & Leading acoustic eigenvalue-gap coefficient, critical frequency-gap prefactor, and first nonzero spectral splitting order. & Eqs.~\eqref{eq:ap}, \eqref{eq:pbeta}\\
$\Delta\omega,\ \Delta c_T^2$ & Frequency separation and absolute difference of the two squared transverse speeds. & Secs.~\ref{sec:spectrum}, \ref{sec:models}\\
$\mathsf K_{uu},\ \mathsf K_{\eta\eta}$ & Mechanical and structural diagonal blocks of the coupled frequency-domain operator. & Eq.~\eqref{eq:Schur}\\
$\mathsf K_{u\eta},\ \mathsf K_{\eta u}$ & Off-diagonal blocks coupling displacement and internal-state perturbations. & Eq.~\eqref{eq:Schur}\\
$\mathsf K_{\rm eff}$ & Effective mechanical operator obtained by eliminating the internal-state perturbation. & Eq.~\eqref{eq:Schur}\\

$R_0,\ P_\perp,\ B_1,B_2$ & Reduced resolvent, complementary polarization projector, and first two effective doublet coefficients. & Eqs.~\eqref{eq:reduced-resolvent-derivation}, \eqref{eq:doublet-second-order-derivation}\\
$\mathbf e,\mathbf z,\ C_\infty,B,H,M,\ L_\omega$ & Strain and structural increments, frozen stiffness, cross coupling, structural Hessian and mobility, and dynamic structural block. & Eqs.~\eqref{eq:coupled-quadratic-derivation}--\eqref{eq:dynamic-passivity-derivation}\\
$\mathbf D_E,\ \gamma_E,\ W_{\rm an}$ & Diagonal representative of the canonical $E_g$ doublet, explicit cubic elastic coupling, and its energy. & Eqs.~\eqref{eq:Eg-basis-derivation}, \eqref{eq:Eg-explicit-elastic-coupling}\\
$m_0,\ \Gamma_Q$ & Positive reference energy bound and scalar structural-tensor mobility. & Eqs.~\eqref{eq:Q-stability-bound}, \eqref{eq:Q-projected-evolution}\\

$\lambda_L,\ \mu,\ \gamma_1,\gamma_2$ & Lam\'e moduli and the two structure--elasticity coupling coefficients of the tensor model. & Eq.~\eqref{eq:Qcoupling}\\
$\mathbf D,\ \mathbf V,\ a_D,a_V$ & Dilatation and Voigt contractions of the tangent and coefficients linking their deviators to $\QQ$. & Eq.~\eqref{eq:contractions}\\
$\vartheta,\ \mathbf m,\ q_m$ & Angle between wave normal and ordering axis, unit wave normal in the derivation, and its quadratic contraction with $\QQ$. & App.~\ref{app:Qacoustic}\\
$K_E,\ g,\ \mathcal B_a,\ B_E$ & Bare $E_g$ strain modulus, bilinear strain--order coupling, covariant elastic bilinear forms, and leading shear-stiffness contrast coefficient. & Sec.~\ref{sec:ferroelastic}\\
$C_{11}, C_{12}, C_{44}, C_{66}$ & Voigt normal, cross-normal, out-of-plane shear and in-plane shear stiffnesses in the indicated cubic or tetragonal axes. & Sec.~\ref{sec:ferroelastic}, Fig.~\ref{fig7}\\
$\mathbf y,\ \mathcal O,\ \bm\nu$ & Measured observables, observation map, and nuisance-parameter vector. & Sec.~\ref{sec:inverse}\\
$J_\eta,J_\nu,\ \Sigma$ & Sensitivity matrices for structural and nuisance parameters, and measurement noise covariance. & Sec.~\ref{sec:inverse}\\
$\widetilde J_\eta,\widetilde J_\nu$ & Sensitivity matrices premultiplied by $\Sigma^{-1/2}$ to whiten measurement noise. & Sec.~\ref{sec:inverse}\\
$P_\nu,\ \cI_{\eta\mid\nu}$ & Projector onto the whitened nuisance sensitivity space and structural information after projecting out that space. & Eq.~\eqref{eq:schurFisher}\\

\addlinespace
\multicolumn{3}{@{}l}{\textbf{Numerical benchmarks and NiTi data}}\\*
$s,\ \mathbf B,\ \mathbf q,\ d_j$ & Reference strain scale, orthonormal $E_g$ basis matrix, \rev{dimensionless strain doublet $(e_2,e_3)/s=(\eta_2,-\eta_1)/s$, and reduced diagonal strain} $(\mathbf B\mathbf q)_j$. & Figs.~\ref{fig1}, \ref{fig3}\\
$N,\ E_0,\ U,\ K_j,\ K_0$ & Site count, reference energy, static bond energy, polarization-dependent spring stiffness, and reference stiffness. & Fig.~\ref{fig1}\\
$t,\ t_*,\ F_t,\ P_t,\ I_3$ & Reduced temperature $\kb T/E_0$, equal-depth temperature, reconstructed free energy, strain probability density, and cubic invariant $\sum_j d_j^3$. & Fig.~\ref{fig1}\\
$\mathcal H,\ J,\ L,\ \mathbf q_i$ & Lattice Hamiltonian, nearest-neighbour coupling, linear lattice size, and site $i$ strain doublet in the $E_g$ model. & Figs.~\ref{fig2}, \ref{fig5}\\
$\overline{\mathbf q},\ \mathbf h$ & Lattice-average strain doublet and uniform field in the analytic stratum-crossing example. & Fig.~\ref{fig2}\\
$T_L,\ P_L(E),\ \Delta F_L$ & Equal-basin-weight temperature, energy histogram at that temperature, and barrier inferred from its peaks and valley. & Fig.~\ref{fig2}\\
$E_1,E_2,E_{\rm v},\ \sigma$ & Peak and valley energy coordinates and interfacial free energy per unit area. & Fig.~\ref{fig2}\\
$c_V,\ \chi_m$ & Heat capacity per site and fluctuation susceptibility of the global strain amplitude in the $E_g$ lattice. & Fig.~\ref{fig2}\\
$\mathbf A,\ \mathbf Q,\ \mathbf G,\ \mathbf P$ & Strain projection map, auxiliary acoustic matrix, projected compliance, and projector onto accessible Fourier strain components. & Eq.~\eqref{eq:bench-compat-matrices}\\
$\mathbf U,\ F_{\rm compat},\ d_{4h}$ & Nonlocal compatibility kernel, associated elastic energy, and distance to the closure of the tetragonal stratum. & Fig.~\ref{fig3}\\
$\lambda,\ \mu,\ \mu_0,\ \gamma$ & Auxiliary Lam\'e moduli, frozen reference shear modulus, and linear strain-to-shear coefficient in the wave calculation. & Fig.~\ref{fig3}\\
$\mathbf v,\ \mathbf v_j$ & Scaled Fourier displacement amplitude in the compatibility constraint and unit vectors along the three tetragonal variant directions. & Fig.~\ref{fig3}\\
$\mathbf K,\ \mathcal D$ & Frozen, density-scaled discrete wave operator and discrete symmetric gradient. & Fig.~\ref{fig3}\\
$k_b,\ \alpha,\ \delta_b$ & Harmonic bond stiffness, quartic bond coefficient, and projected relative displacement of bond $b$. & Fig.~\ref{fig4}\\
$\mathbf R_b,\ \mathbf n_b,\ \mathbf R_i$ & Reference bond vector, its unit direction, and reference position of atom $i$ in the assigned cell. & Fig.~\ref{fig4}\\
$\mathbf D_0,\ \mathbf A_0$ & Harmonic lattice dynamical matrix and the acoustic limit of that matrix divided by $k^2$. & Eq.~\eqref{eq:bench-md-harmonic}\\
$S_{\mathbf e},\ \mathbf e,\ w(t)$ & Polarization-resolved velocity power, unit probe polarization, and Hann time window. & Eq.~\eqref{eq:bench-md-power}\\
$s_i,\ m,\ T_c$ & Ising spin, magnetization per site, and critical temperature. & Figs.~\ref{fig5}, \ref{fig6}\\
$U_4,\ S(k),\ \xi_L$ & Binder cumulant, static spin structure factor, and finite-size second-moment correlation length. & Eq.~\eqref{eq:bench-ising-binder}\\
$\mathbf D_{\rm A},\mathbf D_{\rm B},\ \mathbf S$ & Direct and mediated probe matrices and the corresponding spin-inversion operator on mode space. & Fig.~\ref{fig5}\\
$v_T,v_L,\ g$ & Reference transverse and longitudinal speeds and the probe's magnetization-to-stiffness coupling. & Fig.~\ref{fig5}\\
$C_{\rm eff},\ C_\infty,\ H(\omega)$ & Complex dynamic modulus, its frozen limit, and the damped mechanical transfer function. & Fig.~\ref{fig6}\\
$a,\ g,\ \Gamma_\eta,\ \tau_\eta$ & Internal-mode curvature, mechanical--internal coupling, mobility, and structural relaxation time of the scalar probe. & Eq.~\eqref{eq:bench-dynamic-modulus}\\
$\gamma_{\rm m},\ \lambda,\ \delta C,\ \Delta C'(\omega)$ & Mechanical damping, relaxing fraction of the frozen stiffness contrast, that contrast, and the resulting storage-modulus difference. & Fig.~\ref{fig6}\\
$\tau_Q,\ \xi_\times,\ \eta_{\rm I},\ C(r)$ & Quench time scale, crossover length at critical passage, Ising anomalous dimension, and shell-averaged spin correlation. & Fig.~\ref{fig6}\\
$\alpha_\xi,\ \alpha,\ m_B(t)$ & Quench-rate exponents for the crossover length and probe gap, followed by the time-dependent block magnetization. & Fig.~\ref{fig6}\\
$C',\ A,\ \delta c_{[110]}$ & Cubic shear modulus $(C_{11}-C_{12})/2$, Zener anisotropy $C_{44}/C'$, and signed transverse-speed difference along $[110]$. & Eq.~\eqref{eq:niti-christoffel}\\
$A_{\rm cool},A_{\rm heat},\ T_1,T_2$ & Anisotropy on the two thermal branches and endpoints of their common temperature interval. & Eq.~\eqref{eq:niti-history-bound}\\
$\mathrm{RMSE}_{\rm pair}(f)$ & Root-mean-square error of a single-valued anisotropy prediction evaluated against both thermal branches. & Eq.~\eqref{eq:niti-history-bound}\\
$\varphi,\ E_M,E_A,\ E_V,E_R,E_H$ & Martensite volume fraction, phase endmember Young's moduli, and Voigt, Reuss and Hill estimates of the wire modulus. & Eq.~\eqref{eq:niti-mixture}\\

\addlinespace
\multicolumn{3}{@{}l}{\textbf{Operators and statistical conventions}}\\*
$\II,\ \tr,\ \Tr,\ \sym,\ \dev$ & Identity, finite tensor trace, functional trace in the scale-flow equation, symmetric part, and deviatoric part. & Throughout\\*
$:\ ,\ \otimes,\ \|\cdot\|$ & Tensor contraction, tensor product, and norm. & Throughout\\*
\({}^{T},\ {}^\dagger,\ {}^+,\ \widehat{(\cdot)}\) & Transpose, Hermitian adjoint, Moore--Penrose pseudoinverse, and Fourier coefficient for a field. A hat on a direction denotes normalization. & Throughout\\*
$\langle\cdot\rangle,\ \operatorname{Var}$ & Ensemble or sample average in the stated protocol, and variance. & Throughout\\*
$\partial,\ \delta,\ \nabla$ & Partial derivative, variation in the variational equations, and spatial gradient. A dot denotes a time derivative. & Throughout\\*
$O(\cdot),\ o(\cdot),\ \sim$ & Term bounded by a constant times its argument, asymptotically smaller term, and leading asymptotic scaling in the critical laws. & Throughout\\
\end{longtable}
\endgroup










\end{document}